\documentclass[11pt]{article}
\usepackage[margin=1in]{geometry}

\usepackage{placeins}
\usepackage{caption}
\usepackage{subcaption}
\usepackage{graphicx, color, url}
\usepackage{amsmath,amsthm}
\usepackage{dsfont}
\usepackage{amssymb}
\usepackage{amsfonts}
\usepackage{url}
\usepackage{amsbsy}
\usepackage{setspace}
\usepackage{bm}
\usepackage{textcomp}
\usepackage{booktabs}
\usepackage{rotating}

\usepackage[
    backend=biber,
    style=numeric-comp,
    sortcites,
    sorting=none,
    giveninits=true,
    natbib,
    hyperref=false,
    maxbibnames=99,
    doi=false,isbn=false,url=false,eprint=false
]{biblatex}

\usepackage[colorinlistoftodos]{todonotes}

\usepackage{graphicx}
\usepackage{amssymb}
\usepackage{amsmath}
\usepackage{amsbsy}
\usepackage{mathtools}
\usepackage{bbm}
\usepackage{algpseudocode}
\usepackage{algorithm}

\newtheorem{theorem}{Theorem}
\newtheorem{proposition}{Proposition}
\newtheorem{assumption}{Assumption}

\renewcommand{\P}[1]{\mathbb{P}\left[#1\right]}
\newcommand{\E}[1]{\mathbb{E}\left[#1\right]}
\newcommand{\I}[1]{\mathbbm{1}\left[#1\right]}

\makeatletter
\newcommand*\bigcdot{\mathpalette\bigcdot@{.5}}
\newcommand*\bigcdot@[2]{\mathbin{\vcenter{\hbox{\scalebox{#2}{$\m@th#1\bullet$}}}}}
\makeatother

\begin{document}
\title{Integrative conformal p-values for powerful out-of-distribution testing with labeled outliers}

\author{Ziyi Liang\thanks{Department of Mathematics, University of Southern California, Los Angeles, CA, USA.}, Matteo Sesia\thanks{Department of Data Sciences and Operations, University of Southern California, Los Angeles, CA, USA.}, Wenguang Sun\thanks{Center for Data Science, Zhejiang University, Hangzhou, Zhejiang, China.} \footnote{Authors listed alphabetically. For correspondence: \url{sesia@marshall.usc.edu}}}
\maketitle

\begin{abstract}
This paper develops novel conformal methods to test whether a new observation was sampled from the same distribution as a reference set. Blending inductive and transductive conformal inference in an innovative way, the described methods can re-weight standard conformal p-values based on dependent side information from known out-of-distribution data in a principled way, and can automatically take advantage of the most powerful model from any collection of one-class and binary classifiers. The solution can be implemented either through sample splitting or via a novel transductive cross-validation+ scheme which may also be useful in other applications of conformal inference, due to tighter guarantees compared to existing cross-validation approaches. After studying false discovery rate control and power within a multiple testing framework with several possible outliers, the proposed solution is shown to outperform standard conformal p-values through simulations as well as applications to image recognition and tabular data.
\end{abstract}

\section{Introduction}

\subsection{Conformal inference for out-of-distribution testing}

This paper considers the problem of testing whether a new observation was randomly sampled from the same unknown distribution as a reference data set~\cite{grubbs1969procedures}, allowing for high-dimensional features and avoiding parametric assumptions.
This problem---known as {\it novelty detection}, {\it testing for outliers}, or {\em out-of-distribution testing}---arises naturally from many applications in medical diagnostics, security monitoring, fraud detection, and natural language processing, to name a few examples~\cite{pimentel2014review}. 
Different models have been proposed to address this task, ranging from parametric ones~\cite{markou2003novelty} to complex machine learning alternatives, including random forests~\cite{desir2013one}, deep neural networks~\cite{markou2003novelty}, nearest neighbor classifiers~\cite{hautamaki2004outlier}, and support vector machines~\cite{clifton2014probabilistic}.
As opposed to classical parametric models, which rely on assumptions that do not always fit the data well, machine learning algorithms are very flexible and can achieve high power in diverse settings, but they are not designed to estimate uncertainty and they typically do not give any guarantees in finite samples.
The latter is a significant practical limitation, as uncertainty is often inherent in novelty detection applications and needs to be accounted whenever incorrect predictions can have costly consequences.
For instance, medical images are sometimes hard to parse even for experienced physicians, as certain malignant features may be difficult to visually tell apart from benign ones. Machine learning algorithms may help streamline those initial screenings, reducing health care costs, but a lack of precise statistical inferences might make practitioners more hesitant to take advantage of these tools.
Fortunately, the field of conformal prediction~\cite{vovk1999machine,vovk2005algorithmic} offers promising solutions to this challenge.

Conformal prediction can produce rigorous inferences in finite samples for any machine learning model, assuming only that the data are {\em exchangeable}~\cite{vovk1999machine,vovk2005algorithmic}. 
This assumption is similar to saying the test point is sampled i.i.d.~from the same distribution as the reference set, although it is technically weaker.
In the context of testing for outliers, conformal inference is useful to obtain a valid p-value for testing the null hypothesis that the new data point is an {\em inlier}~\cite{bates2021testing,haroush2021statistical}.
The standard procedure for this relies on data splitting and begins by training a one-class classification model~\cite{moya1993one,khan2014one,pimentel2014review,sabokrou2018adversarially} on a random subset of the reference inliers. The conformal p-value is then intuitively determined by the relative rank of the one-class {\em conformity score} (i.e., a measure of similarity between the data point and training inliers) output by the fitted model for the test point among the corresponding scores for the hold-out inliers; e.g.,~\cite{bates2021testing}. 
Although this implicitly assumes only inlier data are available, the same idea can be easily extended to leverage labeled outliers. For example, an appealing solution is to replace the one-class model with a binary classifier trained on all outliers and a subset of inliers~\cite{vovk2003mondrian,haroush2021statistical}, and then proceed as above to rank the binary conformity score (e.g., the estimated probability of being an inlier) for the test point among those for the hold-out inliers.
However, as this paper will explain, that is not necessarily an optimal way of leveraging outlier data.

\subsection{Labeled outlier data: to use or not to use?}

\begin{figure}[!htb]
    \centering
    \includegraphics[width=0.75\linewidth]{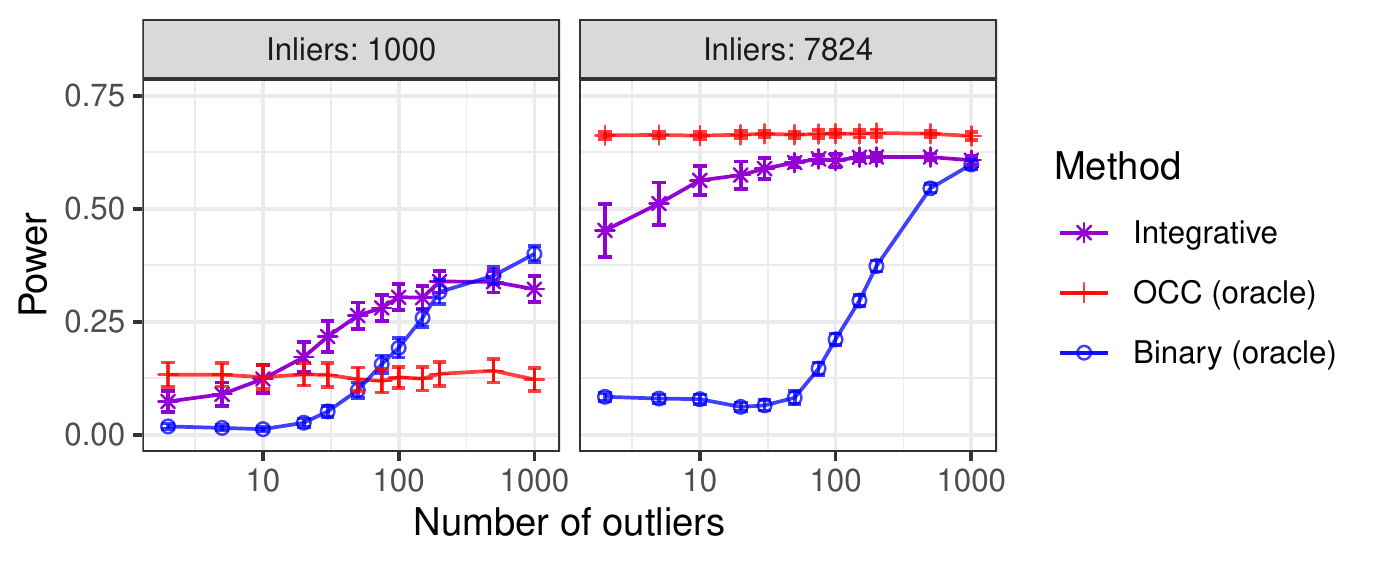}
    \caption{Performance of conformal p-values for out-of-distribution
testing with image data, in terms of power under false discovery rate control within a multiple testing setting. The results are shown as a function of the number of labeled inliers and outliers. The oracles are impractical but are useful to approximate the power of standard conformal p-values based on the optimal one-class or binary classifier from a rich machine learning toolbox. The integrative p-values often perform almost as well as those of the best oracle, and can sometimes even outperform both of them.}
    \label{fig:exp-animals}
\end{figure}

Labeled outliers are often available in real-world applications and may contain valuable information.
For example, large financial institutions likely have numerous records of confirmed illicit transactions. Even if these fraudulent activities only account for a relatively small share of all transactions, many of them may share similar patterns that could be useful to help flag future suspicious activities.
Similarly, physicians are likely to have CT scans from past patients with rare tumors, although healthy samples may be much more abundant.
In both cases, the labeled outliers may be informative about future out-of-distribution samples and thus one should intuitively try to leverage them.
At the same time, however, it is unclear whether: (i) the straightforward solution of computing standard conformal p-values based on binary classification scores is optimal, and (ii) the latter is even preferable to the one-class approach that ignores the labeled outliers.
The underlying issue is that off-the-shelf one-class classifiers can sometimes be much more powerful at separating outliers than binary classifiers, especially if the data are high-dimensional and involve class imbalance.
This phenomenon---visualized in Figure~\ref{fig:exp-animals} through experiments with real data---motivates our proposal of a novel {\it integrative} method for calculating conformal p-values with labeled outlier data that can automatically combine the individual strengths of the two alternative approaches described above, and sometimes even simultaneously outperforms both of them.

It would be premature to anticipate here all details behind Figure~\ref{fig:exp-animals}, and thus those explanations are postponed to Section~\ref{sec:exp-synthetic}. 
For the time being, Figure~\ref{fig:exp-animals} serves three purposes: (i) to highlight some perhaps surprising limitations of the most intuitive approach for incorporating labeled outliers into conformal p-values, (ii) to emphasize these issues cannot be simply resolved by switching to a different machine learning model, and (iii) to preview the novel {\em integrative} solution described in this paper.
The crucial aspect of Figure~\ref{fig:exp-animals} is that it compares the performance of integrative p-values to those of two {\em imaginary oracles} based on standard conformal p-values.
These oracles are designed to pick for each setting the {\em best model} leading to the highest power over 100 independent experiments, choosing from a rich machine learning toolbox, or collection, of one-class or binary classification algorithms; see Section~\ref{sec:exp-synthetic} for details. 
The oracles are obviously only imaginary, as they require knowing the ground truth, but they are useful benchmarks in simulations because they give us some idea of the highest power practically achievable by standard conformal p-values.
By contrast, the proposed integrative method does not enjoy unfair advantages here compared to any other real data application.

The relative strength of integrative conformal p-values partly derives from their ability to leverage outlier data to automatically select the most powerful model from any machine learning toolbox. However, Figure~\ref{fig:exp-animals} already suggests this is not the only innovation presented in this paper, because the integrative method can sometimes outperform standard conformal p-values even if the latter are based on the best possible model in the available toolbox.
Yet, integrative conformal p-values are able to carry out data-driven model selection without losing their theoretical guarantees, and this would already be a valuable contribution by itself. In fact, different machine learning models (e.g., a neural network vs.~a random forest, or variations thereof based on diverse hyper-parameters) may exhibit inconsistent relative performances on distinct data sets (this is also demonstrated explicitly in Section~\ref{sec:exp-synthetic}), and there is no practical way of knowing in advance which one will work best for a particular outlier testing application.
This behaviour makes the flexibility of the conformal inference framework look somewhat under-utilized if one has to commit in advance to a fixed model.
Further, flexibility without clear guidance tends to be confusing, and it may even involuntarily encourage dangerous ``data snooping'' heuristics, such as naively trying different models and then picking that leading to the discovery of the largest number of likely out-of-distribution samples, for example.
Unlike the aforementioned heuristic, which we will demonstrate empirically to yield severely inflated type-I errors, our solution can select near-optimal models while remaining theoretically valid in finite samples regardless of the size or variety of the toolbox of models considered.

\subsection{Outline of this paper}

Section~\ref{sec:weighted} contains our main methodological contributions. To begin, Section~\ref{sec:standard-conformal} defines useful technical notation and reviews the relevant background on standard conformal p-values based on sample splitting.
Then, Section~\ref{sec:adaptive-weighting} introduces the first core idea of our proposed method, which integrates in a principled manner a {\em dependent} pair of standard conformal p-values based on one-class classifiers respectively trained on labeled inliers and outliers.
Ensuring the finite-sample validity of these integrative p-values requires an innovative conformal calibration technique, which goes beyond the standard recipe based on fixed conformity score functions~\cite{vovk2005algorithmic,bates2021testing}.
Section~\ref{sec:auto-tuning} extends the integrative method to take advantage of an entire machine learning toolbox of one-class and binary classifiers, automatically selecting in a data driven way the most promising model that approximately maximizes power. 
Section~\ref{sec:adaptive-weighting-cv+} develops a more sophisticated version of these method based on a novel type of cross-validation, which is more computationally expensive than data splitting but often also more powerful, while remaining provably valid in finite samples.

Sections~\ref{sec:fdr} and~\ref{sec:asymptotic} study from a more theoretical perspective two interesting aspects of integrative p-values.
In Section~\ref{sec:fdr}, we consider a multiple testing problem in which many unlabeled data points are available and our goal is to discover which of them are likely outliers~\cite{bates2021testing,mary2021semi}, controlling the false discovery rate (FDR)~\cite{benjamini1995controlling}. This problem is interesting because integrative conformal p-values for different test points are not independent of one another, and their relatively complicated correlation structure makes it unclear whether powerful FDR control methods such as the prominent Benjiamini-Hochberg procedure (BH)~\cite{benjamini1995controlling} remain valid~\cite{benjamini2001control}.
We show the FDR control results of~\cite{bates2021testing} do not immediately extend to integrative conformal p-values but this challenge can be overcome with the conditional FDR calibration strategy of~\cite{fithian2020conditional}. The latter method can be hard to implement in general but applies naturally in our context.
In Section~\ref{sec:asymptotic}, we compare theoretically the power of integrative conformal p-values under FDR control to that of their standard alternatives, focusing for simplicity on the data splitting implementation based on one-class classifiers.
In particular, we prove that integrative p-values can be more powerful if the labeled outliers are sufficiently informative in an appropriate sense.
Finally, in Sections~\ref{sec:exp-synthetic} and~\ref{sec:exp-real}, we explore empirically the performance of integrative p-values and compare it to that of standard conformal p-values through extensive numerical experiments with simulated and real data.

\subsection{Related work} \label{sec:preview}


This paper builds upon prior work on {\em inductive} conformal p-values for out-of-distribution
testing~\cite{laxhammar2015inductive,smith2015conformal,ishimtsev2017conformal,guan2019prediction,cai2020real,haroush2021statistical,bates2021testing,haroush2021statistical}, expanding the broader conformal inference literature~\cite{vovk1999machine, vovk2005algorithmic,angelopoulos2021gentle}.
Our main innovation consists of allowing the available data to include labeled outliers and leveraging this information in a principled way to improve power by gathering strength from an toolbox of different classification models through an entirely novel calibration method. 
This solution is inspired by prior work on weighted hypothesis testing~\cite{genovese2006false,hu2010false,efron2012large,ignatiadis2016data,li2019multiple,ignatiadis2021covariate,Caietal20,liang2022locally,basu2018weighted,benjamini1997weights,roquain2008optimal}, but it departs from that classical framework as it deals with non-independent side information. Further, this solution also differs from prior approaches based on adaptive sequential tests with side information~\cite{li2017accumulation,lei2018adapt} as it can produce valid inferences for one test point at a time.
Two versions of our method are developed: one based on sample splitting~\cite{lei2018distribution} and one based on cross-validation+~\cite{barber2019predictive}. The latter is a more sophisticated and parsimonious data hold-out scheme originally developed by~\cite{barber2019predictive} for regression and later extended to multi-class classification by~\cite{romano2020classification}. In this paper, we modify their approach to address certain technical challenges arising in our problem, specifically by borrowing previously unrelated ideas from {\em transductive} conformal inference~\cite{vovk2013transductive}.
Interestingly, we will show that our transductive cross-validation+ can also be useful to guarantee tighter coverage in other types of conformal inference applications, including regression~\cite{lei2014distribution,lei2018distribution,chernozhukov2019distributional, kivaranovic2019adaptive, kuchibhotla2019nested,izbicki2019flexible,romano2019conformalized,sesia2021conformal} and multi-class classification~\cite{vovk2003mondrian,hechtlinger2018cautious,sadinle2019least,cauchois2020knowing,romano2020classification,angelopoulos2020uncertainty}.

Conformal inference was originally intended to calibrate a pre-trained machine learning model~\cite{vovk2005algorithmic}, but a few recent works have also proposed algorithms for conformalized model training; e.g., \cite{colombo2020training,einbinder2022training}. We similarly move beyond relying on a single pre-trained model, but we study how to perform model selection and hyper-parameter tuning with exact guarantees as opposed to training black-box classifiers.
Several other papers have utilized model ensembles to improve the performance of conformal predictors \cite{lofstrom2013effective,beganovic2018ensemble,linusson2020efficient,kim2020predictive,gupta2022nested,fannjiang2022conformal} or conformal out-of-distribution tests~\cite{xu2021conformal}, but they took different approaches. Those works focused on {\it aggregating} the output of several simpler models and evaluated conformity scores based on out-of-bag predictions, in the fashion of a random forest for example, while we aim to {\em select} and {\em tune} the best performing model from any given toolbox.
Other lines of prior work have developed solutions to strengthen the marginal validity guarantees for conformal inferences, conditioning for example on the hold-out calibration data~\cite{vovk2012conditional,bates2021testing}, on the observed features of the test point~\cite{barber2019limits}, or on certain protected attributes~\cite{romano2019malice}. Those papers focus on orthogonal issues but they are relevant insofar as their results could in principle be combined with our method to strengthen its theoretical guarantees. We leave it to future research to explore the technical details and performance trade-offs related to that idea.

Finally, there exist other statistical frameworks for obtaining rigorous inferences from the output of complex machine learning models in the context of testing for outliers, including in particular the Neyman-Pearson classification paradigm~\cite{rigollet2011neyman,tong2018neyman,tong2013plug,zhao2016neyman}.
However, the latter takes a different perspective compared to conformal inference as it focuses on controlling both type-I and type-II errors, and it is typically restricted to specific algorithms~\cite{rigollet2011neyman}, assumes large sample sizes~\cite{tong2018neyman}, or requires other assumptions such as low data dimensions~\cite{tong2013plug} or feature independence~\cite{zhao2016neyman}.

\section{Integrative conformal p-values} \label{sec:weighted}

\subsection{Review of standard conformal p-values} \label{sec:standard-conformal}

Consider a data set containing $n$ observations, $(X_i,Y_i)$ for $i \in [n] = \{1,\ldots,n\}$, sampled from an arbitrary and unknown distribution $P_{X,Y}$. Each $X_i \in \mathbb{R}^d$ is a vector of $d$ features for the $i$-th individual, whose label is $Y_i \in \{0,1\}$.
In this paper, we refer to individuals with $Y_i=0$ as {\it inliers} and to those with $Y_i=1$ a {\it outliers}.
The problem is to test whether a new observation with features $X_{n+1}$ is an inlier, in the sense that it was drawn exchangeably from the same distribution $P_{X \mid Y =0}$ as $\mathcal{D}_0 = \{ i \in [n] : Y_i = 0\}$. In short, we will refer to this null hypothesis as $\mathcal{H}_0: Y_{n+1}=0$.
Note that the inliers are assumed to be sampled exchangeably with one another (or, for simplicity, one could think of them a being i.i.d.), but no assumptions are needed about the exchangeability of the outliers.
Yet, it may sometimes be useful to also think of the data with $Y_i=1$ as being {\em approximately} exchangeable, including with the test point under the alternative hypothesis $\mathcal{H}_1: Y_{n+1}=1$, because that explains why leveraging them may help increase power.
Formally, our goal is to compute a powerful and {\em marginally} valid p-value $\hat{u}(X_{n+1})$ that is guaranteed to be super-uniform under $\mathcal{H}_0$, i.e., for all $\alpha \in (0,1)$,
\begin{align} \label{eq:super-uniform}
  \P{\hat{u}(X_{n+1}) \leq \alpha \mid Y_{n+1} = 0} \leq \alpha.
\end{align}
The p-value $\hat{u}(X_{n+1})$ is said to be {\em marginally} valid because the probability above is taken over $X_{n+1}$ as well as the labeled inliers indexed by $\mathcal{D} = \mathcal{D}_0 \cup \mathcal{D}_1$.
Although treating the labeled inliers as random does not lead to the strongest possible guarantees, especially from a multiple testing perspective~\cite{bates2021testing}, marginal p-values are undoubtedly useful as a first principled solution for calibrating the output of complex machine learning algorithms. Further, they enable rigorous {\em average} control of different type-I errors, including the FDR~\cite{bates2021testing}.

Assuming the labeled data contain only inliers, the standard sample-splitting approach randomly partitions $\mathcal{D}_0$ into two disjoint subsets: $\mathcal{D}_0^{\mathrm{train}}$ and $\mathcal{D}_0^{\mathrm{cal}}$.
The data in $\mathcal{D}_0^{\mathrm{train}}$ are utilized to train a one-class classifier~\cite{moya1993one,pimentel2014review}, whose purpose is to learn a suitable {\em conformity score} function $\hat{s}_0 : \mathbb{R}^d \to \mathbb{R}$. This function is designed to approximately capture the observed distribution of inliers and thus indirectly highlight possible inliers seen at test time. In particular, a smaller value of $\hat{s}_0(X)$ should suggest $X$ is a likely outlier.
Importantly, this model may involve complex machine learning algorithms and does not need to offer any precise guarantees in finite samples.
This is where conformal inference comes in: by carefully comparing the score $\hat{s}_0(X_{n+1})$ of a new test point to the empirical distribution of $\hat{s}_0(X_i)$ for the hold-out calibration data indexed by $i \in \mathcal{D}_0^{\mathrm{cal}}$, it becomes possible to obtain rigorous statistical inferences about the unknown $Y_{n+1}$ label.
More precisely, a conformal p-value function $\hat{u}_0$ is defined such that, for any $x \in \mathbb{R}^d$,
\begin{align} \label{eq:p-value-0}
  \hat{u}_0(x) = \frac{|i\in \{n+1\} \cup \mathcal{D}_{0}^{\mathrm{cal}} : \hat{s}_0(X_i) \leq  \hat{s}_0(x)|}{1+|\mathcal{D}_{0}^{\mathrm{cal}}|}.
\end{align}
Intuitively, $\hat{u}_0(X_{n+1})$ is the normalized rank of $\hat{s}_0(X_{n+1})$ among the scores $\hat{s}_0(X_i)$ for the hold-out data points with indices $i \in \mathcal{D}_0^{\mathrm{cal}}$.
It follows quite easily from the inlier exchangeability assumption that $\hat{u}_0(X_{n+1})$ is a valid conformal p-value for testing $\mathcal{H}_0: Y_{n+1}=0$.
\begin{proposition}[e.g., from~\cite{vovk2003mondrian} or \cite{bates2021testing}] \label{prop:standard-p-values}
  If the inliers in $\mathcal{D}_0^{\mathrm{cal}}$ are exchangeable among themselves and with $X_{n+1}$, then $\P{\hat{u}_0(X_{n+1}) \leq \alpha \mid Y_{n+1} = 0} \leq \alpha$ for all $\alpha \in (0,1)$.
\end{proposition}

As anticipated in the introduction, an important question arises at this point: if some labeled outliers are available, how should they be utilized to increase power?
The most intuitive solution is to include them in the training set, which thus becomes $\mathcal{D}_0^{\mathrm{train}} \cup \mathcal{D}_1$, and then fit a binary classification model instead of a one-class classifier. Conformal p-values can then be constructed by applying~\eqref{eq:p-value-0} with conformity scores $\hat{s}_0(x)$ based on the output of the binary classifier~\cite{vovk2003mondrian}; e.g., the estimated probability of a new sample with features $X=x$ having label $Y=0$.
That approach is technically sound---the output p-value still satisfies~\eqref{prop:standard-p-values} under the same inlier exchangeability assumption---and quite appealing in its simplicity, but it does not always work well, as previewed earlier in Figure~\ref{fig:exp-animals}. 
This limitation motivates the novel {\it integrative} method presented below.
 

\subsection{Integrative conformal p-values with data-driven weighting} \label{sec:adaptive-weighting}

This section introduces a first version of our integrative method based on a single pair of one-class classifiers separately trained on the labeled inlier and outlier data.
This solution will later be extended to automatically take advantage of any number of one-class or binary classifiers.
Let us begin by randomly splitting the labeled inliers into $\mathcal{D}_0^{\mathrm{train}}$ and $\mathcal{D}_0^{\mathrm{cal}}$, and the labeled outliers indexed by $\mathcal{D}_1$ into two disjoint subsets $\mathcal{D}_1^{\mathrm{train}}$ and $\mathcal{D}_1^{\mathrm{cal}}$.
The inliers in $\mathcal{D}_0^{\mathrm{train}}$ and $\mathcal{D}_0^{\mathrm{cal}}$ are utilized to compute a preliminary standard conformal p-value $\hat{u}_0(X_{n+1})$ for the test point $X_{n+1}$. 
In particular, the first one-class classifier is fitted on $\mathcal{D}_0^{\mathrm{train}}$, giving a conformity score function $\hat{s}_0$.
The function $\hat{s}_0$ is calibrated on $\{n+1\} \cup \mathcal{D}_0^{\mathrm{cal}}$ as usual, yielding $\hat{u}_0(X_{n+1})$ via~\eqref{eq:p-value-0}.
Similarly, the outliers in $\mathcal{D}_1^{\mathrm{train}}$ and $\mathcal{D}_1^{\mathrm{cal}}$ are utilized to train a score function $\hat{s}_1$ and compute a preliminary standard conformal p-value $\hat{u}_1(X_{n+1})$ for the opposite null hypothesis that $Y_{n+1}=1$, by applying
\begin{align} \label{eq:p-value-1}
  \hat{u}_1(x) = \frac{1 + |j \in \mathcal{D}_{1}^{\mathrm{cal}} : \hat{s}_1(X_j) \leq  \hat{s}_1(x)|}{1+|\mathcal{D}_{1}^{\mathrm{cal}}|}.
\end{align}

Note that $\hat{u}_1(X_{n+1})$ is not a provably valid conformal p-value for testing whether $Y_{n+1}=1$ unless the outlier data are also exchangeable.
However, this additional assumption is not required for any of our formal results. That being said, we will continue to informally call $\hat{u}_1(X_{n+1})$ a ``conformal p-value'' because we find it helpful to think of it as such.
In fact, despite being generally dependent on $\hat{u}_0(X_{n+1})$ and not exactly valid, $\hat{u}_1(X_{n+1})$ should intuitively be useful to strengthen or weaken the evidence against the target null hypothesis. Further, $\hat{u}_1(X_{n+1})$ may not be very far from being uniformly distributed in practice if $Y_{n+1}=1$ and the outliers do not experience too much distributional shift~\cite{tibshirani2019conformal}.
It must then be emphasized that it is reasonable to compare the values of $\hat{u}_0(X_{n+1})$ and $\hat{u}_1(X_{n+1})$ to one another when weighting the evidence provided by the inlier and outlier data, because both of these statistics are principled measures of statistical confidence. In particular, they are (at least approximately valid) conformal p-values, which can only take values between 0 and 1 and are known to follow (approximately) a uniform distribution under their respective null hypotheses. By contrast, it is not clear at all how one could otherwise directly compare the raw conformity scores $\hat{s}_0(X_{n+1})$ and $\hat{s}_1(X_{n+1})$ output by different black-box machine learning models trained on different data sets.

Having distilled separate sources of information into $\hat{u}_0(X_{n+1})$ and $\hat{u}_1(X_{n+1})$, we will integrate the evidence contained in these conformal p-values through a combination function $\hat{r}$ that returns a scalar statistic.
This function could take any form in principle, but for we will focus on the intuitive
\begin{align} \label{eq:ratio-stats}
  \hat{r}(x) = \frac{\hat{u}_0(x)}{\hat{u}_1(x)}.
\end{align}
This choice is inspired by the classical strategies for p-value weighting~\cite{genovese2006false,hu2010false}, but $\hat{r}(X_{n+1})$ cannot be directly treated as a legitimate weighted p-value~\cite{ignatiadis2021covariate} because $\hat{u}_0(X_{n+1})$ is not independent of the data-driven weight $1/\hat{u}_1(X_{n+1})$.
Therefore, we need to develop an innovative weighting approach designed to deal with the specific type of dependence arising in our conformal framework.
In a nutshell, we will use $r(x)$ as a score function to re-calibrate a new conformal p-value, in such a way that exchangeability arguments can be applied to prove the validity of the final output. Fortunately, this solution will not introduce any more significant computational burdens.

More precisely, let us go back to the computation of the standard conformal p-value $\hat{u}_0(X_{n+1})$ based on the pre-trained score function $\hat{s}_0$ and let us also calculate analogous p-values $\hat{u}_0(X_{i})$ by applying~\eqref{eq:p-value-0} after swapping $X_{n+1}$ with $X_i$, for all $i \in \mathcal{D}_0^{\mathrm{cal}}$.
Similarly, we also go back to the computation of the standard conformal p-value $\hat{u}_1(X_{n+1})$ based on the pre-trained score function $\hat{s}_1$ and calculate analogous p-values $\hat{u}_1(X_{i})$ by applying~\eqref{eq:p-value-1} after swapping $X_{n+1}$ with $X_i$, for all $i \in \mathcal{D}_0^{\mathrm{cal}}$.
Next, we can apply~\eqref{eq:ratio-stats} to define $\hat{r}(X_i)$ for all $i \in \mathcal{D}_0^{\mathrm{cal}}$.
Finally, the same conformalization strategy as in~\eqref{eq:p-value-0} is deployed to re-calibrate an {\em integrative} conformal p-value $\hat{u}(X_{n+1})$:
\begin{align} \label{eq:p-value-ratio}
  \hat{u}(X_{n+1}) = \frac{1 + |i\in \mathcal{D}_{0}^{\mathrm{cal}} : \hat{r}(X_i) \leq \hat{r}(X_{n+1})|}{1+|\mathcal{D}_{0}^{\mathrm{cal}}|}.
\end{align}
This procedure, summarized in Algorithm~\ref{alg:weighted-pvalues}, provably yields a valid conformal p-value.
\begin{theorem} \label{theorem:ratio-exchangeable}
  If $X_{n+1}$ is an inlier and is exchangeable with the inliers in $\mathcal{D}_0^{\mathrm{cal}}$ and $\hat{u}(X_{n+1})$ is computed by Algorithm~\ref{alg:weighted-pvalues}, then $\P{\hat{u}(X_{n+1}) \leq \alpha \mid Y_{n+1} = 0} \leq \alpha$ for all $\alpha \in (0,1)$.
\end{theorem}

Theorem~\ref{theorem:ratio-exchangeable} is a novel result and it does not follow from Proposition~\ref{prop:standard-p-values}, although its proof (in Appendix~\ref{appendix:proofs}, with all other proofs) involves similar exchangeability arguments.
The main novelty is that the statistics $\hat{r}(X_i)$ in~\eqref{eq:p-value-ratio} do not fit the typical definition of conformity scores because they make use of a combination function $\hat{r}$ that depends not only on the training data but also on $\mathcal{D}_0^{\mathrm{cal}}$ and $X_{n+1}$.
This solution shares some similarity with transductive conformal inference \cite{vovk2013transductive}, which can also be utilized to obtain conformal p-values. However, the two approaches are different because we do not look at $X_{n+1}$ while training the machine-learning models $\hat{s}_0$ and $\hat{s}_1$, and this makes our method computationally efficient even if the number of test points is very large.

\begin{algorithm}[H]
    \caption{Integrative conformal p-values}
    \label{alg:weighted-pvalues}
    \begin{algorithmic}[1]
        \State \textbf{Input}: inlier data $\mathcal{D}_0$, outlier data $\mathcal{D}_1$, one-class classifiers $\mathcal{A}_0, \mathcal{A}_1$ and test point $X_{n+1}$.
        \State Randomly split $\mathcal{D}_0$ into two disjoint subsets, $\mathcal{D}_0^{\mathrm{train}}, \mathcal{D}_0^{\mathrm{cal}}$.
        \State Randomly split $\mathcal{D}_1$ into two disjoint subsets, $\mathcal{D}_1^{\mathrm{train}}, \mathcal{D}_1^{\mathrm{cal}}$.
        \State Train $\mathcal{A}_0$ on $\mathcal{D}_0^{\mathrm{train}}$ and $\mathcal{A}_1$ on $\mathcal{D}_1^{\mathrm{train}}$.
        \State Calculate conformity scores $\hat{s}_0(X_i)$ based on $\mathcal{A}_0$, for all $i \in \{n+1\} \cup \mathcal{D}^{\mathrm{cal}}_0$.
        \State Compute standard conformal p-values $\hat{u}_0(X_i)$ with~\eqref{eq:p-value-0}, for all $i \in \{n+1\} \cup \mathcal{D}^{\mathrm{cal}}_0$.
        \State Calculate conformity scores $\hat{s}_1(X_i)$ based on $\mathcal{A}_1$, for all $i \in \{n+1\} \cup \mathcal{D}^{\mathrm{cal}}_0 \cup \mathcal{D}^{\mathrm{cal}}_1$.
        \State Compute standard conformal p-values $\hat{u}_1(X_i)$ with~\eqref{eq:p-value-1}, for all $i \in \{n+1\} \cup \mathcal{D}^{\mathrm{cal}}_0$.
        \State Combine $\hat{u}_0(X_i)$ and $\hat{u}_1(X_i)$ into $\hat{r}(X_i)$ with~\eqref{eq:ratio-stats}, for all $i \in \{n+1\} \cup \mathcal{D}^{\mathrm{cal}}_0$.
        \State \textbf{Output}: conformal p-value $\hat{u}(X_{n+1})$ computed with~\eqref{eq:p-value-ratio}.
    \end{algorithmic}
\end{algorithm}

\subsection{Integrative conformal p-values with automatic model selection} \label{sec:auto-tuning}

This section further develops the above method for computing integrative conformal p-values in order to endow it with automatic model selection and tuning capabilities.
This is a useful extension which can significantly boost power without invalidating the theoretical guarantee of Theorem~\ref{theorem:ratio-exchangeable}.
The idea is to train an {\em toolbox} of different classifiers on $\mathcal{D}_0^{\mathrm{train}}$ and $\mathcal{D}_1^{\mathrm{train}}$, and then to evaluate integrative conformal p-values through~\eqref{eq:ratio-stats} and~\eqref{eq:p-value-ratio} based on the most promising model based on its out-of-sample performance on $\mathcal{D}_{1}^{\mathrm{cal}}$ or $\mathcal{D}_{0}^{\mathrm{cal}}$, respectively.
For example, one may intuitively want to select the pair of models leading to the smallest (e.g., median) values of the calibration scores $\hat{s}_0(X_i)$ for $i \in \mathcal{D}_{1}^{\mathrm{cal}}$ and $\hat{s}_1(X_i)$ for $i \in \mathcal{D}_{0}^{\mathrm{cal}}$, although this is of course not the only possible criterion one could follow.
Our method can deal with arbitrarily diverse toolbox of models, which may include various families of classifiers (e.g., random forests, neural networks, Bayesian models, etc.) as well as multiple instances of the same model trained with different hyper-parameters.
The key to rigorously retain the marginal validity of the output p-values is simply to utilize a model selection criterion that is invariant to the order of the observations in $\{n+1\} \cup \mathcal{D}_{0}^{\mathrm{cal}}$.

\begin{figure}[!htb]
    \centering
    \includegraphics[width=\linewidth]{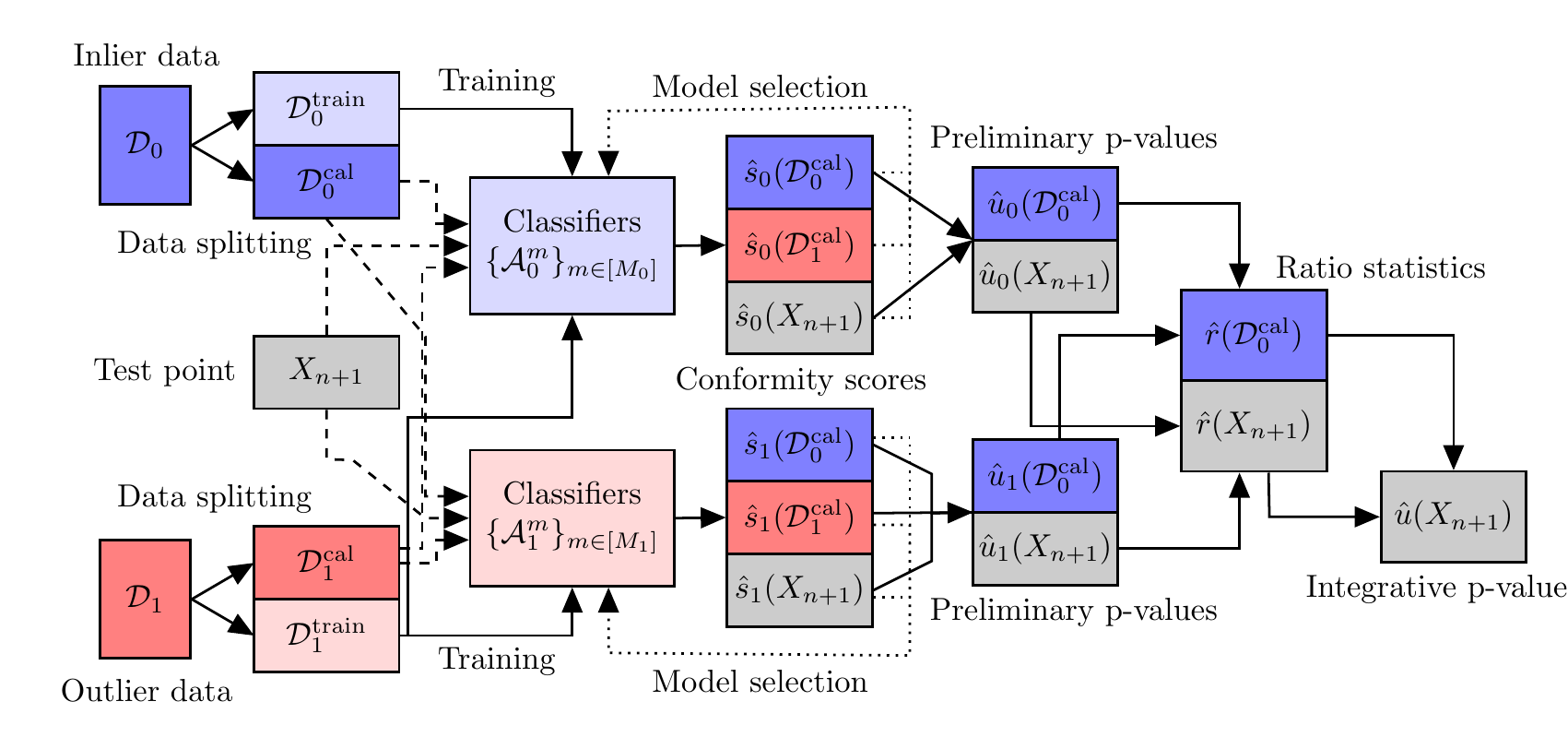}
    \caption{Schematic representation of Algorithm~\ref{alg:weighted-pvalues-tuning}, which computes integrative conformal p-values with automatic model selection. The dashed lines represent the flow of information utilized for the evaluation of the machine-learning models, but not for training. The dotted lines indicate the flow of information utilized for model selection and tuning.}
    \label{fig:weighted-pvals-diagram}
\end{figure}

Concretely, for each model trained on $\mathcal{D}_0^{\mathrm{train}}$, we evaluate $\hat{s}_0(X_i)$ for all $i \in \mathcal{D}_{1}^{\mathrm{cal}}$ as well as for all  $i \in \{n+1\} \cup \mathcal{D}_{0}^{\mathrm{cal}}$; then, we pick the model that maximizes the median difference in $\hat{s}_0(X_i)$ across the two aforementioned groups of data points.
The intuition is that a powerful $\hat{s}_0$ must be able to separate well inliers from outliers, and $\mathcal{D}_{1}^{\mathrm{cal}}$ contains only outliers while all but possibly one element of $\{n+1\} \cup \mathcal{D}_{0}^{\mathrm{cal}}$ are inliers.
The model trained on $\mathcal{D}_1^{\mathrm{train}}$ is selected similarly, by maximizing the median difference in $\hat{s}_1(X_i)$ across the same two groups of observations: $\mathcal{D}_{1}^{\mathrm{cal}}$ and $\{n+1\} \cup \mathcal{D}_{0}^{\mathrm{cal}}$.
Note that this method does not need to be implemented by maximizing the median difference in conformity scores; any alternative criteria may be utilized to identify the most promising model, as long its decision is invariant to permutations of $\{n+1\} \cup \mathcal{D}_{0}^{\mathrm{cal}}$.
The approach described above was adopted in this paper simply because it is easy to explain and works quite well in practice, but it is unlikely to be optimal.
Further, note that that our method is not limited to one-class classifiers.
To the contrary, any available labeled outliers can be easily incorporated into the computation of integrative conformal p-values conformity scores $\hat{s}_0$ and $\hat{s}_1$ based on a binary classifier trained on the combined data set $\mathcal{D}_0^{\mathrm{train}} \cup \mathcal{D}_1^{\mathrm{train}}$.
This flexibility eliminates the dilemma of whether one should rely on one-class or binary classification algorithms in order to obtain the most powerful conformal p-values for a particular data set.
The complete methodology described here is outlined in Algorithm~\ref{alg:weighted-pvalues-tuning} and a schematic representation of it is provided in Figure~\ref{fig:weighted-pvals-diagram}.
The validity of this approach is formally established in Theorem~\ref{theorem:ratio-exchangeable-tuning}.

\begin{theorem} \label{theorem:ratio-exchangeable-tuning}
  If $X_{n+1}$ is an inlier and is exchangeable with the inliers in $\mathcal{D}_0^{\mathrm{cal}}$, and $\hat{u}(X_{n+1})$ is computed as outlined in Algorithm~\ref{alg:weighted-pvalues-tuning}, then $\P{\hat{u}(X_{n+1}) \leq \alpha \mid Y_{n+1} = 0} \leq \alpha$ for all $\alpha \in (0,1)$.
\end{theorem}

\begin{algorithm}[H]
    \caption{Integrative conformal p-values with automatic model selection}
    \label{alg:weighted-pvalues-tuning}
    \begin{algorithmic}[1]
        \State \textbf{Input}: inlier data $\mathcal{D}_0$, outlier data $\mathcal{D}_1$, one-class and binary classifiers $\{\mathcal{A}_0^m\}_{m \in [M_0]}$, one-class classifiers $\{\mathcal{A}_1^m\}_{m \in [M_1]}$, and test point $X_{n+1}$.
        \State Randomly split $\mathcal{D}_0$ into two subsets, $\mathcal{D}_0^{\mathrm{train}}, \mathcal{D}_0^{\mathrm{cal}}$.
        \State Randomly split $\mathcal{D}_1$ into two subsets, $\mathcal{D}_1^{\mathrm{train}}, \mathcal{D}_1^{\mathrm{cal}}$.
        \State For each $m \in M_0$, train $\mathcal{A}_0^m$ on $\mathcal{D}^{\mathrm{train}}_0$ and, if the model is binary, also on $\mathcal{D}^{\mathrm{train}}_0$.
        \State Calculate scores $\hat{s}^m_0(X_i)$ based on $\mathcal{A}^m_0$ for all $i \in \{n+1\} \cup \mathcal{D}^{\mathrm{cal}}_0 \cup \mathcal{D}^{\mathrm{cal}}_1$ and for all $m \in M_0$.
        \State \label{select}Select the model $m_0^*$ that maximizes the median difference between the conformity scores $\hat{s}^m_0(X_i)$ evaluated on $\{n+1\} \cup \mathcal{D}^{\mathrm{cal}}_0 $ and those evaluated on $\mathcal{D}^{\mathrm{cal}}_1 $.
        \State For each $m \in M_1$, train $\mathcal{A}_1^m$ on $\mathcal{D}^{\mathrm{train}}_1$.
        \State Calculate scores $\hat{s}^m_1(X_i)$ based on $\mathcal{A}^m_1$ for all $i \in \{n+1\} \cup \mathcal{D}^{\mathrm{cal}}_0 \cup \mathcal{D}^{\mathrm{cal}}_1$.
        \State \label{select}Select the model $m_1^*$ that maximizes the median difference between the conformity scores $\hat{s}^m_1(X_i)$ evaluated on $\{n+1\} \cup \mathcal{D}^{\mathrm{cal}}_0 $ and those evaluated on $\mathcal{D}^{\mathrm{cal}}_1 $.
        \State Compute conformal p-values $\hat{u}_0(X_i)$ with (\ref{eq:p-value-0}) for all $i \in \{n+1\} \cup \mathcal{D}^{\mathrm{cal}}_0$, using model $m_0^*$.
        \State Compute conformal p-values $\hat{u}_1(X_i)$ with (\ref{eq:p-value-1}) for all $i \in \{n+1\} \cup \mathcal{D}^{\mathrm{cal}}_0$, using model $m_1^*$.
        \State Combine $\hat{u}_0(X_i)$ and $\hat{u}_1(X_i)$ into $\hat{r}(X_i)$ with (\ref{eq:ratio-stats}) for all $i \in \{n+1\} \cup \mathcal{D}^{\mathrm{cal}}_0$.
        \State \textbf{Output} conformal p-value $\hat{u}(X_{n+1})$ computed with (\ref{eq:p-value-ratio}).
    \end{algorithmic}
\end{algorithm}

A compelling demonstration of the advantages of integrative conformal p-values arises from a sometimes strange behaviour of one-class classifiers which we highlight here.
Recall from Section~\ref{sec:standard-conformal} that it is typically assumed a one-class classifier trained on inlier data should produce smaller scores $\hat{s}_0(X)$ if $X$ is an outlier.
However, this is not always the case; for instance, Figure~\ref{fig:auto-tuning} shows the results of simulations in which a one-class classifier effectively separates outliers from inliers but assigns smaller conformity scores to the latter, resulting in powerless standard conformal p-values. Intuitively, this ``inversion'' of the conformity scores occurs if the model happens to learn a low-dimensional representation of the data that captures well the distribution of inliers, but results in unusually concentrated (i.e., less extreme) scores for the outliers.
Algorithm~\ref{alg:weighted-pvalues-tuning} completely solves this problem because it can automatically decide for the data at hand whether $\hat{s}_0$ or $\hat{s}_1$ should be transformed into $-\hat{s}_0$ or $-\hat{s}_1$ prior to computing $\hat{u}_0$ and $\hat{u}_1$, respectively.
This particular type of tuning will be applied throughout this paper for all integrative conformal p-values based on one-class classifiers.

\subsection{Integrative p-values with transductive cross-validation+} \label{sec:adaptive-weighting-cv+}

Data splitting requires training the machine-learning models only once, which is computationally efficient even if the number of test points is large.
However, this approach is not always optimal because it reduces the effective sample size, and it introduces more randomness into the calibration procedure. Intuitively, these issues tend to be more accentuated in applications with fewer labeled data.
To overcome the above limitation, we develop here an alternative method for computing integrative conformal p-values that is more computationally expensive but makes a more efficient use of the available data. This solution is inspired by the existing techniques of cross-validation+~\cite{barber2019predictive} and transductive conformal inference~\cite{vovk2013transductive}.
To facilitate the exposition, we begin by focusing on a single pair of one-class classifiers, thus postponing the discussion of automatic model selection.

Given a desired number of folds $K_1 \in [|\mathcal{D}_1|]$, we randomly split the labeled outliers in $\mathcal{D}_1$ into $K_1$ disjoint subsets: $\mathcal{D}_1^1, \dots, \mathcal{D}_1^{K_1}$.
For any $j \in \mathcal{D}_1$, we define $k_1(j)$ as the fold to which $X_j$ is assigned, and for any $k \in [K_1]$, we let $\hat{s}_1^{k}$ denote the conformity score function computed by the one-class classifier trained on the data in $\mathcal{D}_1 \setminus \mathcal{D}_1^{k}$.
Then, for any possible feature vector $x$, we define the {\em preliminary} p-value function $\hat{u}_1(x)$ as follows:
\begin{equation}\label{eq:cv++_pval_1}
  \hat{u}_1(x) = \frac{1+| \{j \in \mathcal{D}_1:  \hat{s}_1^{k_1(j)}(X_{j}) \leq \hat{s}_1^{k_1(j)}(x) \}| } {1+|\mathcal{D}_1|}.
\end{equation}
Note that $\hat{u}_1(X)$ may be seen as an intuitive cross-validation+~\cite{barber2019predictive} version of a standard conformal p-value~\cite{bates2021testing} for testing whether $X$ is exchangeable with the outliers in $\mathcal{D}_1$. However, as in Section~\ref{sec:weighted}, we do not claim $\hat{u}_1(X)$ is super-uniform under any hypothesis because we make no explicit assumptions about the exchangeability of the outliers.
Instead, we will simply utilize $\hat{u}_1$ to distill potentially useful side information to reweigh the conformal p-values calibrated with inlier data.

Given a number of folds $K_0 \in [|\mathcal{D}_0|+1]$, which may differ from $K_1$, we randomly split $\mathcal{D}_0 \cup \{n+1\}$, into $K_0$ disjoint subsets: $\mathcal{D}_0^1, \dots, \mathcal{D}_0^{K_0}$.
For any $j \in \mathcal{D}_0 \cup \{n+1\}$, we define $k_0(j)$ as the fold to which $X_j$ is assigned, and for any $k \in [K_0]$, we let $\hat{s}_0^{k}$ denote the score function computed by the one-class classifier trained on $\mathcal{D}_0 \setminus \mathcal{D}_0^{k}$.
Then, for any feature vector $x$ and any $l \in [K_0]$, we define the {\em preliminary} p-value function $\hat{u}_0(x,k)$ as follows:
\begin{equation}\label{eq:cv++_pval_0}
  \hat{u}_0(x; l) = \frac{|j\in \{n+1\} \cup \mathcal{D}_{0}^{\mathrm{cal}} : \hat{s}_0^{k_0(j)}(X_j) \leq \hat{s}_0^{l}(x) |}{1+|\mathcal{D}_{0}^{\mathrm{cal}}|}.
\end{equation}
Note that $\hat{u}_0(X_{n+1}; k_0(n+1))$ may be intuitively seen as a cross-validation+~\cite{barber2019predictive} conformal p-value for testing whether $X_{n+1}$ is exchangeable with the inliers in $\mathcal{D}_0$. Yet, there is an important difference between our approach, which we call {\em transductive cross-validation+} (TCV+), and the method of~\cite{barber2019predictive}: we treat $X_{n+1}$ as a training data point for $K_0-1$ out of $K_0$ models, while the original cross-validation+ procedure does not look at the test point at all during training.
As it will become clear in the proof of the next theorem, our transductive approach has nice theoretical advantages because it is completely symmetric with respect to the test point and the labeled inliers, although at the cost of being more computationally expensive when many different test points are involved~\cite{vovk2013transductive}.

\begin{figure}[!htb]
    \centering
    \includegraphics[width=\linewidth]{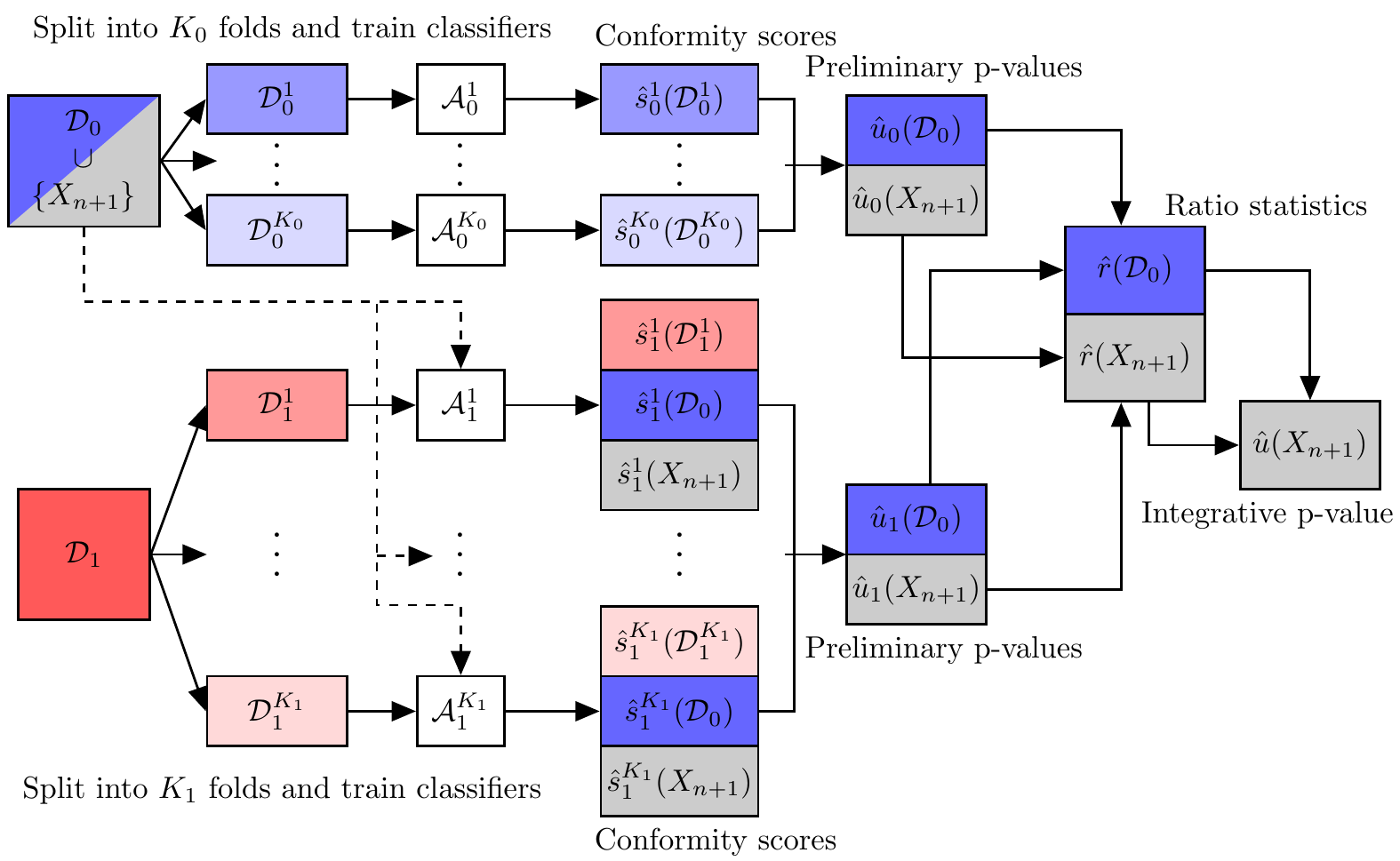}
    \caption{Schematic representation of Algorithm~\ref{alg_clra_cv+}, computing integrative conformal p-values via TCV+ instead of data splitting. Other details are as in Figure~\ref{fig:weighted-pvals-diagram}.}
    \label{fig:cvpp_diagram}
\end{figure}

The information contained in $\hat{u}_0(X_{n+1}; k_0(n+1))$ and $\hat{u}_1(X_{n+1})$ must now be combined as to obtain a valid integrative conformal p-value satisfying~\eqref{eq:super-uniform}. Our approach is similar to that of Section~\ref{sec:adaptive-weighting}.
First, we compute the preliminary p-values $\hat{u}_1(X_{i})$ and $\hat{u}_0(X_{i}; k_0(i))$ via~\eqref{eq:cv++_pval_1} and~\eqref{eq:cv++_pval_0}, respectively, for all $i \in \mathcal{D}_0 \cup \{n+1\}$. Then, for each $i \in \mathcal{D}_0 \cup \{n+1\}$, we combine $\hat{u}_1(X_{i})$ and $\hat{u}_0(X_{i}; k_0(i))$ with a function $\hat{r}$ such as the one defined earlier in~\eqref{eq:ratio-stats}: $\hat{r}(X_{i}) = \hat{u}_0(X_{i}) / \hat{u}_1(X_{i})$. Finally, we output an integrative conformal p-value $\hat{u}(X_{n+1})$ based on~\eqref{theorem:ratio-exchangeable}. This procedure is outlined in Algorithm~\ref{alg_clra_cv+} and sketched graphically in Figure~\ref{fig:cvpp_diagram}.
As long as the machine-learning algorithms utilized to compute the conformity score functions $\hat{s}_0$ and $\hat{s}_1$ are invariant to permutations of their input data points, $\hat{u}(X_{n+1})$ is guaranteed to be a valid conformal p-value.

\begin{theorem} \label{theorem:ratio-cv+}
  Assume $X_{n+1}$ is an inlier and is exchangeable with the inliers in $\mathcal{D}_0$, and $\hat{u}(X_{n+1})$ is computed by Algorithm~\ref{alg_clra_cv+}, using machine-learning training algorithms that are invariant to the order of the input data points.
  Then, $\P{\hat{u}(X_{n+1}) \leq \alpha \mid Y_{n+1} = 0} \leq \alpha$ for all $\alpha \in (0,1)$.
\end{theorem}

\begin{algorithm}[H]
    \caption{Integrative conformal p-values via transductive cross-validation+}
    \label{alg_clra_cv+}
    \begin{algorithmic}[1]
        \State \textbf{Input}: inlier data $\mathcal{D}_0$, outlier data $\mathcal{D}_1$, permutation-invariant one-class classifiers $\mathcal{A}_0, \mathcal{A}_1$, test point $X_{n+1}$, numbers of folds $K_0$ and $K_1$.
        \State Split $\mathcal{D}_1$ into $K_1$ folds, $\mathcal{D}_1^1, \dots, \mathcal{D}_1^{K_1}$.
       \For{$k = 1, \dots, K_1$}
        \State Fit a model $\mathcal{A}^k_1$ by training $\mathcal{A}_1$ on $\mathcal{D}_1 \setminus \mathcal{D}_1^k$; this yields a score function $\hat{s}_1^{k}$.
        \EndFor
        \State Split $\mathcal{D}_0 \cup \{n+1\}$ into $K_0$ folds, $\mathcal{D}_0^1, \dots, \mathcal{D}_0^{K_0}$; let $k_0(i)$ denote the fold for $i$.
       \For{$k = 1, \dots, K_0$}
        \State Fit a model $\mathcal{A}^k_0$ by training $\mathcal{A}_0$ on $\mathcal{D}_0 \setminus \mathcal{D}_0^k$; this yields a score function $\hat{s}_0^{k}$.
        \EndFor
        \State For all $i \in \mathcal{D}_0 \cup \{n+1\}$, compute $\hat{u}_1(X_i)$ with~\eqref{eq:cv++_pval_1}.
        \State For all $i \in \mathcal{D}_0 \cup \{n+1\}$, compute $\hat{u}_0(X_i; k_0(i))$ with~\eqref{eq:cv++_pval_0}.
        \State For all $i \in \mathcal{D}_0 \cup \{n+1\}$, compute $\hat{r}(X_i) = \hat{u}_0(X_i; k_0(i)) / \hat{u}_1(X_i)$.
        \State \textbf{Output}: conformal p-value $\hat{u}(X_{n+1})$ computed with~\eqref{eq:p-value-ratio}.
    \end{algorithmic}
\end{algorithm}

Although Algorithm~\ref{alg_clra_cv+} was introduced assuming the availability of a single pair of one-class classifiers for ease of exposition, TCV+ can accommodate automatic model selection to compute integrative conformal p-values based on a toolbox of different machine-learning algorithms.
The key idea is the same as that in Section~\ref{sec:auto-tuning}: data-driven model selection does not invalidate our integrative conformal p-values as long as the selection criterion is invariant to permutations of the data points indexed by $\mathcal{D}_0 \cup \{n+1\}$.
In particular, one can implement Algorithm~\ref{alg_clra_cv+} by training with cross-validation a toolbox of classifiers, as in Algorithm~\ref{alg:weighted-pvalues-tuning}, and then compute the integrative p-value based on the most promising pair of machine-learning algorithms.
Concretely, the software package accompanying this paper selects the algorithm $\mathcal{A}_0$ maximizing the median difference in $\hat{u}_0(X_i)$ across the two groups $\mathcal{D}_{1}$ and $\mathcal{D}_{0}^{\mathrm{cal}} \cup \{n+1\}$, and the algorithm $\mathcal{A}_1$ minimizing the corresponding group median difference in $\hat{u}_1(X_i)$.

Finally, TCV+ can also be applied to compute conformal p-values if no labeled outliers are available, although without automatic model selection. In that context, a benefit of this approach is that it leads to p-values satisfying the desired super-uniformity property~\eqref{eq:super-uniform} exactly, while standard cross-validation+ in theory requires an additional conservative correction~\cite{barber2019predictive} which tends to decrease power; see Appendix~\ref{sec:cv++_general} for details. 
Further, TCV+ p-values can also be useful beyond the scope of testing for outliers. For example, the same technique can be applied to construct conformal prediction sets for multi-class classification, as explained in Appendix~\ref{sec:cv++_general}.


\section{FDR control via conditional calibration} \label{sec:fdr}

\subsection{Dependence structure of integrative conformal p-values} \label{sec:dependencies}

Consider now a multiple testing version of our problem with several unlabeled data points, namely $X_{n+1}, X_{n+2}, \ldots, X_{n+m}$, and suppose the goal is to discover which of them are likely outliers. In this context, a meaningful notion of type-I errors is the FDR~\cite{benjamini1995controlling}, which is especially reasonable if the expected number of outliers in the test set is large.
Recall that typical methods for FDR control, including BH, assume the p-values to be independent or positively dependent according to a specific {\it PRDS} property~\cite{benjamini2001control}. However, the integrative p-values output by Algorithm~\ref{alg:weighted-pvalues} or Algorithm~\ref{alg:weighted-pvalues-tuning} for different test points cannot be independent of one another if they share the same calibration data.
Standard conformal p-values have been proved to satisfy the PRDS property~\cite{bates2021testing}, but the same does not necessarily extend to integrative p-values.
The issue is that the combination function $\hat{r}$ in~\eqref{eq:p-value-ratio} depends not only on the training data but also on the observations in $\mathcal{D}_0^{\mathrm{cal}}$ and on $X_{n+1}$.
Integrative p-values for different test points are thus based on random conformity score functions that may differ from one another, and this makes their dependence structure more complicated compared to that studied by~\cite{bates2021testing}.
Indeed, our simulations will show that pairwise correlations between integrative conformal p-values may be affected by the data distribution, although they similarly tend to decrease as the number of calibration data points increases.
Consequently, it remains unclear whether the p-values output by Algorithm~\ref{alg:weighted-pvalues} can be safely utilized with BH.
Similar concerns also apply to the TCV+ p-values output by Algorithm~\ref{alg_clra_cv+}.

The numerical experiments presented later in Sections~\ref{sec:exp-synthetic} and~\ref{sec:exp-real} suggest that BH is quite robust to the dependencies of integrative conformal p-values. 
While this is reassuring, it may still be useful to present a different approach that can theoretically guarantee FDR control with integrative conformal p-values, without excessive loss of power.
The following solution is enabled by~\cite{2020conditionalcalibration}, which recently proposed a powerful and flexible {\it conditional calibration} framework for FDR control under any p-value dependence structure. 
Conditional FDR calibration can be much more powerful than more drastic approaches, such as the Benjamini-Yekutieli procedure for FDR control under arbitrary dependence~\cite{benjamini2001control}, but this advantage comes at a cost in practicality. In fact, conditional FDR calibration is more of a versatile theoretical blueprint than a practical algorithm. It can be implemented successfully in specific cases~\cite{2020conditionalcalibration}, but it is computationally unfeasible in general.
Interestingly, however, conditional FDR calibration turns out to be practical and intuitive to implement in the case of integrative conformal p-values, especially if the latter are obtained via Algorithm~\ref{alg:weighted-pvalues}.

\subsection{FDR control with integrative conformal p-values} \label{sec:fdr-3steps}

The general conditional FDR calibration strategy of~\cite{2020conditionalcalibration} involves three main steps, which can be recalled intuitively as follows.
In the first step, one computes a data-adaptive rejection threshold $\tau_i$ (as a function of the desired FDR level $\alpha$) separately for each test point $i \in \mathcal{D}^{\mathrm{test}} = \{n+1,\ldots,n+m\}$, in such a way as to ensure $\tau_i$ only depends on the data through a suitable sufficient statistic $\Phi_i$. The choice of  $\Phi_i$ is delicate and has to follow two key principles: (1) the p-value for the $i$-th test point must be super-uniform conditional on $\Phi_i$; (2) the total number of rejections obtained by thresholding each p-value $u(X_i)$ at level $\tau_i$ should be bound from below by a known function of $\Phi_i$. Further, to achieve high power, the bound should hold if $u(X_i) \leq \tau_i$ and be as tight as possible.
In the second step, one applies the individual rejection thresholds $\tau_i$ separately to each p-value $u(X_i)$ and verifies whether the aforementioned lower-bound constraints on the total number of rejections are satisfied.
If all constraints are satisfied, that set of rejections is guaranteed to control the FDR below $\alpha$. Otherwise, if some constraints are violated, the rejection set must be randomly thinned (or pruned) through a final third step, and this will tend to decrease power.

It must be emphasized that there may be infinitely many different ways of computing the individual thresholds $\tau_i$ in the first step of the generic FDR calibration recipe described above. In fact, it is generally unclear how to translate this sequence of high-level ideas into a practical algorithm than can be both powerful and computationally feasible for the specific problem at hand. This difficulty makes the specialized conditional FDR calibration method for integrative conformal p-values developed here  technically non-trivial.
For simplicity, we will focus on controlling the FDR with the p-values output by Algorithm~\ref{alg:weighted-pvalues}, but the same idea is in principle also applicable to Algorithm~\ref{alg:weighted-pvalues-tuning}, although at higher computational cost.
The three steps of our solution are outlined below and summarized graphically in Figure~\ref{fig:split_fdr_diagram}.

\textbf{Step 1: Calibration.} For each of the $m$ test points, $i \in \mathcal{D}^{\mathrm{test}} = \{n+1,\ldots,n+m\}$, let $\Phi_i$ be the following collection of random variables: $\hat{s}_0(X_j)$ and $\hat{s}_0(X_j)$ for all $j \in \mathcal{D}^{\mathrm{test}} \setminus \{i\}$; $\hat{s}_1(X_j)$ and $\hat{s}_1(X_j)$ for all $j \in \mathcal{D}_1^{\mathrm{cal}}$; and the {\em unordered} set of $\{(\hat{s}_0(X_j), \hat{s}_1(X_j))\}_j$ for all $j \in \mathcal{D}_0^{\mathrm{cal}} \cup \{i\}\}$.
In other words, $\Phi_i$ contains information on all conformity scores calculated by Algorithm~\ref{alg:weighted-pvalues}, up to a random permutation of those involving the test point $i$ and the labeled calibration inliers in $\mathcal{D}_0^{\mathrm{cal}}$. In short, one may write $\Phi_i$ as:
\begin{align}\label{eq:sigma_algebra}
    \Phi_i=
  \begin{rcases}
    \begin{dcases}
      \hat{s}_0(X_j), \hat{s}_1(X_j), \hspace{0.2cm}  \forall j \in \mathcal{D}^{\mathrm{test}} \backslash \{i\}, \\
      \hat{s}_1(X_j), \hspace{0.2cm}  \forall j \in \mathcal{D}^{\mathrm{cal}}_1, \\
      \text{unordered set } \{(\hat{s}_0(X_j), \hat{s}_1(X_j))\}_j, \text{ for } j \in \mathcal{D}^0_{\mathrm{cal}} \cup \{i\}.
    \end{dcases}
  \end{rcases}
\end{align}
For each test point $i \in \mathcal{D}^{\mathrm{test}}$, compute a vector of $m$ random variables $\tilde{u}_{i}(X_j)$, for all $j \in \mathcal{D}^{\mathrm{test}} \setminus \{i\}$, by applying Algorithm~\ref{alg:weighted-pvalues} with $\mathcal{D}^{\mathrm{cal}}_0$ replaced by $\mathcal{D}^{\mathrm{cal}}_0 \cup \{i\}$.
Note that  $\tilde{u}_{i}(X_j)$ may not be a valid conformal p-value because it is based on a perturbed inlier calibration set, $\mathcal{D}^{\mathrm{cal}}_0 \cup \{i\}$, which may contain an outlier, but it is a reasonable approximation of the valid p-value $\hat{u}(X_j)$ if $\mathcal{D}^{\mathrm{cal}}_0$ is large.
Next, let $\tilde{R}_i$ indicate the number of rejections obtained by applying BH at level $\alpha$, for some fixed $\alpha \in (0,1)$, to the approximate p-values $(\tilde{u}_i(X_{n+1}), \dots, \tilde{u}_i(X_{i-1}), 0, \tilde{u}_i(X_{i+1}), \tilde{u}_i, \dots, \tilde{u}_i(X_{n+m}))$.

\textbf{Step 2: Preliminary rejection.} Define the preliminary rejection set $\mathcal{R}_+$ as:
\begin{align*}
    \mathcal{R}_+ = \bigg\{i \in \mathcal{D}^{\mathrm{test}}: \hat{u}(X_i) \leq \frac{\alpha\tilde{R}_i}{m}\bigg\},
\end{align*}
$\tilde{R}_+ = |\mathcal{R}_+|$. 
This implicitly means the individual rejection thresholds $\tau_i$ are defined as $\tau_i = \alpha\tilde{R}_i/m$.
If $\tilde{R}_+ \geq \tilde{R}_i$ for all $i \in \mathcal{R}_+$, then return the final rejection set $\mathcal{R} = \mathcal{R}_+$. Otherwise, proceed to the next step.

\textbf{Step 3: Pruning.} Generate independent standard uniform random variables $\epsilon_i$ for each $i \in \mathcal{R}_+$, and define $R$ as:
\begin{align*}
    R = \max \big\{r : \big| i \in \mathcal{R}_+: \epsilon_i \leq r/\tilde{R}_i \big| \geq r \big\}.
\end{align*}
The pruned rejection set $\mathcal{R}$ is that containing the $R$ indices $i \in \mathcal{R}_+$ such that $\epsilon_i < R / \tilde{R}_i$.

It is important to emphasize that this conditional FDR calibration algorithm is not computationally expensive. In fact, as it only operates with pre-computed conformity scores, its cost will typically be negligible compared to that of training the underlying machine learning classifiers.
Thus, the only meaningful disadvantage of this approach compared to BH is that it may sometimes be less powerful, as our numerical experiments will show. However, it enjoys the benefit of guaranteeing rigorous FDR control without being as conservative as the Benjamini-Yekutieli procedure~\cite{benjamini2001control}.

\begin{theorem} \label{theorem:fdr}
Assume $X_{i}$ is an inlier and is exchangeable with the inliers in $\mathcal{D}_0^{\mathrm{cal}}$ for all $i \in \mathcal{D}^{\mathrm{test}}$ with $Y_i=0$.
Then, the expected proportion of inliers in the set $\mathcal{R}$ output by the above three-step procedure is smaller than $\alpha m_0/m$, where $m = |\mathcal{D}^{\mathrm{test}}|$ and $m_0$ is the number of inliers in $\mathcal{D}^{\mathrm{test}}$.
\end{theorem}

\subsection{FDR control with TCV+ integrative conformal p-values} \label{sec:fdr-3steps-cv++}

The conditional FDR calibration method described above can also be adapted to deal with the TCV+ integrative p-values computed by Algorithm~\ref{alg_clra_cv+}, although at higher computational cost.
Our three-step solution is outlined below and summarized graphically in Figure~\ref{fig:cv_loo_diagram}.

\textbf{Step 1: Calibration.} For each of the $m$ test points, $i \in \mathcal{D}^{\mathrm{test}} = \{n+1,\ldots,n+m\}$, let $\Phi_i$ be the following collection of variables:
\begin{align}\label{eq:sigma_algebra-cv++}
    \Phi_i=
  \begin{rcases}
    \begin{dcases}
      \hat{s}_1^{k(j)}(X_j), \hspace{0.2cm}  \forall j \in \mathcal{D}_1, \\
      \text{unordered set } \{\hat{s}^{k}_1(X_j)\}_j, \text{ for } j \in \mathcal{D}^0 \cup \{i\}, \text{ and for all } j \in [K] \\
      \text{unordered set } \{X_j\}_j, \text{ for } j \in \mathcal{D}^0 \cup \{i\}.
    \end{dcases}
  \end{rcases}
\end{align}
For each test point $i \in \mathcal{D}^{\mathrm{test}}$, compute a vector of $m$ random variables $\tilde{u}_{i}(X_j)$, for all $j \in \mathcal{D}^{\mathrm{test}} \setminus \{i\}$, by applying Algorithm~\ref{alg_clra_cv+} with $\mathcal{D}_0$ replaced by $\mathcal{D}_0 \cup \{i\}$.
It must be highlighted that, in contrast with Section~\ref{sec:fdr-3steps}, now evaluating each $\tilde{u}_{i}(X_j)$ involves re-fitting $K$ machine-learning models, and is therefore quite expensive.
Then, as in Section~\ref{sec:fdr-3steps}, let $\tilde{R}_i$ indicate the number of rejections obtained by applying BH at level $\alpha$, for some fixed $\alpha \in (0,1)$, to the approximate conformal p-values $(\tilde{u}_i(X_{n+1}), \dots, \tilde{u}_i(X_{i-1}), 0, \tilde{u}_i(X_{i+1}), \tilde{u}_i, \dots, \tilde{u}_i(X_{n+m}))$.

\textbf{Steps 2--3.} Same as the corresponding steps 2--3 in Section~\ref{sec:fdr-3steps}.

Theorem~\ref{theorem:fdr-cv++} below establishes that this solution rigorously controls the FDR.
Note however that the computational cost of this guarantee is now much higher than it was in Section~\ref{sec:fdr-3steps} for the integrative conformal p-values output by Algorithm~\ref{alg:weighted-pvalues}.
In fact, the conditional calibration algorithm needs to re-train a potentially large number of machine learning classifiers in the case of the TCV+ p-values, and therefore this approach may not be very practical when working with large data sets. Fortunately, pairwise correlations between integrative conformal p-values naturally tend to decrease as the number of labeled data points increases, and therefore it is reasonable to expect that BH can be safely relied upon in the limit of large sample sizes.

\begin{theorem} \label{theorem:fdr-cv++}
Assume $X_{i}$ is exchangeable with the inliers in $\mathcal{D}_0$ for all $i \in \mathcal{D}^{\mathrm{test}}$ with $Y_i=0$.
Then, the expected proportion of inliers in the set $\mathcal{R}$ output by the above three-step procedure is smaller than $\alpha m_0/m$, where $m = |\mathcal{D}^{\mathrm{test}}|$ and $m_0$ is the number of inliers in $\mathcal{D}^{\mathrm{test}}$.
\end{theorem}


\section{Asymptotic power analysis} \label{sec:asymptotic}

This section presents an asymptotic power analysis which makes rigorous the intuitive idea that integrative p-values output by Algorithm~\ref{alg:weighted-pvalues} can be more powerful than standard conformal p-values, as long as the data-driven weights are sufficiently informative. This analysis is inspired by the rich literature on multiple testing with side information~\cite{lei2018adapt} and weighted p-values~\cite{genovese2006false,hu2010false,efron2012large,ignatiadis2016data,li2019multiple,ignatiadis2021covariate,Caietal20,basu2018weighted, benjamini1997weights, roquain2008optimal}, in which several works have theoretically studied power under FDR with BH. For instance, \cite{genovese2006false} performs a power analysis for generic p-value weights, and \cite{Caietal20,liang2022locally} show specific choices of weights can in some cases outperform unweighted procedures. In particular, we focus on an approach similar to that of ~\cite{Caietal20,liang2022locally}. As detailed below, our analysis requires a few additional technical assumptions, and it begins from a convenient simplification which allows our conformal inference problem to be directly connected to the classical covariate-assisted multiple testing framework considered by~\cite{Caietal20,liang2022locally}.
In particular, we imagine that the ranking of the integrative p-values $\hat{u}(X_i)$ output by Algorithm~\ref{alg:weighted-pvalues} is the same as that of the statistics $\hat{r}(X_i) = \hat{u}_0(X_i) / \hat{u}_1(X_i)$.
This may not always be exactly true, because the definition of the function $\hat{u}_0$ in~\eqref{eq:p-value-0} is such that the values of $\hat{u}_0(X_j)$ for all $j \in \mathcal{D}_0^{\mathrm{cal}}$ are dependent on the current test point, but it is not unreasonable if both $\mathcal{D}_0^{\mathrm{cal}}$ and $\mathcal{D}_1^{\mathrm{cal}}$ are large.
The reason why this simplification is convenient is that it allows us to see $\hat{r}(X_i)$ as standard conformal p-value weighted by $1/ \hat{u}_1(X_i)$.

Without loss of generality, let us think of the one-class classifier $\mathcal{A}_1$ utilized to compute the conformity scores $\hat{s}_1(X_i)$ in Algorithm~\ref{alg:weighted-pvalues} as learning from the data in $\mathcal{D}_1^{\mathrm{train}}$ a function that maps $X_i$ to a low-dimensional representation $Z_{1,i}$, for each $i \in \mathcal{D}^{\mathrm{test}}$.
In the following, we will treat as fixed the training data in $\mathcal{D}_0^{\mathrm{train}}$ and $\mathcal{D}_1^{\mathrm{train}}$, as well as the fitted functions $\hat{s}_0$ and $\hat{s}_1$. 
With this premise, one may then say in the language of weighted multiple testing~\cite{liang2022locally, Caietal20} that $Z_{1,i}$ represents the ``side information'' associated with the p-value $\hat{u}_0(X_i)$, and $1/\hat{u}_1(X_i)$ is the associated weight computed from $Z_{1,i}$.
Despite the analogies in this setup, an important difference remains between our problem and the typical framework of weighted hypothesis testing: in our case the side information in $Z_{1,i}$ is not independent of $\hat{u}_0(X_i)$.
Fortunately, the same analysis strategy of~\cite{liang2022locally, Caietal20} can be repurposed to overcome this challenge.

Define $\mathcal{F}$ as the collection of random variables $\{Z_{1,i}\}_{i\in \mathcal{D}^{\mathrm{test}}}$, $\mathcal{D}_0$, and $\mathcal{D}_1$, and let $F_{1,i}(t)$ be the cumulative distribution function of $\hat{u}_0(X_i)$ conditional on $Y_i=1$ as well as on $\mathcal{F}$.
Let also $\delta(t; v)=\{\delta_i(t; v_i): 1\leq i \leq m\}$ be a family of rejection rules for $m$ hypotheses with p-values $\hat{u}_0(X_1),\ldots,\hat{u}_0(X_m)$ and generic inverse weights $v=(v_1,\ldots,v_m)$. By definition, these rules are such that the $i$-th null hypotheses is rejected if and only if $p_i / v_i \leq t$, in which case $\delta_i(t; v_i)=1$.
By applying Theorem 2 from~\cite{Caietal20}, one can show under relatively mild \emph{weak dependence} conditions~\cite{storey2004strong}\footnote{Sufficient conditions for this large-$m$ approximation are provided in~\cite{basu2018weighted} and~\cite{Caietal20}.} that the FDR of $\delta(t; v)$ can be written in the large-$m$ limit as
\begin{align*}
  \mathrm{FDR}(\delta(t; v))
  & = Q(t; v) + o(1),
\end{align*}
where
\begin{align*}
  Q(t; v)
  & = \frac{\sum_{i \in \mathcal{D}^{\mathrm{test}}} v_i t \cdot \P{Y_i = 0 \mid Z_{1,i}} }{ \sum_{i \in \mathcal{D}^{\mathrm{test}}} v_i t \cdot \P{Y_i = 0 \mid Z_{1,i}} + \sum_{i \in \mathcal{D}^{\mathrm{test}}} F_{1,i}(v_it \mid Z_{1,i}) \cdot \P{Y_i = 1 \mid Z_{1,i}}}.
\end{align*}
Next, define $\Psi(t;v)$ as the expected number of true discoveries produced by $\delta(t; v)$:
\begin{align} \label{eq:true-discoveries}
  \Psi(t;v)
  &= \E{\sum_{i \in \mathcal{D}^{\mathrm{test}}} Y_i \cdot \delta_i(t; v_i) }.
\end{align}
Further, for any $v=(v_1,\ldots,v_m)$, define the oracle significance threshold $t^{\alpha}_{\mathrm{oracle}}(v)$ as the largest possible threshold $t$ controlling the approximate large-$m$ FDR $Q(t; v)$ below a fixed level $\alpha$:
\begin{align*}
  t^{\alpha}_{\mathrm{oracle}}(v) = \sup \{ t: Q(t; v) \leq \alpha\}.
\end{align*}
Below, the FDR and the power of the oracle rule $\delta(t^{\alpha}_{\mathrm{oracle}}(v); v)$ will be compared under two alternative choices of weights: $v = 1/\hat{u}_1(X_i)$, which corresponds to the integrative method proposed in this paper, and $v=(1,\ldots,1)$, which corresponds to standard conformal p-values. For convenience, we sometimes write $v=1$ instead of $v=(1,\ldots,1)$.

\begin{theorem}[Adapted from~\cite{liang2022locally}] \label{theorem:power}
Under Assumptions~\ref{assumption:informative-weights}--\ref{assumption:cdf}, stated below, for any $\alpha \in (0,1)$,
\begin{align}
  \Psi(t^{\alpha}_{\mathrm{oracle}}(\hat{u}_1); \hat{u}_1) \geq \Psi(t^{\alpha}_{\mathrm{oracle}}(1); 1).
\end{align}
\end{theorem}

\begin{assumption} \label{assumption:informative-weights}
The inverse weights $\hat{u}_1(X_{i})$, for $i \in \mathcal{D}^{\mathrm{test}}$, satisfy:
\begin{align} \label{eq:informative-weights}
  \frac{\sum_{i \in \mathcal{D}^{\mathrm{test}}}{\P{ Y_i = 0 \mid Z_{1,i}} }}{\sum_{i \in \mathcal{D}^{\mathrm{test}}}\P{ Y_i = 0 \mid Z_{1,i}} \hat{u}_1(X_{i})} \cdot
  \frac{\sum_{i \in \mathcal{D}^{\mathrm{test}}}{\P{ Y_i = 1 \mid Z_{1,i}}}}{\sum_{i \in \mathcal{D}^{\mathrm{test}}}\P{Y_i = 1 \mid Z_{1,i}} \hat{u}_1(X_{i})^{-1}}
  \geq 1.
\end{align}
\end{assumption}

\begin{assumption} \label{assumption:cdf}
The cumulative distribution function $F_{1,i}(t)$ satisfies:
\begin{align*}
  \sum_{i \in \mathcal{D}^{\mathrm{test}}} a_i F_{1,i}(t/x_i)
  \geq
  \sum_{i \in \mathcal{D}^{\mathrm{test}}} a_i F_{1,i}\left( \frac{t \sum_{j=1}^m a_j}{\sum_{j=1}^m a_jx_j} \right),
\end{align*}
for any $0 \leq a_i\leq 1$, $\min_{1 \leq i \leq m} \tilde{w}_i^{-1} \leq x_i \leq \max_{1 \leq i \leq m} \tilde{w}_i^{-1}$, and $t > 0$, where $\tilde{w}^*_i$ is defined as:
\begin{align} \label{eq:power-tilde-w}
  \tilde{w}^*_i = \hat{u}_1(X_i) \cdot \frac{\sum_{j \in \mathcal{D}^{\mathrm{test}}} \P{ Y_j = 0 \mid Z_j} }{ \sum_{j \in \mathcal{D}^{\mathrm{test}}} w_j \cdot \P{Y_j = 0 \mid Z_j} }.
\end{align}

\end{assumption}

Assumption~\ref{assumption:informative-weights} intuitively says the weights must be ``informative'', in the sense that larger values of $\P{Y_i = 1 \mid Z_{1,i}}$ should correspond to larger $\hat{u}_1(X_{i})$; similar assumptions were utilized in~\cite{Genetal06,liang2022locally}.
Assumption~\ref{assumption:cdf} is a more technical condition on the shape of the alternative distribution of $\hat{u}_0(X_i)$ given $Z_{1,i}$.
If these distributions are homogeneous, i.e., $F_{1,i}(t) = F_{1,n+1}(t)$ for all $i \in \mathcal{D}^{\mathrm{test}}$, it reduces to saying that the function $x \rightarrow F_{1}(t/x)$ is convex~\cite{liang2022locally}.

The take-away message from Theorem~\ref{theorem:power} is that tests based on integrative conformal p-values asymptotically dominate those based on unweighted p-values, as long as the weights $1/\hat{u}_1$ are sufficiently informative (Assumption~\ref{assumption:informative-weights}) and some regularity conditions hold (Assumption~\ref{assumption:cdf}). In fact, $t^{\alpha}_{\mathrm{oracle}}(\hat{u}_1)$ and $t^{\alpha}_{\mathrm{oracle}}(1)$ are the most powerful thresholds (approximately) controlling the FDR below $\alpha$, in the two respective cases.
However, this result is not fully satisfactory for our purposes because it is not very specific to Algorithm~\ref{alg:weighted-pvalues}.
Fortunately, the specific structure of our inverse weights $\hat{u}_1(X_i)$ can be leveraged to recast Assumption~\ref{assumption:informative-weights} into a more easily interpreted form.
To this end, assume further that the low-dimensional representation $Z_{1,i}$ of $X_i$ learnt by the machine-learning algorithm $\mathcal{A}_1$ is perfectly informative in theory about the true $Y_i$; that is, $\P{Y_i=1 \mid Z_{1,i}} = Y_i$.
This extra assumption does not make our power analysis trivial, as it does not limit how informative the conformal p-value weights $1/\hat{u}_1(X_i)$ might be in practice, but it is useful to simplify the expression in~\eqref{eq:informative-weights} into:
\begin{align*}
  \frac{\sum_{i \in \mathcal{D}^{\mathrm{test}}} (1-Y_i) \hat{u}_1(X_{i})}{\sum_{i \in \mathcal{D}^{\mathrm{test}}}{ (1-Y_i) }} \cdot
  \frac{\sum_{i \in \mathcal{D}^{\mathrm{test}}} Y_i  [\hat{u}_1(X_{i})]^{-1}}{\sum_{i \in \mathcal{D}^{\mathrm{test}}}{ Y_i}}
  \leq 1.
\end{align*}
In the large-$m$ limit, assuming different test points $X$ are independent of one another and of the labeled data, the left-hand-side above becomes:
\begin{align*}
  & \E{\hat{u}_1(X) \mid Y = 0} \cdot \E{\frac{1}{\hat{u}_1(X)} \mid Y = 1} + o_{\mathbb{P}}(1).
\end{align*}
Then, as $\hat{u}_1(X)$ is a valid conformal p-value for testing the null hypothesis that $Y=1$,
\begin{align*}
  \E{\frac{1}{\hat{u}_1(X)} \mid Y = 1}
  & = \frac{1}{n_1+1} \sum_{k=1}^{n_1+1} \left( \frac{k}{n_1+1} \right)^{-1} \approx \log(n_1),
\end{align*}
if $n_1 = |\mathcal{D}_1|$ is sufficiently large.
Therefore, in the limit of large $m$ and large $n_1$, Assumption~\ref{assumption:informative-weights} is approximately equivalent to:
\begin{align} \label{eq:informative-weights-simple}
  & \E{\hat{u}_1(X) \mid Y = 0} \leq \frac{1}{\log(n_1+1)}.
\end{align}
Combined with Theorem~\ref{theorem:power}, this result suggests the integrative p-values produced by Algorithm~\ref{alg:weighted-pvalues} tend to be more powerful than standard conformal p-values if the inverse weights $\hat{u}_1(X_i)$ are sufficiently small under the null hypothesis that $Y_i=0$.
This condition matches the intuitive idea that $\hat{u}_1(X_i)$ should be sufficiently ``informative'' for our integrative method to be beneficial, and it has the desirable property of being empirically verifiable.

\section{Numerical experiments with synthetic data} \label{sec:exp-synthetic}

\subsection{Setup} \label{sec:exp-setup}

In this section, integrative conformal p-values are computed with Algorithm~\ref{alg:weighted-pvalues-tuning} for synthetic data and their performance is compared to that of standard conformal p-values.
The latter are calculated based on two different families of classifiers: one-class classification models (which only look at the labeled inliers) and binary classification models (which also utilize the labeled outliers).
Six different one-class classification models are considered: a support vector machine with four alternative choices of kernel (linear, radial basis function, sigmoid, and polynomial of degree 3), an isolation forest, and a ``local outlier factor'' nearest neighbor method. These models are taken off-the-shelf, as implemented in the popular Python package {\em scikit-learn} \cite{scikit-learn}.
Further, six different binary classification models are considered: a random forest classifier, a $K$-nearest neighbor classifier, a support vector classifier, a Gaussian naive Bayes model, a quadratic discriminant analysis model, and a multi-layer perceptron. These models are also applied as implemented in {\em scikit-learn}, with default parameters.
The integrative method has access to all of the above models and automatically selects on a case-by-case basis the most promising one as detailed in Section~\ref{sec:auto-tuning}. The performance of our integrative p-values is compared to that of standard conformal p-values, after giving to the latter the unfair advantage of having the {\em best} performing model from each family selected by an imaginary oracle. 
Of course, such oracle would be unavailable in real applications, but it provides an informative and challenging benchmark within the scope of these experiments.

A simpler variation of our integrative conformal method is also considered as an additional benchmark.
This utilizes the labeled outlier data only to automatically select the most promising classification model from the available toolbox, as well as to tune any relevant hyper-parameters, but it does not re-weigh the standard conformal p-values. In other words, this {\em ensemble} benchmark consists of applying Algorithm~\ref{alg:weighted-pvalues-tuning} with $\hat{r}(X_i) = \hat{u}_0(X_i)$ instead of $\hat{r}(X_i) = \hat{u}_0(X_i) / \hat{u}_1(X_i)$, and it may be thought of as a practical and theoretically valid approximation of the aforementioned imaginary oracle. 
To help emphasize the relative advantages of different families of classification models under different data distributions, the aforementioned ensemble benchmark and the imaginary oracle will be applied separately with the one-class and binary classification models.
Finally, note that there is an additional subtle but sometimes practically important difference between the imaginary oracle and the ensemble benchmark.
The latter takes as input the raw conformity scores output by a one-class classifier, while the former also has the ability to automatically flip their signs if that increases power, as explained in Section~\ref{sec:auto-tuning}.

\subsection{Split-conformal integrative p-values} \label{sec:exp-split}

We begin the numerical experiments by simulating synthetic data from a model analogous to that utilized by~\cite{bates2021testing}.
Each sample $X_i \in \mathbb{R}^{1000}$ is generated from a multivariate Gaussian mixture model $P_X^{a}$, such that $X_i = \sqrt{a} \, V_i + W_i$, for some constant $a\geq 1$ and appropriate random vectors $V_i,W_i \in \mathbb{R}^{1000}$.
The vector $V_i$ has independent standard Gaussian components, and each coordinate of $W_i$ is independent and uniformly distributed on a discrete set $\mathcal{W} \subseteq \mathbb{R}^{1000}$ with cardinality $|\mathcal{W}|=1000$.
The vectors in $\mathcal{W}$ are sampled independently from the uniform distribution on $[-3,3]^{1000}$ prior to the first experiment.
The number of labeled inliers and outliers in the simulated data set is varied as a control parameter. All methods are applied to calculate conformal p-values for an independent test set containing 1000 data points, half of which are outliers. Then, the performance of each method is evaluated in terms of the empirical FDR and power obtained by applying to the corresponding p-values BH \cite{benjamini1995controlling} with Storey's correction~\cite{storey2002direct,storey2004strong}, at the nominal 10\% level. The results are averaged over 100 independent experiments.
Note that all methods are applied using half of the labeled inliers for training and half for calibration. The methods leveraging binary classification models also utilize half of the labeled outliers for training.

\begin{figure}[!htb]
    \centering
    \includegraphics[width=0.9\linewidth]{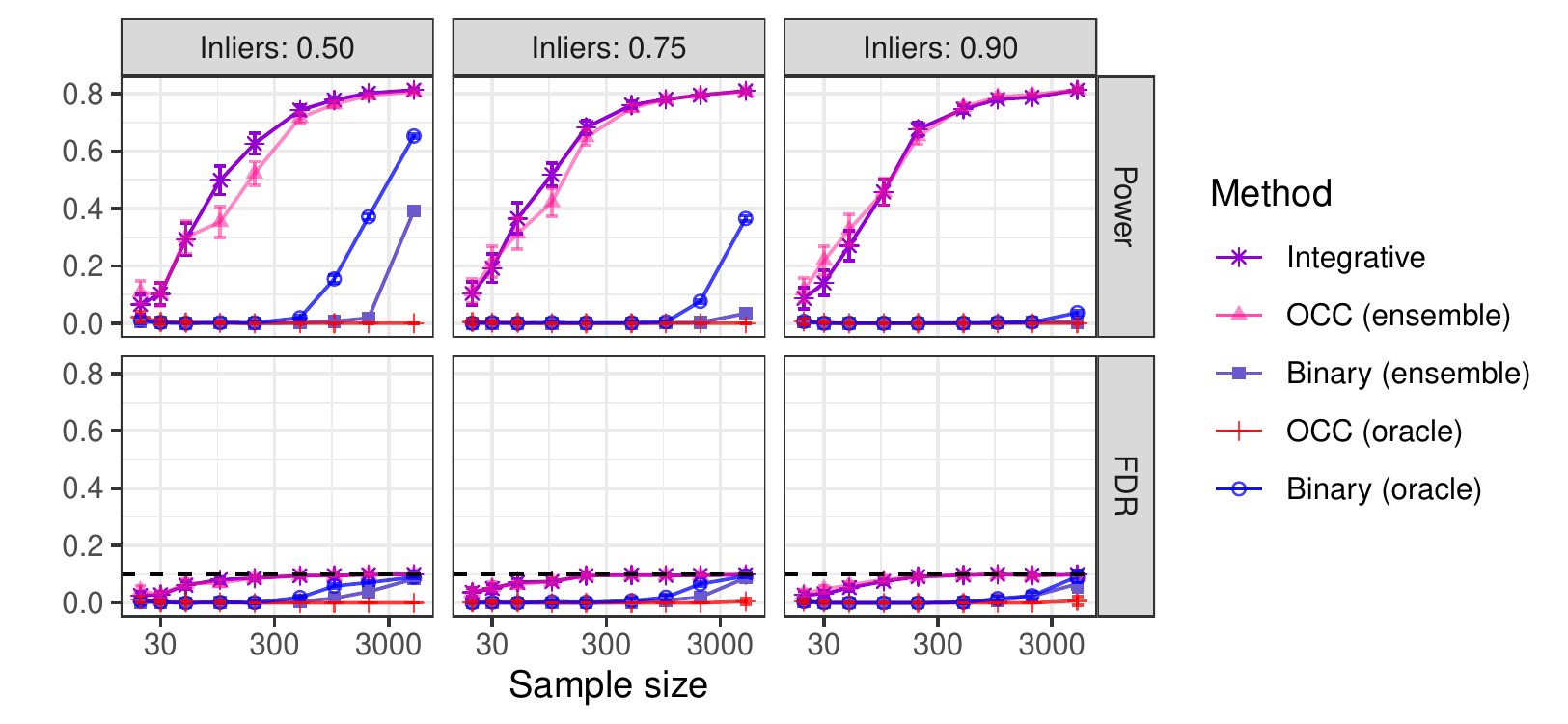}
    \caption{Performance of Storey's BH applied to conformal p-values computed with different methods, on simulated data. The results are shown as a function of the sample size, for calibration data sets with different proportions of inliers. The horizontal dashed line corresponds to the nominal 10\% FDR level.}
    \label{fig:exp-1-n}
\end{figure}

Figure~\ref{fig:exp-1-n} shows the empirical power and FDR obtained with conformal p-values computed by different methods, as a function of the sample size and of the fraction of outliers in the data. The data are generated from the model described above with parameter $a=0.7$.
All methods considered here produce theoretically valid conformal p-values, although FDR control is not theoretically guaranteed with integrative and ensemble p-values, as explained in Section~\ref{sec:dependencies}. Nonetheless, the empirical results suggest all methods lead to valid FDR control in these experiments, consistently with the general robustness of BH. 
The integrative and one-class ensemble approaches achieve the highest power, with the former being slightly more powerful when the proportion of outliers in the data is large.
It is interesting to note that the data distribution is such that the outliers tend to receive less extreme one-class classification scores than the inliers for all models considered in these experiments, which explains why the one-class oracle has no power.
It should not be too surprising that all binary classification models have no power here unless the sample size is large and the class imbalance is small, because the data are high dimensional and intrinsically difficult to separate.

Figure~\ref{fig:exp-2-bh} in Appendix~\ref{app:figures} reports on analogous experiments with different sample sizes and proportions of outliers.
Figures~\ref{fig:exp-1-fixed} and~\ref{fig:exp-2-fixed} show results from the same experiments as in Figures~\ref{fig:exp-1-n} and~\ref{fig:exp-2-bh}, respectively, with the difference that now possible outliers in the test data are identified by rejecting the null hypotheses for which the conformal p-values are below the fixed threshold 0.1, instead of seeking to control the FDR.
Figure~\ref{fig:exp-1-qq} visualizes the empirical distribution on test data of conformal p-values computed with the alternative methods, separately for inliers and outliers, under the same experimental setting as in Figure~\ref{fig:exp-1-n}. These results indicate all methods considered here produce uniformly distributed p-values for the inliers, as expected, but the integrative method leads to smaller p-values for the outliers.

Figure~\ref{fig:exp-3-n} reports on experiments similar to those in Figure~\ref{fig:exp-1-n}, with the difference that now the data are simulated with parameter $a=1.25$ instead of $a=0.7$. Under this distribution, the outlier detection problem becomes intrinsically easier and the outliers naturally tend to receive more extreme one-class conformity scores compared to the inliers, as one would typically expect. Consequently, the one-class oracle benchmark is able to achieve much higher power compared to the previous experiments, although its performance does not surpass that of the integrative method unless the sample size is very small.
Figure~\ref{fig:exp-3-oracle} compares under the same setting as in Figure~\ref{fig:exp-3-n}  the performance of integrative conformal p-values to that of standard conformal p-values based on two alternative one-class classification models, as well as to those based on the most powerful model selected by an imaginary oracle. These results show that distinct models may perform very differently from one another depending on the data at hand, which strengthens the practical motivation for our integrative approach and highlights the unfairness of the oracle advantage.

Figures~\ref{fig:exp-4-n}--\ref{fig:exp-4-n-fixed} report on similar experiments with data from a different distribution. For each observation, a feature vector $X \in \mathbb{R}^{100}$ is sampled from a standard multivariate Gaussian with diagonal covariance.
The conditional distribution of the outlier label $Y \in \{0,1\}$ given $X=x$ is binomial with weights
$w_k(x) = z_k(x) / \sum_{k'=0}^{1} z_{k'}(x)$, for $k \in \{0,1\}$, where $z_k(x) = \exp(\gamma + x^T \beta_k)$ and each $\beta_k \in \mathbb{R}^{p}$ is sampled from an independent Gaussian distribution with mean 0 and variance $3$.
The intercept $\gamma$ is determined by an independent Monte Carlo simulation with binary search in order to ensure the desired expected proportion of outliers in the data.
Under this setting, binary classification models tend to perform better than one-class models, but the integrative method remains competitive with the oracle binary benchmark throughout a wide range of sample sizes and proportions of labeled outliers.

\subsection{Naive model selection yields invalid p-values} \label{sec:experiments-naive}

\begin{figure}[!htb]
    \centering
    \includegraphics[width=0.95\linewidth]{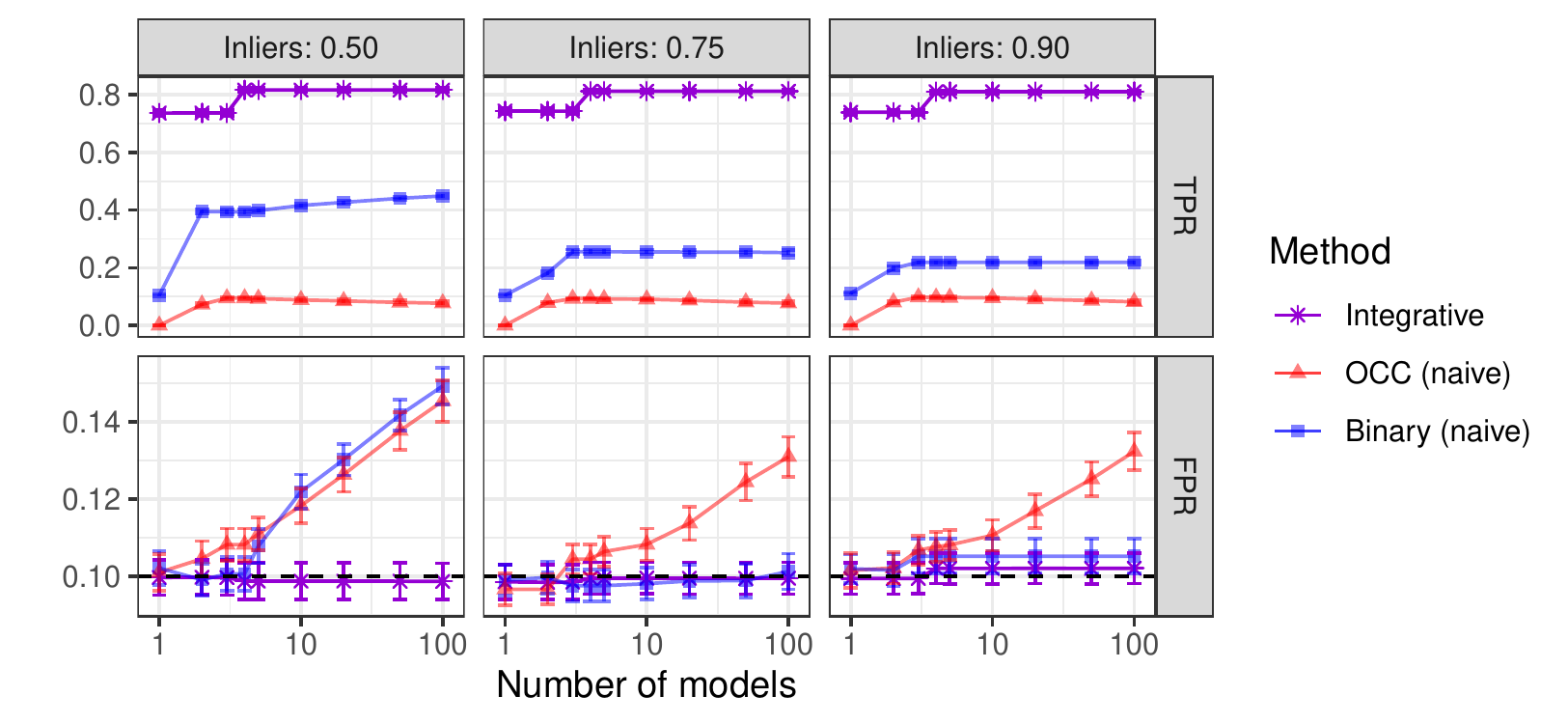}
    \caption{Performance of integrative conformal p-values and naive benchmarks on simulated data, as a function of the number of machine learning models considered.
The performance is measured in terms of true positive rate (TPR) and false positive rate (FPR). The horizontal dashed line corresponds to the nominal 10\% FPR level. Other details are as in Figure~\ref{fig:exp-1-n}.}
    \label{fig:exp-greedy}
\end{figure}

A key advantage of integrative conformal p-values is their ability to automatically select in a data-driven way the best performing model from an ensemble of different options.
This strength is emphasized here by experiments that demonstrate empirically the invalidity of a naive alternative approach which consists of computing standard conformal p-values separately for each model and then {\em greedily} picking the model reporting the largest number of potential outliers in the test set.
These simulations are conducted using independent data samples of size 1000 generated from the same Gaussian mixture model as in Section~\ref{sec:exp-split}. Conformity scores are calculated using the same families of one-class and binary classification models as in Section~\ref{sec:exp-split}, with the only difference that now several independent instances of the isolation forest and random forest classifier models are utilized with different random seeds; the number of possible models in each family is thus varied as a control parameter and increased up to 100.
Figure~\ref{fig:exp-greedy} compares the performance of integrative conformal p-values to that of the aforementioned naive benchmark based on one-class or binary models, as a function of the number of models considered.
In this experiment, possible outliers in the test data are identified by rejecting the null hypotheses for which the conformal p-values are below the fixed threshold 0.1, and the performance is measured in terms of true positive rate and false positive rate (i.e., the probabilities of rejecting the null hypothesis for a true outlier or a true inlier, respectively).
The results demonstrate that the integrative method is more powerful and always controls the type-I errors, unlike the naive benchmark.

\subsection{Pairwise correlation between integrative conformal p-values}

\begin{figure}[!htb]
    \centering
    \includegraphics[width=0.7\linewidth]{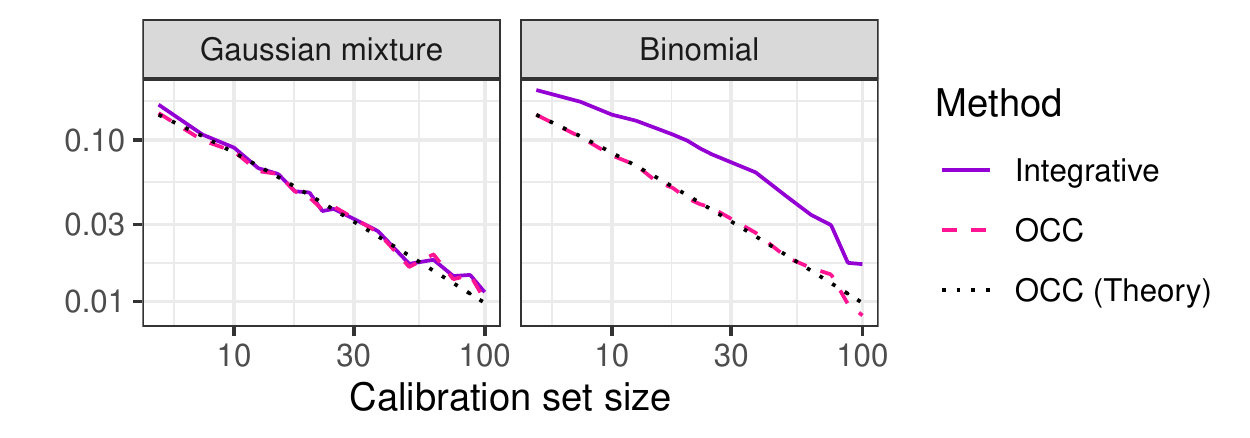}
    \caption{Pairwise correlation between integrative conformal p-values for independent test points, on synthetic data from two different distributions. The results are shown as a function of the size of the shared calibration data set. The dotted line indicates the theoretical pairwise correlation between standard one-class classification (OCC) conformal p-values.}
    \label{fig:exp-corr}
\end{figure}

In this section, we investigate empirically the correlation structure of integrative conformal p-values calculated for different test points using the same calibration data. It was discussed in Section~\ref{sec:dependencies} that integrative p-values may have more complicated dependencies compared to standard conformal p-values~\cite{bates2021testing};  Figure~\ref{fig:exp-corr} demonstrates this empirically. Synthetic data are generated separately from the same Gaussian mixture and binomial distributions as in Section~\ref{sec:exp-split}, with $p=1000$ features per observation. The sample size is varied as a control parameter.
To reduce the computational cost of these experiments, the integrative method is applied with a narrower selection of possible one-class classification models compared to Section~\ref{sec:exp-split}: a support vector machine with radial basis function, sigmoid, or third-degree polynomial kernel.
Figure~\ref{fig:exp-corr} shows the average pairwise correlation between the integrative p-values for different test points as a function of the number of inliers in the calibration data set, separately for the two data-generating distributions. These empirical correlations are averaged over 200,000 independent experiments.
The results show the correlation between conformal p-values for different test points intuitively decreases as the number of calibration data points $n_0$ increases. However, unlike in the case of standard conformal p-values~\cite{bates2021testing}, this relation appears to be data-dependent and the correlation is not exactly equal to $1/n_0$.
In particular, integrative conformal p-values appear to experience generally stronger correlations compared to standard conformal p-values. This should not be surprising given that our integrative method extracts more information from the shared calibration data.

\subsection{Integrative conformal p-values via TCV+}

\begin{figure}[!htb]
    \centering
    \includegraphics[width=0.95\linewidth]{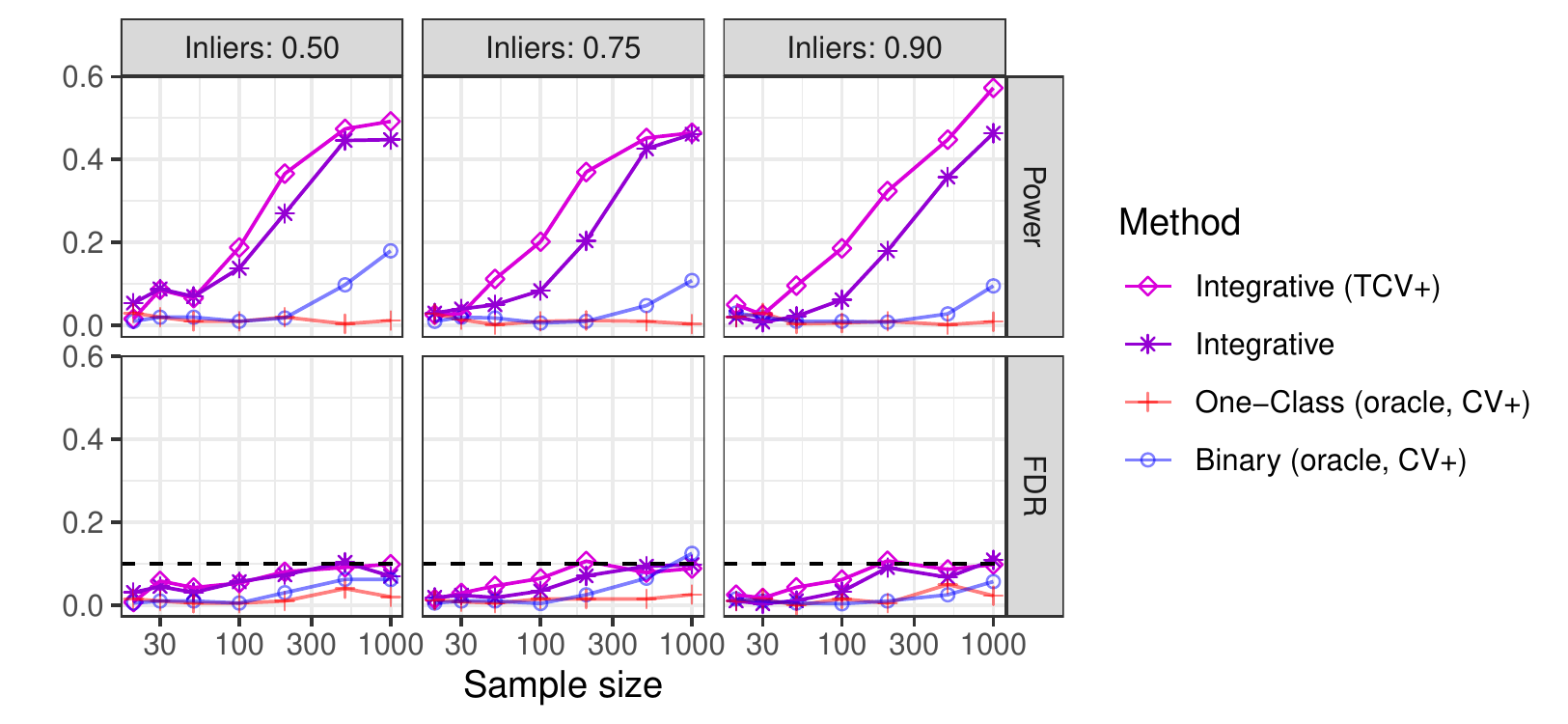}
    \caption{Performance on simulated data of Storey's BH applied to integrative conformal p-values computed with either TCV+ or sample splitting. To serve as benchmarks, standard conformal p-values based on one-class or binary classification models tuned by an ideal oracle are computed using cross-validation+. Other details are as in Figure~\ref{fig:exp-1-n}.}
    \label{fig:exp-cv}
\end{figure}

This section investigates the performance of TCV+ integrative conformal p-values.
These experiments are conducted using the same data-generating Gaussian mixture model and the same families of one-class and binary classifiers as in Section~\ref{sec:exp-split}.
The performance of integrative conformal p-values calculated via 5-fold TCV+ is compared to that of integrative p-values based on sample splitting, as well as to that of p-values produced by the imaginary oracles described in Section~\ref{sec:adaptive-weighting-cv+}.
The performance of each method is evaluated in terms of FDR and power obtained by applying to the corresponding p-values Storey's BH, as in Section~\ref{sec:adaptive-weighting-cv+}.
Figure~\ref{fig:exp-cv} reports the average results of 100 independent experiments, as a function of the sample size and of the fraction of outliers in the labeled data. These results suggest all methods control the FDR, while TCV+ leads to higher power compared to sample splitting, consistently with its higher data parsimony.

Additional related results are reported in Appendix~\ref{app:figures-cv}.
In particular, Figure~\ref{fig:exp-cv-fixed} shows results analogous to those in Figure~\ref{fig:exp-cv}, although here the performances of different methods are quantified in terms of true positive rate and false positive rate after thresholding the p-values at the fixed level 0.1.
Figures~\ref{fig:exp-cv-binomial} and~\ref{fig:exp-cv-fixed-binomial} report on results similar to those in Figures~\ref{fig:exp-cv} and~\ref{fig:exp-cv-fixed}, respectively, from experiments in which the data are generated from the binomial model described in Section~\ref{sec:adaptive-weighting-cv+} instead of the Gaussian mixture model.

\subsection{FDR control via conditional calibration}

\begin{figure}[!htb]
    \centering
    \includegraphics[width=0.95\linewidth]{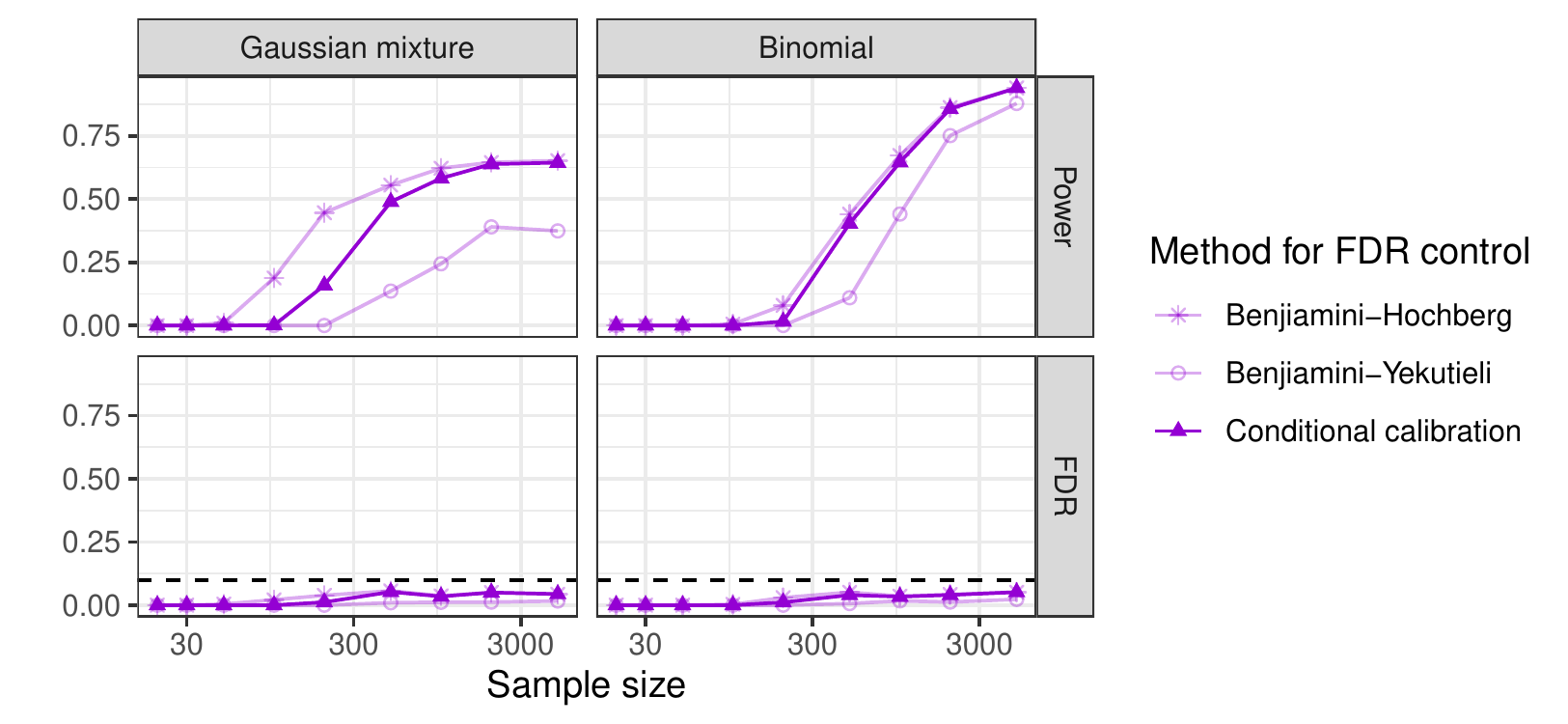}
    \caption{Performance on simulated data of different methods for FDR control with integrative conformal p-values, as a function of the sample size. Other details are as in Figure~\ref{fig:exp-1-n}. }
    \label{fig:exp-cv-fdr}
\end{figure}

This section investigates the performance of conditional calibration for FDR control with integrative conformal p-values.
These experiments are based on the same data-generating models and the same families of one-class or binary classifiers as in Section~\ref{sec:exp-split}.
Integrative conformal p-values are calibrated via sample splitting as in Section~\ref{sec:exp-split}. Then, we aim to control the FDR over a test set of size 10, containing 5 inliers and 5 outliers on average, utilizing three alternative methods: BH \cite{benjamini1995controlling}, the Benjamini-Yekutieli procedure~\cite{benjamini2001control}, and the conditional calibration method described in Section~\ref{sec:fdr}.
Figure~\ref{fig:exp-cv-fdr} reports on these experiments as a function of the number of labeled data points.
All methods empirically control the FDR, and BH appears to be the most powerful one, especially if the sample size is moderate.
Conditional calibration can be seen to be often almost as powerful as the BH benchmark, and it typically significantly outperforms the Benjamini-Yekutieli method.

\subsection{Asymptotic power analysis} \label{sec:exp-power}

\begin{figure}[!htb]
    \centering
    \includegraphics[width=0.95\linewidth]{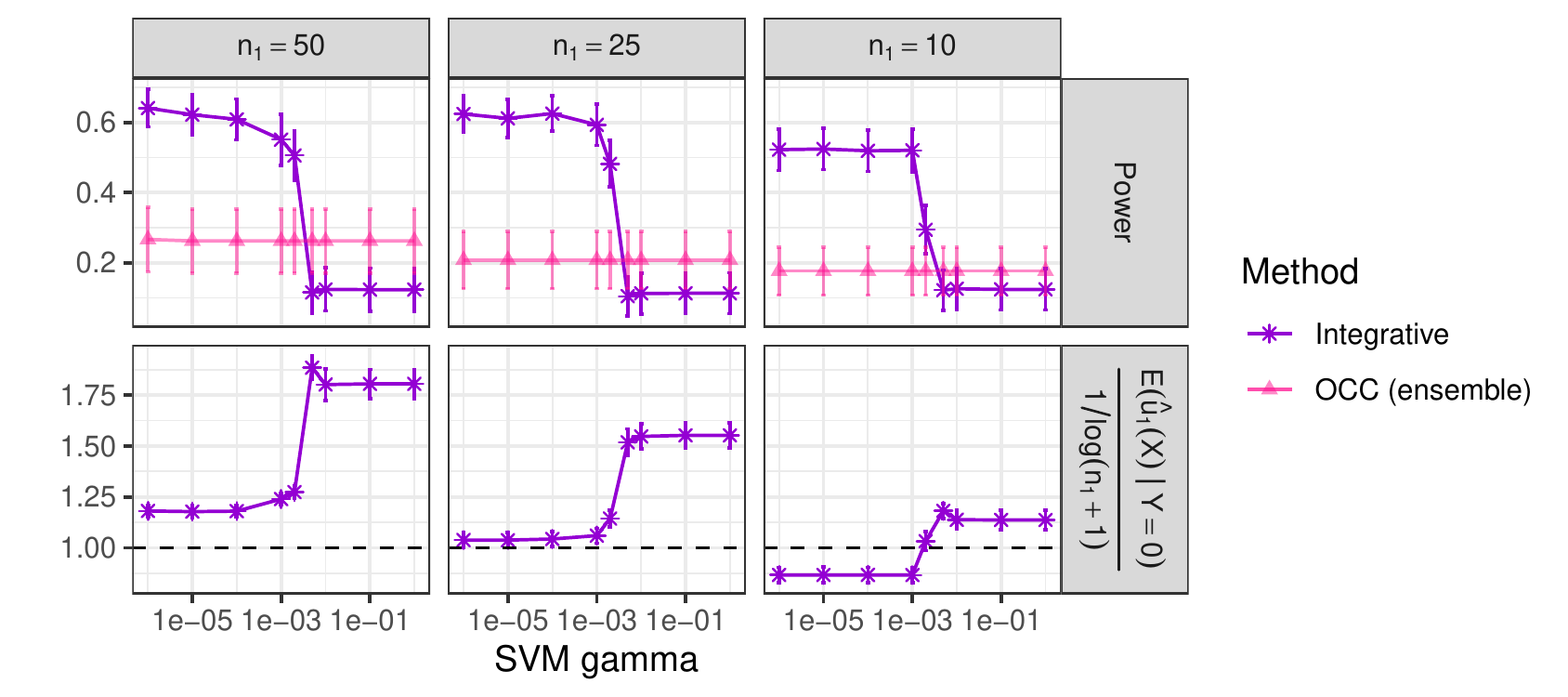}
    \caption{Validation with synthetic data of an asymptotic power analysis for integrative p-values. Top: power of BH applied to integrative and standard conformal p-values, as a function of the scale parameter of the underlying support vector machine. Bottom: average ``informativeness'' ratio for the p-values calculated taking the outlier distribution as the null.}
    \label{fig:exp-power}
\end{figure}

In this section, we validate empirically the asymptotic power analysis presented in Section~\ref{sec:asymptotic}.
For this purpose, independent data samples of size 200 are generated from the same Gaussian mixture model as in Section~\ref{sec:exp-split}.
Integrative conformal p-values are obtained as in Section~\ref{sec:exp-split}, with the only difference that now a single model is utilized to compute conformity scores: a one-class support vector machine with radial basis kernel, for different values of the scale parameter. This parameter affects the informativeness of the resulting conformal p-values, which allows us to probe empirically the theoretical relation between power and the ``informativeness'' ratio $\E{\hat{u}_1(X) \mid Y = 0} / (1/\log(n_1+1))$  in~\eqref{eq:informative-weights-simple}, where $n_1$ is the size of the outlier calibration set, which is varied from 10 to 50.
Figure~\ref{fig:exp-power} summarizes the performance in these experiments of integrative conformal p-values, with and without adaptive weighting, as a function of the support vector machine scale parameter. Here, performance is evaluated in terms of the power of BH applied to these p-values at the nominal FDR level 10\%. The results show that re-weighting the p-values tends lead to higher power if $\E{\hat{u}_1(X) \mid Y = 0} / (1/\log(n_1+1))$ is small, as anticipated in Section~\ref{sec:asymptotic}.

\section{Numerical experiments with real data} \label{sec:exp-real}

\subsection{Animal classification from image data} \label{sec:real-data-animals}

The first real data set we consider is the Animal-10N resource \cite{song2019selfie}, which contains 50,000 images for 10 different animal species: cat, lynx, jaguar, cheetah, wolf, coyote, chimpanzee, orangutan, hamster, and guinea pig.
The images of hamsters and guinea pigs are taken to be the inliers, while the other animals are treated as outliers.
Our goal is to compute p-value for testing the null hypothesis that a new image is an inlier.
To avoid the high computational cost of fitting a complex classification model from scratch, we pre-process the 50,000 raw images with a standard pre-trained ResNet50~\cite{he2016deep} deep neural network, available through the Tensorflow~\cite{tensorflow2015-whitepaper} Python package.
Each image in Animal-10N is thus fed to the ResNet50 model and the predicted values in the final hidden layer of the network are utilized as input features for the subsequent analysis; this gives a vector of 2048 pre-processed features for each image.
To further reduce the data dimensions, we proceed with only the top 512 principal components for each image.
Note that ResNet50 was trained on over 14 million images in the separate Imagenet~\cite{deng2009imagenet} data set, which we assume to be independent of Animal-10N.
After such pre-processing, the experiments are carried out as in Section~\ref{sec:exp-setup}.

A subset of 1000 inliers and 1000 outliers is chosen at random to serve as test set, while the remaining images are randomly assigned to the training and calibration data sets.
The training and calibration data sets have the same cardinality and contain equal proportions of inliers; both of the aforementioned quantities are varied as control parameters.
Integrative conformal p-values and standard conformal p-values tuned by an imaginary oracle are calculated based on the same six one-class and six binary classification machine learning models as in Section~\ref{sec:exp-setup}.
The performance of the p-values produced by each method is quantified in terms of the power and FDR obtained by applying BH \cite{benjamini1995controlling} with Storey's correction, at the nominal 10\% level. All results are averaged over 100 experiments based on independent choices of the training, calibration, and test sets.

\begin{figure}[!htb]
    \centering
    \includegraphics[width=0.75\linewidth]{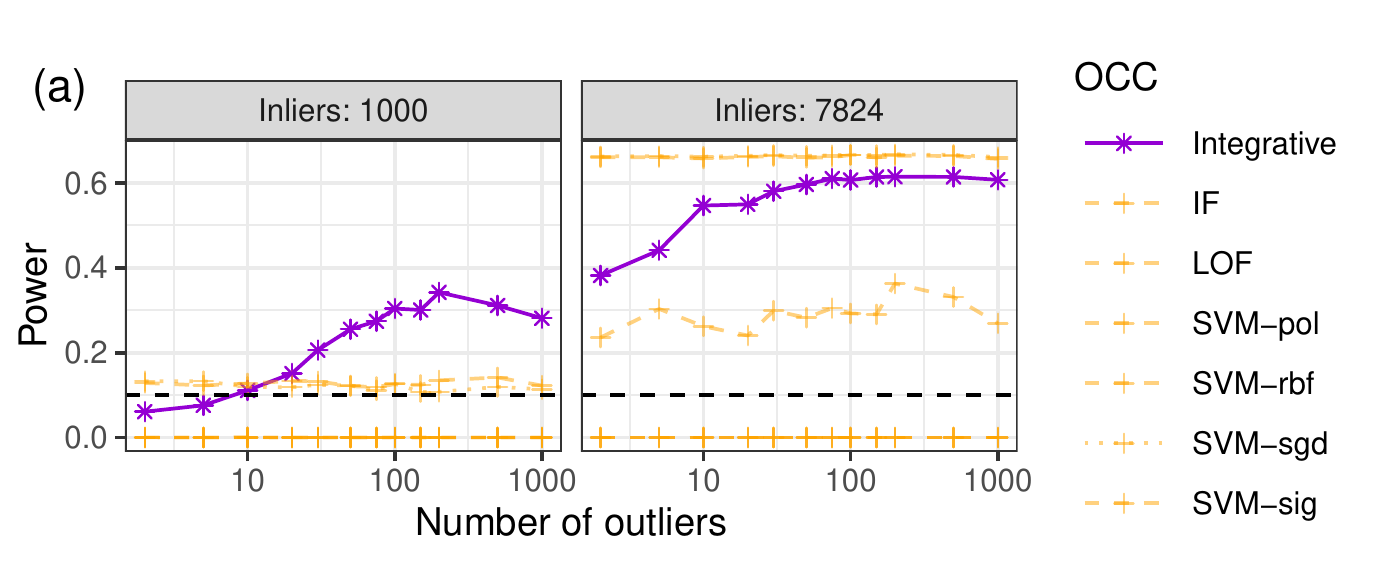}
    \includegraphics[width=0.75\linewidth]{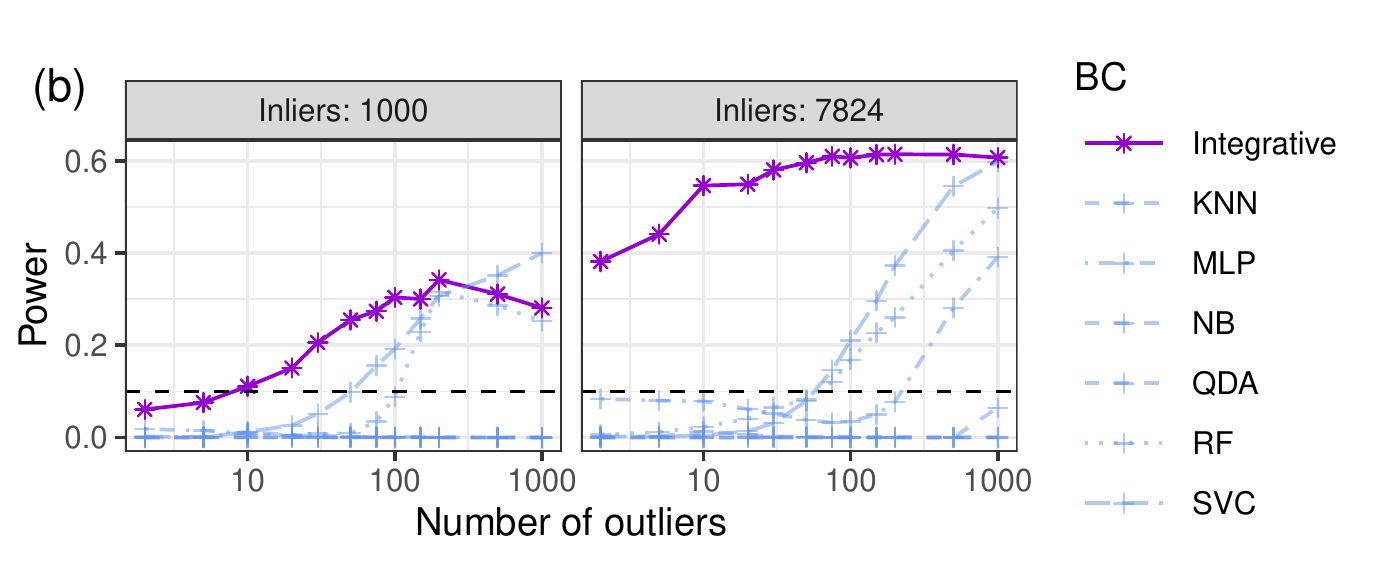}
    \caption{Power of Storey's BH applied to conformal p-values based on different underlying machine-learning models, on animal image classification data. Integrative conformal p-values are compared to standard conformal p-values based on six different one-class classification models (a) and six different binary classification models (b). Other details are as in Figure~\ref{fig:exp-animals}.}
    \label{fig:exp-animals-oracle}
\end{figure}

Figure~\ref{fig:exp-animals}, previewed in Section~\ref{sec:preview}, compares the power achieved by the integrative method to that of the ideal oracles which compute standard conformal p-values based on the best-performing machine learning model from each class, as a function of the number of inliers and outliers in the training/calibration data.
The results show that standard conformal p-values based on one-class classifiers tend to be relatively more powerful compared to the alternative approaches when the number of labeled outliers is small, and their absolute performance improves when the number of inliers is large, as expected. Further, standard conformal p-values based on binary classifiers perform poorly when the labeled inliers greatly outnumber the outliers, but otherwise they can be preferable to their one-class counterparts. By contrast, integrative conformal p-values are able to gather strength from both approaches: their power is relatively high under all settings considered here, and they sometimes outperform both alternatives.
This is remarkable because integrative conformal p-values are at an intrinsic disadvantage compared to the oracle benchmarks because they do not make use of any unrealistic external knowledge about which model will perform best on a given data set.
The ability of the integrative method to automatically gather strength from a broad collection of classifiers is especially useful given that standard conformal p-values based on different models may perform very differently from one another, as mentioned previously in Section~\ref{sec:exp-split} and confirmed here in Figure~\ref{fig:exp-animals-oracle}.
Note that the FDR is not shown explicitly in Figures~\ref{fig:exp-animals} and~\ref{fig:exp-animals-oracle} because all methods considered here empirically control it below the nominal level, consistently with the results on synthetic data presented earlier in Section~\ref{sec:exp-split}.

\subsection{Other image classification data sets} \label{sec:real-data-images}

\begin{figure}[!htb]
    \centering
    \includegraphics[width=0.75\linewidth]{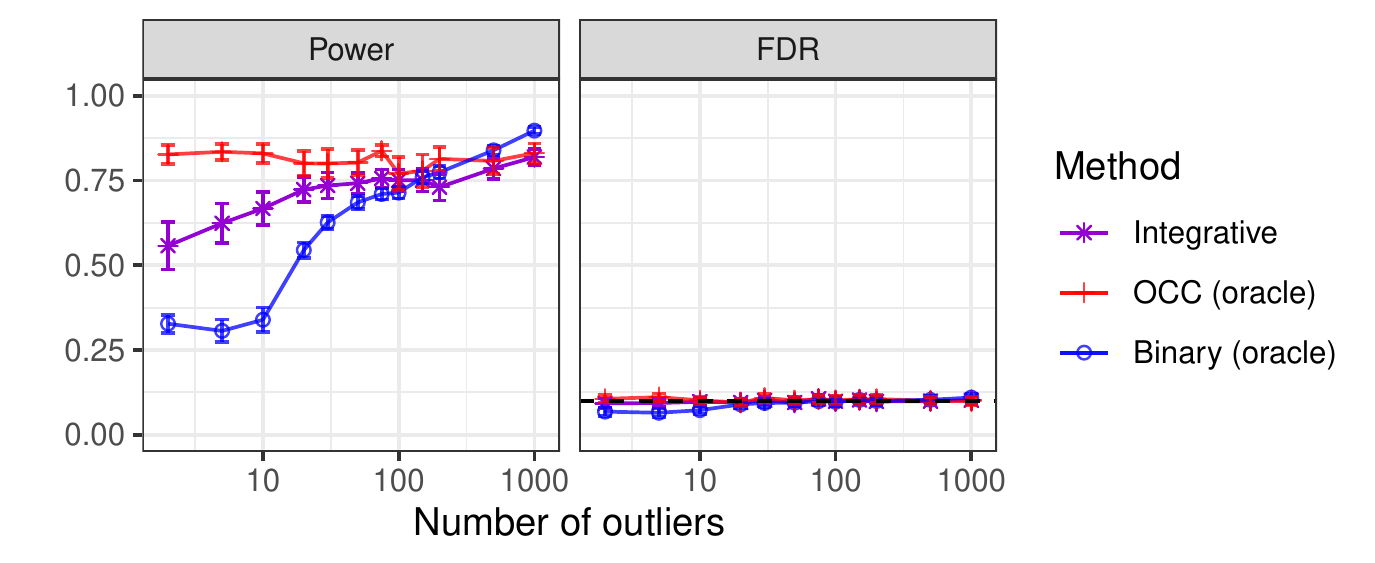}
    \caption{Performance of Storey's BH applied to conformal p-values based on different underlying machine-learning models, on flower classification data. Other details are as in Figure~\ref{fig:exp-animals}.}
    \label{fig:exp-flowers}
\end{figure}

Additional experiments based on different image classification data sets are carried out following the same setup and pre-processing steps as in Section~\ref{sec:real-data-animals}.
Figure~\ref{fig:exp-flowers} reports on experiments based on a standard flower classification data set available through the Tensorflow Python package \cite{tfflowers}. The raw data consist of 3,670 labeled images of five different types of flower: daisies, dandelions, roses, sunflowers, tulips. Our goal is to compute conformal p-values for the null hypothesis that an unlabeled image depicts a rose; other types of flowers are treated as outliers.
The results in Figure~\ref{fig:exp-flowers} illustrate that, in this case, one-class classifiers tend to yield more powerful standard conformal p-values than binary classifiers, unless the number of labeled outliers is large. However, Figure~\ref{fig:exp-flowers-oracle} in Appendix~\ref{app:figures} suggest that the performance of different types of one-class classifiers can vary substantially depending on both the data at hand and the sample size, which makes the self-tuning ability of our integrative approach even more appealing. Figure~\ref{fig:exp-flowers-cv} reports on similar experiments in which integrative conformal p-values are computed via transductive 5-fold cross validation+ instead of sample splitting; these results confirm that cross-validation further increases the power of our approach, as expected, although at the cost of higher computational cost.

Analogous results from similar experiments based on car classification data are reported in Appendix~\ref{app:figures}.
These experiments are based on data obtained by merging the following two data sets: 16,185 images of cars from \cite{krause20133d} and 6,899 labeled images of 8 different types of objects (airplane, car, cat, dog, flower, fruit, motorbike and person) from \cite{roy2018effects}. In our analysis, we pre-process the raw images as in the previous experiments, and then we treat cars as inliers and all other objects as outliers.
Figure~\ref{fig:exp-cars} shows that this outlier detection problem is relatively easy, as integrative conformal p-values and standard conformal p-values based on a one-class classifier achieve almost perfect power. However, standard conformal p-values based on binary classifiers have low power if the number of labeled inliers is small.

\subsection{Tabular data}

Finally, additional experiments based on tabular data sets are carried out following the same setup as in Section~\ref{sec:real-data-animals}, without pre-processing.
For this purpose, two standard data sets are utilized. The one is available from the UCI machine learning repository~\cite{Dua:2019} and pertains a study of thyroid disease~\cite{quinlan1987inductive}. This data set contains 7,200 labeled observations of 6 real-valued features; 6,666 observations are from healthy patients (inliers) and 534 are patients with thyroid disease (outliers).
From these observations, a test set is assembled by randomly picking 267 inliers and 267 outliers. The remaining 6,399 inliers are assigned to the training/calibration set, which also contains a variable number of the remaining outliers.
The second data set analyzed in this section is available from OpenML~\cite{OpenML2013} and contains 11,183 observations of 6 real-valued features; 10,923 samples are labeled as ``non-calcification'' (inlier) and 260 samples are labeled as ``calcification'' (outlier).
From these observations, a test set is assembled by randomly picking 130 inliers and 130 outliers.
The training/calibration set contains all remaining inliers and a variable number of the remaining outliers.
Integrative conformal p-values, as well as standard conformal p-values tuned by an imaginary oracle, are calculated based on six one-class and six binary classifiers, as in Section~\ref{sec:exp-setup}.

\begin{figure}[!htb]
    \centering
    \includegraphics[width=0.75\linewidth]{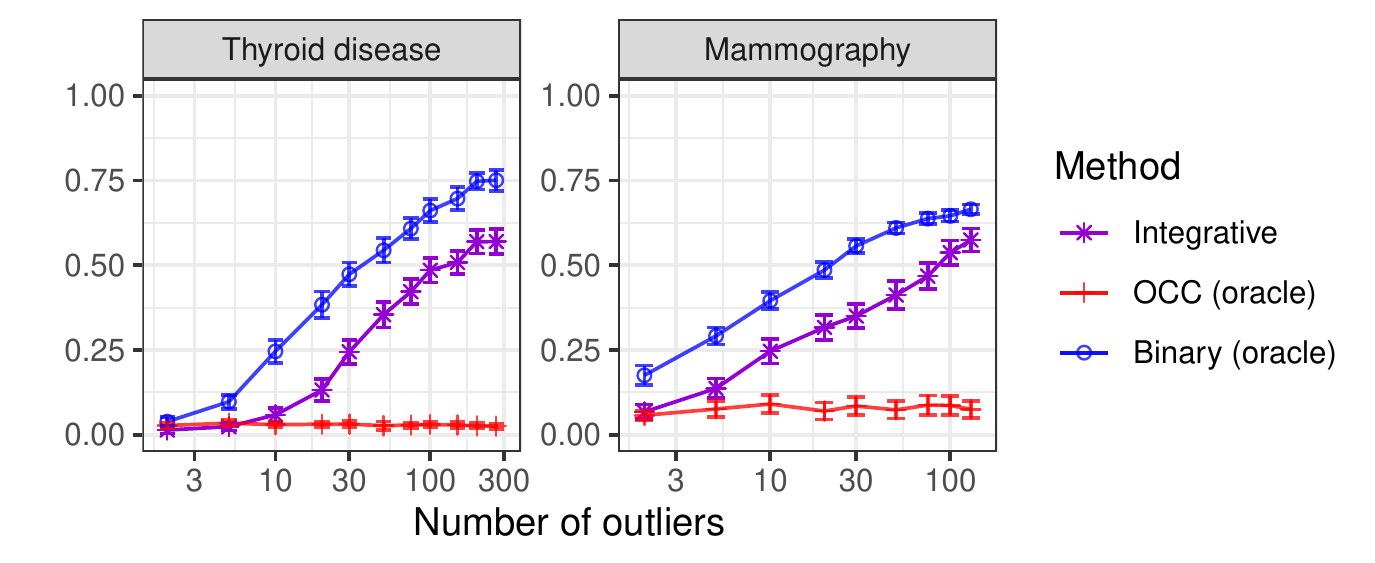}
    \caption{Performance of Storey's BH applied to conformal p-values based on different underlying machine-learning models, on two medical diagnostics data sets. Other details are as in Figure~\ref{fig:exp-animals}.}
    \label{fig:exp-tabular}
\end{figure}

The results in Figure~\ref{fig:exp-tabular} show that binary classifiers tend to yield more powerful standard conformal p-values than one-class classifiers, in contrast with the previous sections. However, Figure~\ref{fig:exp-tabular-oracle} in Appendix~\ref{app:figures} suggest that different binary classifiers can vary in performance substantially  from one another, as also noted in the previous sections for one-class classifiers. The general difficulty of knowing a priori which model will work best for a particular data set makes our integrative approach potentially quite useful even though its power is not as high as that of the binary oracle in these experiments.

\section{Discussion}

This paper has brought together in a natural way several ideas from quite different fields, including inductive~\cite{vovk2005algorithmic} and transductive~\cite{vovk2013transductive} conformal inference, weighted hypothesis testing~\cite{genovese2006false,Caietal20}, and conditional calibration for FDR under dependence~\cite{2020conditionalcalibration}. 
The product is a principled and effective method to compute conformal p-values for out-of-distribution testing, which can leverage the information contained in known outliers to tune and select the most promising model from any imaginable toolbox of black-box classifiers.
A key benefit of this approach is that it removes the typical ambiguity around the optimal choice of machine learning model when computing conformal p-values, thereby allowing practitioners to safely take full advantage of the intrinsic flexibility of the conformal inference framework.
Further, this work opens several opportunities for future research.

First, we have noted that the transductive cross-validation+ technique described in this paper may also be of independent interest in other areas of conformal prediction. We believe such approach may bring practical gains in applications such as multi-class classification~\cite{romano2020classification}.
In fact, the typical inductive version of cross-validation+~\cite{barber2019predictive} requires an often excessively conservative adjustment to guarantee exact inferences in finite samples, which can only be relaxed asymptotically if the underlying machine-learning models satisfy certain stability properties~\cite{steinberger2018conditional}.
By contrast, transductive cross-validation+ is simpler to analyze theoretically and does not require such conservative corrections to achieve exact label-conditional coverage~\cite{vovk2003mondrian,romano2020classification}. Although it can be computationally expensive, it may be useful in applications with small sample sizes, for which cross-validation is especially preferable over data splitting. It is also worth noting that transductive cross-validation+ has one important advantage over typical transductive conformal inference techniques~\cite{vovk2005algorithmic,vovk2013transductive,lei2018distribution}: it always evaluates the conformity scores on hold-out data that were not used for training, and thus it may be less susceptible to overfitting in high-dimensional applications. That being said, it is clear that further work is needed to carefully investigate the possible theoretical and empirical benefits of transductive cross-validation+ in multi-class classification and beyond.

Second, our methods could be re-purposed to compute more powerful conformal prediction sets with label-conditional coverage for multi-class classification~\cite{vovk2003mondrian,romano2020classification}, leveraging the general duality between hypothesis testing and confidence sets.
This paper only took a small step in that direction within Appendix~\ref{sec:cv++_general} for lack of space, but the idea is promising because label-conditional coverage is often practically desirable, including in applications involving fairness concerns~\cite{romano2019malice,rava2021burden}. Constructing label-conditional prediction sets by inverting our integrative conformal p-values would bring all benefits of automatic model selection and tuning to multi-class classification applications.
Third, it may be possible in the future to relax our exchangeability assumptions for the inlier data following an approach similar to that of~\cite{barber2022conformal}, or to improve the performance of our integrative conformal p-values by leveraging possible covariate shift in the outlier distribution with techniques inspired by~\cite{tibshirani2019conformal}.
Fourth, conformal inference techniques have recently started to find new applications to very different types of learning problems (e.g.,~\cite{angelopoulos2022image,sesia2022conformalized,angelopoulos2022conformal}) that are not directly related to out-of-distribution testing, but which may also benefit from some of the ideas developed in this paper, including transductive cross-validation+.
Finally, this paper may help inspire future research into other multiple testing problems for which conditional FDR calibration~\cite{2020conditionalcalibration} can be effectively deployed, adding for example to the recent contributions of~\cite{luo2022improving} in the context of conditional independence testing with knockoffs.

\subsection*{Software availability}

Software implementing the methods and analyses described in this paper is available online at
\url{https://github.com/ZiyiLiang/weighted_conformal_pvalues}.

\subsection*{Acknowledgements}

M.~S.~was partly supported by NSF grant DMS-2210637.

\FloatBarrier

\printbibliography

@article{mary2021semi,
  title={Semi-supervised multiple testing},
  author={Mary, David and Roquain, Etienne},
  journal={arXiv preprint arXiv:2106.13501},
  year={2021}
}

@article{li2017accumulation,
  title={Accumulation tests for FDR control in ordered hypothesis testing},
  author={Li, Ang and Barber, Rina Foygel},
  journal={Journal of the American Statistical Association},
  volume={112},
  number={518},
  pages={837--849},
  year={2017},
  publisher={Taylor \& Francis}
}

@article{lei2018adapt,
  title={{AdaPT}: an interactive procedure for multiple testing with side information},
  author={Lei, Lihua and Fithian, William},
  journal={Journal of the Royal Statistical Society: Series B (Statistical Methodology)},
  volume={80},
  number={4},
  pages={649--679},
  year={2018},
  publisher={Wiley Online Library}
}

@article{ignatiadis2021covariate,
  title={Covariate powered cross-weighted multiple testing},
  author={Ignatiadis, Nikolaos and Huber, Wolfgang},
  journal={Journal of the Royal Statistical Society: Series B (Statistical Methodology)},
  volume={83},
  number={4},
  pages={720--751},
  year={2021},
  publisher={Wiley Online Library}
}

@article{li2019multiple,
  title={Multiple testing with the structure-adaptive Benjamini--Hochberg algorithm},
  author={Li, Ang and Barber, Rina Foygel},
  journal={Journal of the Royal Statistical Society: Series B (Statistical Methodology)},
  volume={81},
  number={1},
  pages={45--74},
  year={2019},
  publisher={Wiley Online Library}
}

@article{hu2010false,
  title={False discovery rate control with groups},
  author={Hu, James X and Zhao, Hongyu and Zhou, Harrison H},
  journal={Journal of the American Statistical Association},
  volume={105},
  number={491},
  pages={1215--1227},
  year={2010},
  publisher={Taylor \& Francis}
}

@article{tibshirani2019conformal,
  title={Conformal prediction under covariate shift},
  author={Tibshirani, Ryan J and Foygel Barber, Rina and Candes, Emmanuel and Ramdas, Aaditya},
  journal={Advances in neural information processing systems},
  volume={32},
  year={2019}
}

@article{barber2022conformal,
  title={Conformal prediction beyond exchangeability},
  author={Barber, Rina Foygel and Cand{\`e}s, Emmanuel J and Ramdas, Aaditya and Tibshirani, Ryan J},
  journal={arXiv preprint arXiv:2202.13415},
  year={2022}
}

@article{tong2013plug,
  title={A plug-in approach to neyman-pearson classification},
  author={Tong, Xin},
  journal={The Journal of Machine Learning Research},
  volume={14},
  number={1},
  pages={3011--3040},
  year={2013},
  publisher={JMLR. org}
}

@article{zhao2016neyman,
  title={Neyman-Pearson classification under high-dimensional settings},
  author={Zhao, Anqi and Feng, Yang and Wang, Lie and Tong, Xin},
  journal={The Journal of Machine Learning Research},
  volume={17},
  number={1},
  pages={7469--7507},
  year={2016},
  publisher={JMLR. org}
}

@article{rigollet2011neyman,
  title={Neyman-pearson classification, convexity and stochastic constraints},
  author={Rigollet, Philippe and Tong, Xin},
  journal={Journal of Machine Learning Research},
  year={2011}
}

@article{tong2018neyman,
  title={Neyman-Pearson classification algorithms and NP receiver operating characteristics},
  author={Tong, Xin and Feng, Yang and Li, Jingyi Jessica},
  journal={Science advances},
  volume={4},
  number={2},
  pages={eaao1659},
  year={2018},
  publisher={American Association for the Advancement of Science}
}

@article{angelopoulos2022conformal,
  title={Conformal Risk Control},
  author={Angelopoulos, Anastasios N and Bates, Stephen and Fisch, Adam and Lei, Lihua and Schuster, Tal},
  journal={arXiv preprint arXiv:2208.02814},
  year={2022}
}

@article{angelopoulos2021gentle,
  title={A gentle introduction to conformal prediction and distribution-free uncertainty quantification},
  author={Angelopoulos, Anastasios N and Bates, Stephen},
  journal={arXiv preprint arXiv:2107.07511},
  year={2021}
}

@inproceedings{angelopoulos2022image,
  title={Image-to-image regression with distribution-free uncertainty quantification and applications in imaging},
  author={Angelopoulos, Anastasios N and Kohli, Amit Pal and Bates, Stephen and Jordan, Michael and Malik, Jitendra and Alshaabi, Thayer and Upadhyayula, Srigokul and Romano, Yaniv},
  booktitle={International Conference on Machine Learning},
  pages={717--730},
  year={2022},
  organization={PMLR}
}

@article{angelopoulos2020uncertainty,
  title={Uncertainty sets for image classifiers using conformal prediction},
  author={Angelopoulos, Anastasios and Bates, Stephen and Malik, Jitendra and Jordan, Michael I},
  journal={arXiv preprint arXiv:2009.14193},
  year={2020}
}

@article{luo2022improving,
  title={Improving knockoffs with conditional calibration},
  author={Luo, Yixiang and Fithian, William and Lei, Lihua},
  journal={unpublished manuscript},
  year={2022}
}

@article{steinberger2018conditional,
  title={Conditional predictive inference for stable algorithms},
  author={Steinberger, Lukas and Leeb, Hannes},
  journal={arXiv preprint arXiv:1809.01412},
  year={2018}
}

@article{clifton2014probabilistic,
  title={Probabilistic novelty detection with support vector machines},
  author={Clifton, Lei and Clifton, David A and Zhang, Yang and Watkinson, Peter and Tarassenko, Lionel and Yin, Hujun},
  journal={IEEE Transactions on Reliability},
  volume={63},
  number={2},
  pages={455--467},
  year={2014},
  publisher={IEEE}
}

@inproceedings{hautamaki2004outlier,
  title={Outlier detection using k-nearest neighbour graph},
  author={Hautamaki, Ville and Karkkainen, Ismo and Franti, Pasi},
  booktitle={Proceedings of the 17th International Conference on Pattern Recognition, 2004. ICPR 2004.},
  volume={3},
  pages={430--433},
  year={2004},
  organization={IEEE}
}

@article{markou2003novelty,
  title={Novelty detection: a review—part 1: statistical approaches},
  author={Markou, Markos and Singh, Sameer},
  journal={Signal Process.},
  volume={83},
  number={12},
  pages={2481--2497},
  year={2003},
  publisher={Elsevier}
}

@article{desir2013one,
  title={One class random forests},
  author={D{\'e}sir, Chesner and Bernard, Simon and Petitjean, Caroline and Heutte, Laurent},
  journal={Pattern Recognition},
  volume={46},
  number={12},
  pages={3490--3506},
  year={2013},
  publisher={Elsevier}
}

@article{pimentel2014review,
  title={A review of novelty detection},
  author={Pimentel, Marco AF and Clifton, David A and Clifton, Lei and Tarassenko, Lionel},
  journal={Signal Process.},
  volume=99,
  pages={215--249},
  year=2014,
  publisher={Elsevier}
}

@article{grubbs1969procedures,
  title={Procedures for detecting outlying observations in samples},
  author={Grubbs, Frank E},
  journal={Technometrics},
  volume={11},
  number={1},
  pages={1--21},
  year={1969},
  publisher={Taylor \& Francis}
}

@inproceedings{haroush2021statistical,
  title={A Statistical Framework for Efficient Out of Distribution Detection in Deep Neural Networks},
  author={Haroush, Matan and Frostig, Tzviel and Heller, Ruth and Soudry, Daniel},
  booktitle={International Conference on Learning Representations},
  year={2021}
}

@article{kim2020predictive,
  title={Predictive inference is free with the jackknife+-after-bootstrap},
  author={Kim, Byol and Xu, Chen and Barber, Rina},
  journal={Advances in Neural Information Processing Systems},
  volume={33},
  pages={4138--4149},
  year={2020}
}

@article{fannjiang2022conformal,
  title={Conformal prediction for the design problem},
  author={Fannjiang, Clara and Bates, Stephen and Angelopoulos, Anastasios and Listgarten, Jennifer and Jordan, Michael I},
  journal={arXiv preprint arXiv:2202.03613},
  year={2022}
}

@article{xu2021conformal,
  title={Conformal Anomaly Detection on Spatio-Temporal Observations with Missing Data},
  author={Xu, Chen and Xie, Yao},
  journal={arXiv preprint arXiv:2105.11886},
  year={2021}
}

@article{gupta2022nested,
  title={Nested conformal prediction and quantile out-of-bag ensemble methods},
  author={Gupta, Chirag and Kuchibhotla, Arun K and Ramdas, Aaditya},
  journal={Pattern Recognition},
  volume={127},
  pages={108496},
  year={2022},
  publisher={Elsevier}
}

@article{linusson2020efficient,
  title={Efficient conformal predictor ensembles},
  author={Linusson, Henrik and Johansson, Ulf and Bostr{\"o}m, Henrik},
  journal={Neurocomputing},
  volume={397},
  pages={266--278},
  year={2020},
  publisher={Elsevier}
}

@inproceedings{beganovic2018ensemble,
  title={Ensemble cross-conformal prediction},
  author={Beganovic, Dorian and Smirnov, Evgueni},
  booktitle={2018 IEEE International Conference on Data Mining Workshops (ICDMW)},
  pages={870--877},
  year={2018},
  organization={IEEE}
}

@inproceedings{lofstrom2013effective,
  title={Effective utilization of data in inductive conformal prediction},
  author={L{\"o}fstr{\"o}m, Tuve and Johansson, Ulf and Bostr{\"o}m, Henrik},
  booktitle={International Joint Conference on Neural Networks, Dallas, TX, USA, August 4-9, 2013.},
  year={2013},
  organization={IEEE}
}

@inproceedings{colombo2020training,
  title={Training conformal predictors},
  author={Colombo, Nicolo and Vovk, Vladimir},
  booktitle={Conformal and Probabilistic Prediction and Applications},
  pages={55--64},
  year={2020},
  organization={PMLR}
}

@article{einbinder2022training,
  title={Training Uncertainty-Aware Classifiers with Conformalized Deep Learning},
  author={Einbinder, Bat-Sheva and Romano, Yaniv and Sesia, Matteo and Zhou, Yanfei},
  journal={arXiv preprint arXiv:2205.05878},
  year={2022}
}

@article{OpenML2013,
      author = {Joaquin Vanschoren and Jan N. van Rijn and Bernd Bischl and Luis Torgo},
      title = {OpenML: networked science in machine learning},
      journal = {SIGKDD Explorations},
      volume = {15},
      number = {2},
      year = {2013},
      pages = {49-60},
      url = {http://doi.acm.org/10.1145/2641190.264119},
      doi = {10.1145/2641190.2641198},
      publisher = {ACM}
    }

@misc{Dua:2019,
author = "Dua, Dheeru and Graff, Casey",
year = "2017",
title = "{UCI} Machine Learning Repository",
url = "http://archive.ics.uci.edu/ml",
institution = "University of California, Irvine, School of Information and Computer Sciences" }

@inproceedings{quinlan1987inductive,
  title={Inductive knowledge acquisition: a case study},
  author={Quinlan, John Ross and Compton, Paul J and Horn, KA and Lazarus, Leslie},
  booktitle={Proceedings of the Second Australian Conference on Applications of expert systems},
  pages={137--156},
  year={1987}
}

@article{roy2018effects,
title={Effects of Degradations on Deep Neural Network Architectures},
author={Roy, Prasun and Ghosh, Subhankar and Bhattacharya, Saumik and Pal, Umapada},
journal={arXiv preprint arXiv:1807.10108},
year=2018
}

@inproceedings{krause20133d,
  title={3d object representations for fine-grained categorization},
  author={Krause, Jonathan and Stark, Michael and Deng, Jia and Fei-Fei, Li},
  booktitle={Proceedings of the IEEE international conference on computer vision workshops},
  pages={554--561},
  year={2013}
}

@ONLINE {tfflowers,
author = "The TensorFlow Team",
title = "Flowers",
month = "jan",
year = "2019",
url = "http://download.tensorflow.org/example_images/flower_photos.tgz" }

@inproceedings{he2016deep,
  title={Deep residual learning for image recognition},
  author={He, Kaiming and Zhang, Xiangyu and Ren, Shaoqing and Sun, Jian},
  booktitle={Proceedings of the IEEE conference on computer vision and pattern recognition},
  pages={770--778},
  year={2016}
}

@inproceedings{deng2009imagenet,
  title={Imagenet: A large-scale hierarchical image database},
  author={Deng, Jia and Dong, Wei and Socher, Richard and Li, Li-Jia and Li, Kai and Fei-Fei, Li},
  booktitle={2009 IEEE conference on computer vision and pattern recognition},
  pages={248--255},
  year={2009},
  organization={Ieee}
}

@misc{tensorflow2015-whitepaper,
title={ {TensorFlow}: Large-Scale Machine Learning on Heterogeneous Systems},
url={https://www.tensorflow.org/},
note={Software available from tensorflow.org},
author={
    Mart\'{i}n~Abadi and
    Ashish~Agarwal and
    Paul~Barham and
    Eugene~Brevdo and
    Zhifeng~Chen and
    Craig~Citro and
    Greg~S.~Corrado and
    Andy~Davis and
    Jeffrey~Dean and
    Matthieu~Devin and
    Sanjay~Ghemawat and
    Ian~Goodfellow and
    Andrew~Harp and
    Geoffrey~Irving and
    Michael~Isard and
    Yangqing Jia and
    Rafal~Jozefowicz and
    Lukasz~Kaiser and
    Manjunath~Kudlur and
    Josh~Levenberg and
    Dandelion~Man\'{e} and
    Rajat~Monga and
    Sherry~Moore and
    Derek~Murray and
    Chris~Olah and
    Mike~Schuster and
    Jonathon~Shlens and
    Benoit~Steiner and
    Ilya~Sutskever and
    Kunal~Talwar and
    Paul~Tucker and
    Vincent~Vanhoucke and
    Vijay~Vasudevan and
    Fernanda~Vi\'{e}gas and
    Oriol~Vinyals and
    Pete~Warden and
    Martin~Wattenberg and
    Martin~Wicke and
    Yuan~Yu and
    Xiaoqiang~Zheng},
  year={2015},
}

@inproceedings{song2019selfie,
  title={Selfie: Refurbishing unclean samples for robust deep learning},
  author={Song, Hwanjun and Kim, Minseok and Lee, Jae-Gil},
  booktitle={International Conference on Machine Learning},
  pages={5907--5915},
  year={2019},
  organization={PMLR}
}

@inproceedings{vovk2013transductive,
  title={Transductive conformal predictors},
  author={Vovk, Vladimir},
  booktitle={IFIP International Conference on Artificial Intelligence Applications and Innovations},
  pages={348--360},
  year={2013},
  organization={Springer}
}

@article{liang2022locally,
  title={Locally Adaptive Transfer Learning Algorithms for Large-Scale Multiple Testing},
  author={Liang, Ziyi and Cai, T Tony and Sun, Wenguang and Xia, Yin},
  journal={arXiv preprint arXiv:2203.11461},
  year={2022}
}

@article{ignatiadis2016data,
  title={Data-driven hypothesis weighting increases detection power in genome-scale multiple testing},
  author={Ignatiadis, Nikolaos and Klaus, Bernd and Zaugg, Judith B and Huber, Wolfgang},
  journal={Nature methods},
  volume={13},
  number={7},
  pages={577--580},
  year={2016},
  publisher={Nature Publishing Group}
}

@book{efron2012large,
  title={Large-scale inference: empirical Bayes methods for estimation, testing, and prediction},
  author={Efron, Bradley},
  volume={1},
  year={2012},
  publisher={Cambridge University Press}
}

@article{genovese2006false,
  title={False discovery control with p-value weighting},
  author={Genovese, Christopher R and Roeder, Kathryn and Wasserman, Larry},
  journal={Biometrika},
  volume={93},
  number={3},
  pages={509--524},
  year={2006},
  publisher={Oxford University Press}
}

@article{sesia2022conformalized,
  title={Conformalized Frequency Estimation from Sketched Data},
  author={Sesia, Matteo and Favaro, Stefano},
  journal={arXiv preprint arXiv:2204.04270},
  year={2022}
}

@article{lei2014distribution,
  title={Distribution-free prediction bands for non-parametric regression},
  author={Lei, Jing and Wasserman, Larry},
  journal={Journal of the Royal Statistical Society: Series B (Statistical Methodology)},
  volume={76},
  number={1},
  pages={71--96},
  year={2014},
  publisher={Wiley Online Library}
}

@inproceedings{kivaranovic2019adaptive,
  title={Adaptive, distribution-free prediction intervals for deep networks},
  author={Kivaranovic, Danijel and Johnson, Kory D and Leeb, Hannes},
  booktitle={International Conference on Artificial Intelligence and Statistics},
  pages={4346--4356},
  year={2020},
  organization={PMLR}
}

@article{kuchibhotla2019nested,
  title={Nested Conformal Prediction and the Generalized Jackknife+},
  author={Kuchibhotla, Arun K and Ramdas, Aaditya K},
  journal={arXiv preprint arXiv:1910.10562},
  year={2019}
}

@inproceedings{vovk2012conditional,
  title={Conditional validity of inductive conformal predictors},
  author={Vovk, Vladimir},
  booktitle={Asian conference on machine learning},
  pages={475--490},
  year={2012},
}

@inproceedings{romano2019conformalized,
  title={Conformalized quantile regression},
  author={Romano, Yaniv and Patterson, Evan and Cand{\`e}s, Emmanuel J},
  booktitle={Advances in Neural Information Processing Systems},
  pages={3538--3548},
  year={2019},
}

@article{barber2019limits,
  title={The limits of distribution-free conditional predictive inference},
  author={Barber, Rina Foygel and Cand{\`e}s, Emmanuel J and Ramdas, Aaditya and Tibshirani, Ryan J},
  journal={arXiv preprint arXiv:1903.04684},
  year={2019},
}

@article{sadinle2019least,
  title={Least ambiguous set-valued classifiers with bounded error levels},
  author={Sadinle, Mauricio and Lei, Jing and Wasserman, Larry},
  journal={Journal of the American Statistical Association},
  volume={114},
  number={525},
  pages={223--234},
  year={2019},
  publisher={Taylor \& Francis},
}

@techreport{vovk2003mondrian,
title = {Mondrian Confidence Machine},
author = {Vladimir Vovk and David Lindsay and Ilia Nouretdinov and Alex Gammerman},
note = {On-line Compression Modelling project},
year = {2003},
series = {On-line Compression Modelling project},
type = {Technical Report},
institution = {Royal Holloway, University of London},
}

@article{hechtlinger2018cautious,
  title={Cautious deep learning},
  author={Hechtlinger, Yotam and P{\'o}czos, Barnab{\'a}s and Wasserman, Larry},
  journal={arXiv preprint arXiv:1805.09460},
  year={2018},
}

@article{lei2018distribution,
	title={Distribution-free predictive inference for regression},
	author={Lei, Jing and G'Sell, Max and Rinaldo, Alessandro and Tibshirani, Ryan J. and Wasserman, Larry},
	journal={Journal of the American Statistical Association},
	volume={113},
	number={523},
	pages={1094--1111},
	year={2018},
	publisher={Taylor \& Francis},
}

@book{vovk2005algorithmic,
	title={Algorithmic learning in a random world},
	author={Vovk, Vladimir and Gammerman, Alex and Shafer, Glenn},
	year={2005},
	publisher={Springer},
}

@article{sesia2021conformal,
  title={Conformal Prediction using Conditional Histograms},
  author={Sesia, Matteo and Romano, Yaniv},
  journal={Advances in Neural Information Processing Systems},
  volume={34},
  year={2021}
}

@article{fithian2020conditional,
  title={Conditional calibration for false discovery rate control under dependence},
  author={Fithian, William and Lei, Lihua},
  journal={arXiv preprint arXiv:2007.10438},
  year={2020}
}

@article{storey2002direct,
  title={A direct approach to false discovery rates},
  author={Storey, John D},
  journal={Journal of the Royal Statistical Society: Series B (Statistical Methodology)},
  volume={64},
  number={3},
  pages={479--498},
  year={2002},
  publisher={Wiley Online Library}
}

@inproceedings{sabokrou2018adversarially,
  title={Adversarially learned one-class classifier for novelty detection},
  author={Sabokrou, Mohammad and Khalooei, Mohammad and Fathy, Mahmood and Adeli, Ehsan},
  booktitle={Proceedings of the IEEE Conference on Computer Vision and Pattern Recognition},
  pages={3379--3388},
  year={2018}
}

@article{khan2014one,
  title={One-class classification: taxonomy of study and review of techniques},
  author={Khan, Shehroz S and Madden, Michael G},
  journal={The Knowledge Engineering Review},
  volume={29},
  number={3},
  pages={345--374},
  year={2014},
  publisher={Cambridge University Press}
}

@inproceedings{cai2020real,
  title={Real-time Out-of-distribution Detection in Learning-Enabled Cyber-Physical Systems},
  author={Cai, Feiyang and Koutsoukos, Xenofon},
  booktitle={2020 ACM/IEEE 11th International Conference on Cyber-Physical Systems (ICCPS)},
  pages={174--183},
  year={2020},
  organization={IEEE}
}

@article{laxhammar2015inductive,
  title={Inductive conformal anomaly detection for sequential detection of anomalous sub-trajectories},
  author={Laxhammar, Rikard and Falkman, G{\"o}ran},
  journal={Annals of Mathematics and Artificial Intelligence},
  volume={74},
  number={1-2},
  pages={67--94},
  year={2015},
  publisher={Springer}
}

@inproceedings{smith2015conformal,
  title={Conformal anomaly detection of trajectories with a multi-class hierarchy},
  author={Smith, James and Nouretdinov, Ilia and Craddock, Rachel and Offer, Charles and Gammerman, Alexander},
  booktitle={International symposium on statistical learning and data sciences},
  pages={281--290},
  year={2015},
  organization={Springer}
}

@article{benjamini2001control,
  title={The control of the false discovery rate in multiple testing under dependency},
  author={Benjamini, Yoav and Yekutieli, Daniel},
  journal={Annals of Statistics},
  pages={1165--1188},
  year={2001},
  publisher={JSTOR}
}

@article{moya1993one,
  title={One-class classifier networks for target recognition applications},
  author={Moya, Mary M and Koch, Mark W and Hostetler, Larry D},
  journal={NASA STI/Recon Technical Report N},
  volume={93},
  pages={24043},
  year={1993}
}

@article{benjamini1995controlling,
  title={Controlling the false discovery rate: a practical and powerful approach to multiple testing},
  author={Benjamini, Yoav and Hochberg, Yosef},
  journal={Journal of the Royal statistical society: series B (Methodological)},
  volume={57},
  number={1},
  pages={289--300},
  year={1995},
  publisher={Wiley Online Library}
}

@article{scikit-learn,
 title={Scikit-learn: Machine Learning in {P}ython},
 author={Pedregosa, F. and Varoquaux, G. and Gramfort, A. and Michel, V.
         and Thirion, B. and Grisel, O. and Blondel, M. and Prettenhofer, P.
         and Weiss, R. and Dubourg, V. and Vanderplas, J. and Passos, A. and
         Cournapeau, D. and Brucher, M. and Perrot, M. and Duchesnay, E.},
 journal={Journal of Machine Learning Research},
 volume={12},
 pages={2825--2830},
 year={2011}
}

@inproceedings{izbicki2019flexible,
  title={Flexible distribution-free conditional predictive bands using density estimators},
  author={Izbicki, Rafael and Shimizu, Gilson and Stern, Rafael},
  booktitle={International Conference on Artificial Intelligence and Statistics},
  pages={3068--3077},
  year={2020},
  organization={PMLR}
}

@article{chernozhukov2019distributional,
  title={Distributional conformal prediction},
  author={Chernozhukov, Victor and W{\"u}thrich, Kaspar and Zhu, Yinchu},
  journal={Proceedings of the National Academy of Sciences},
  volume={118},
  number={48},
  pages={e2107794118},
  year={2021},
  publisher={National Acad Sciences}
}

@article{romano2020classification,
    title={Classification with Valid and Adaptive Coverage},
    author={Yaniv Romano and Matteo Sesia and Emmanuel J. Cand{\`e}s},
    year={2020},
  journal={Advances in Neural Information Processing Systems},
  volume={33},
  year={2020}
}

@article{barber2019predictive,
  title={Predictive inference with the jackknife+},
  author={Barber, Rina Foygel and Cand{\`e}s, Emmanuel J and Ramdas, Aaditya and Tibshirani, Ryan J and others},
  journal={Annals of Statistics},
  volume={49},
  number={1},
  pages={486--507},
  year={2021},
  publisher={Institute of Mathematical Statistics}
}

@article{romano2019malice,
	title = {With Malice Toward None: Assessing Uncertainty via Equalized Coverage},
	author = {Romano, Yaniv and Barber, Rina Foygel and Sabatti, Chiara and Cand{\`e}s, Emmanuel},
	year = {to appear},
	journal = {Harvard Data Science Review},
}

@article{cauchois2020knowing,
  title={Knowing what You Know: valid and validated confidence sets in multiclass and multilabel prediction.},
  author={Cauchois, Maxime and Gupta, Suyash and Duchi, John C},
  journal={J. Mach. Learn. Res.},
  volume={22},
  pages={81--1},
  year={2021}
}

@inproceedings{vovk1999machine,
	title={Machine-learning applications of algorithmic randomness},
	author={Vovk, Vladimir and Gammerman, Alexander and Saunders, Craig},
	booktitle={{International Conference on Machine Learning}},
	pages={444--453},
	year={1999}
}

@article{storey2004strong,
  title={Strong control, conservative point estimation and simultaneous conservative consistency of false discovery rates: a unified approach},
  author={Storey, John D and Taylor, Jonathan E and Siegmund, David},
  journal={Journal of the Royal Statistical Society: Series B (Statistical Methodology)},
  volume={66},
  number={1},
  pages={187--205},
  year={2004},
  publisher={Wiley Online Library}
}

@article{guan2019prediction,
  title={Prediction and outlier detection in classification problems},
  author={Guan, Leying and Tibshirani, Rob},
  journal={arXiv preprint arXiv:1905.04396},
  year={2019}
}

@inproceedings{ishimtsev2017conformal,
  title={Conformal $ k $-NN Anomaly Detector for Univariate Data Streams},
  author={Ishimtsev, Vladislav and Bernstein, Alexander and Burnaev, Evgeny and Nazarov, Ivan},
  booktitle={Conformal and Probabilistic Prediction and Applications},
  pages={213--227},
  year={2017},
  organization={PMLR}
}

@article{rava2021burden,
  title={A Burden Shared is a Burden Halved: A Fairness-Adjusted Approach to Classification},
  author={Rava, Bradley and Sun, Wenguang and James, Gareth M and Tong, Xin},
  journal={arXiv preprint arXiv:2110.05720},
  year={2021}
}

@article{bates2021testing,
  title={Testing for outliers with conformal p-values},
  author={Bates, Stephen and Cand{\`e}s, Emmanuel and Lei, Lihua and Romano, Yaniv and Sesia, Matteo},
  journal={arXiv preprint arXiv:2104.08279},
  year={2021}
}

@article{2020conditionalcalibration,
  title={Conditional calibration for false discovery rate control under dependence},
  author={Fithian, William and Lei, Lihua},
  journal={arXiv preprint arXiv:2007.10438},
  year={2020}
}

@article{Genetal06,
 ISSN = {00063444},
 URL = {http://www.jstor.org/stable/20441304},
 author = {Christopher R. Genovese and Kathryn Roeder and Larry Wasserman},
 journal = {Biometrika},
 number = {3},
 pages = {509--524},
 publisher = {[Oxford University Press, Biometrika Trust]},
 title = {False Discovery Control with p-Value Weighting},
 urldate = {2022-04-27},
 volume = {93},
 year = {2006}
}

@article{Caietal20,
	Author = {Cai, T Tony and Sun, Wenguang and Yin Xia},
	  title={{LAWS}: A Locally Adaptive Weighting and Screening Approach To Spatial Multiple Testing},
  author={Cai, T Tony and Sun, Wenguang and Xia, Yin},
  journal={J. Am. Statist. Assoc.},
  pages={1--30},
  year={2021},
  publisher={Taylor \& Francis}
}

@article{basu2018weighted,
  title={Weighted false discovery rate control in large-scale multiple testing},
  author={Basu, Pallavi and Cai, T Tony and Das, Kiranmoy and Sun, Wenguang},
  journal={J. Am. Statist. Assoc.},
  volume={113},
  number={523},
  pages={1172--1183},
  year={2018},
  publisher={Taylor \& Francis}
}

@article{benjamini1997weights,
 ISSN = {03036898, 14679469},
 URL = {http://www.jstor.org/stable/4616464},
 author = {Yoav Benjamini and Yosef Hochberg},
 journal = {Scandinavian Journal of Statistics},
 number = {3},
 pages = {407--418},
 publisher = {[Board of the Foundation of the Scandinavian Journal of Statistics, Wiley]},
 title = {Multiple Hypotheses Testing with Weights},
 urldate = {2022-08-22},
 volume = {24},
 year = {1997}
}

@article{roquain2008optimal,
author = {Roquain, Etienne and Van de Wiel, Mark},
year = {2008},
month = {07},
pages = {},
title = {Optimal weighting for false discovery rate control},
volume = {3},
journal = {Electronic Journal of Statistics},
doi = {10.1214/09-EJS430}
}

\appendix
\renewcommand{\thesection}{A\arabic{section}}
\renewcommand{\theequation}{A\arabic{equation}}
\renewcommand{\thetheorem}{A\arabic{theorem}}
\renewcommand{\theproposition}{A\arabic{proposition}}
\renewcommand{\thelemma}{A\arabic{lemma}}
\renewcommand{\thetable}{A\arabic{table}}
\renewcommand{\thefigure}{A\arabic{figure}}
\renewcommand{\thealgorithm}{A\arabic{algorithm}}
\setcounter{figure}{0}
\setcounter{table}{0}
\setcounter{proposition}{0}
\setcounter{theorem}{0}
\setcounter{lemma}{0}
\setcounter{algorithm}{0}

\section{Additional methodological details}  \label{appendix:method-details}

\subsection{TCV+ beyond conformal out-of-distribution testing} \label{sec:cv++_general}

The TCV+ method presented in Section~\ref{sec:adaptive-weighting-cv+} can be easily extended to compute prediction sets with guaranteed marginal coverage for multi-class classification, as outlined in Algorithm~\ref{alg_cv++}. This solution has the advantage of enjoying tighter theoretical coverage ($1-\alpha$ instead of $1-2\alpha$) compared to the traditional inductive cross-validation+ approach of~\cite{barber2019predictive} and~\cite{romano2020classification}.

\begin{theorem} \label{theorem:general_tcv+}
  If $(X_{n+1},Y_{n+1})$ is exchangeable with the labeled data in $\mathcal{D}$ and $\hat{\mathcal{C}}_{\alpha}(X_{n+1})$ is computed by Algorithm~\ref{alg_cv++}, then $\P{Y_{n+1} \in \hat{\mathcal{C}}_{\alpha}(X_{n+1}) } \geq 1-\alpha$ for any $\alpha \in (0,1)$.
\end{theorem}

\begin{algorithm}[H]
  \caption{TCV+ prediction sets  for multi-class classification}
  \label{alg_cv++}
  \begin{algorithmic}[1]
    \State \textbf{Input}: labeled data set $\mathcal{D}$, black-box classifier  $\mathcal{A}$, test point $X_{n+1}$, number of folds $K$, number of possible classes $C$, confidence level $\alpha \in (0,1)$.
    \For{$y = 1 \dots C$}
    \State Randomly split $\mathcal{D} \cup \{n+1\}$ into K disjoint folds $\mathcal{D}^1, \dots, \mathcal{D}^K$.
    \For{$k = 1 \dots K$}
    \State Train $\mathcal{A}$ on $\mathcal{D} \setminus \mathcal{D}^k$, pretending $Y_{n+1}=y$.
    \State Calculate conformity scores $s^y(X_j; k)$ for all $j\in \mathcal{D}^k$ based on $\mathcal{A}$; e.g., as in~\cite{romano2020classification}.
    \EndFor
    \State Let $k(i)$ denote the fold of $X_i$ and calculate
    \begin{align*}
      \hat{u}(X_{n+1}; y) = \frac{1+\sum_{i \in \mathcal{D}} \I{ s^y(X_{i}; k(i)) \leq s^y(X_{n+1}; k(n+1))} }{1+|\mathcal{D}|}.
    \end{align*}
    Define $\hat{\mathcal{C}}_{\alpha}(X_{n+1}) \subseteq \{1,\ldots, C\}$  as:
    \begin{align*}
      \hat{\mathcal{C}}_{\alpha}(X_{n+1}) = \left\{ y \in \{1,\ldots, C\} : \hat{u}(X_{n+1}; y) > \alpha \right\}.
    \end{align*}
  \EndFor
    \State \textbf{Output} A prediction set $\hat{\mathcal{C}}_{\alpha}(X_{n+1})$ for the unknown label $Y_{n+1}$.
  \end{algorithmic}
\end{algorithm}

The connection between Algorithm~\ref{alg_clra_cv+} and Algorithm~\ref{alg_cv++} becomes clearer if we consider also Algorithm~\ref{alg_cv++-label-conditional}: a further extension of Algorithm~\ref{alg_cv++} designed to construct multi-class classification prediction sets with guaranteed {\em label-conditional}~\cite{vovk2005algorithmic,sadinle2019least,romano2020classification} coverage.

\begin{theorem} \label{theorem:general_tcv+-label-conditional}
Fix any possible label $y$ and suppose $Y_{n+1}$.
  If $(X_{n+1},Y_{n+1})$ is exchangeable with the labeled data points $i$ in $\mathcal{D}$ with $Y_i=y$, and $\hat{\mathcal{C}}_{\alpha}(X_{n+1})$ is computed by Algorithm~\ref{alg_cv++}, then $\P{Y_{n+1} \in \hat{\mathcal{C}}_{\alpha}(X_{n+1}) \mid Y_{n+1} = y} \geq 1-\alpha$ for any $\alpha \in (0,1)$ and for all possible labels $y$.
\end{theorem}

\begin{algorithm}[H]
  \caption{TCV+ label-conditional prediction sets for multi-class classification}
  \label{alg_cv++-label-conditional}
  \begin{algorithmic}[1]
    \State \textbf{Input}: labeled data set $\mathcal{D}$, black-box classifier  $\mathcal{A}$, test point $X_{n+1}$, number of folds $K$, number of possible classes $C$, confidence level $\alpha \in (0,1)$.
    \For{$y = 1 \dots C$}
    \State Define $\mathcal{D}_y = \{i \in \mathcal{D} : Y_i = y$.
    \State Randomly split $\mathcal{D}_y \cup \{n+1\}$ into K disjoint folds $\mathcal{D}_y^1, \dots, \mathcal{D}_y^K$.
    \For{$k = 1 \dots K$}
    \State Train $\mathcal{A}$ on $\mathcal{D} \setminus \mathcal{D}_y^k$, pretending that $Y_{n+1}=y$.
    \State Calculate conformity scores $s^y(X_j; k)$ for all $j\in \mathcal{D}_y^k$ based on $\mathcal{A}$; e.g., as in~\cite{romano2020classification}.
    \EndFor
    \State Let $k(i)$ denote the fold of $X_i$ and calculate
    \begin{align*}
      \hat{u}(X_{n+1}; y) = \frac{1+\sum_{i \in \mathcal{D}_y} \I{ s^y(X_{i}; k(i)) \leq s^y(X_{n+1}; k(n+1))} }{1+|\mathcal{D}_y|}.
    \end{align*}
    Define $\hat{\mathcal{C}}_{\alpha}(X_{n+1}) \subseteq \{1,\ldots, C\}$  as:
    \begin{align*}
      \hat{\mathcal{C}}_{\alpha}(X_{n+1}) = \left\{ y \in \{1,\ldots, C\} : \hat{u}(X_{n+1}; y) > \alpha \right\}.
    \end{align*}
  \EndFor
    \State \textbf{Output} A prediction set $\hat{\mathcal{C}}_{\alpha}(X_{n+1})$ for the unknown label $Y_{n+1}$.
  \end{algorithmic}
\end{algorithm}

\section{Mathematical proofs} \label{appendix:proofs}

\subsection{Integrative conformal p-values}

\begin{proof}[Proof of Theorem~\ref{theorem:ratio-exchangeable}]
It suffices to show the variables $R_i = \hat{r}(X_i)$ for $i \in \{n+1\} \cup \mathcal{D}_0^{\mathrm{cal}}$ are exchangeable, as that implies the rank of $R_{n+1}$ is uniformly distributed on $\{1,2,\ldots,|\mathcal{D}_0^{\mathrm{cal}}|+1\}$, and hence $\hat{u}(X_{n+1}) \sim \text{Uniform}\left(1/(|\mathcal{D}_0^{\mathrm{cal}}|+1), 2/(|\mathcal{D}_0^{\mathrm{cal}}|+1), \ldots, 1\right)$.
To prove the claim, we start by introducing some helpful notation.
Let us refer to full data set augmented with the unlabeled test point as $\mathcal{D} = \mathcal{D}_0 \cup \mathcal{D}_1 \cup \{n+1\}$, and let $\sigma$ be any permutation of $\{1,2,\ldots,n+1\}$ that leaves invariant all elements of $\{1,2,\ldots,n\} \setminus \mathcal{D}_0^{\mathrm{cal}}$.
Then, let $\sigma(\mathcal{D})$ denote the shuffled data set induced by $\sigma$.
Imagine a ``parallel universe'' in which our method is applied to $\sigma(\mathcal{D})$, with exactly the same recipe and the same random seeds, and let  $\hat{s}'_0$ and $\hat{s}'_1$ denote the conformity score functions obtained in that universe. Clearly, $\hat{s}'_0 = \hat{s}_0$ and $\hat{s}'_1 = \hat{s}_1$ because these functions are determined my machine-learning models trained only on subsets of the data in $\{1,2,\ldots,n\} \setminus \mathcal{D}_0^{\mathrm{cal}}$, which are not altered by $\sigma$.
Further, note the functions $\hat{u}_0$ and $\hat{u}_1$ are also identical to the respective counterparts $\hat{u}'_0$ and $\hat{u}'_1$ in the parallel universe, because the expression in~\eqref{eq:p-value-0} is invariant to permutations of $\{n+1\} \cup \mathcal{D}_{0}^{\mathrm{cal}}$.
This also implies $\hat{r}' = \hat{r}$.
Therefore, the variables $R_i$, for $i \in \{n+1\} \cup \mathcal{D}_{0}^{\mathrm{cal}}$, correspond in the parallel universe to variables $R'_{\sigma(i)} = \hat{r}'(X_{\sigma(i)})$ such that
\begin{align*}
  \{\hat{r}'(X_{\sigma(i)})\}_{i \in \{n+1\} \cup \mathcal{D}_{0}^{\mathrm{cal}}}
  = \bar{\sigma}(\{\hat{r}(X_i)\}_{i \in \{n+1\} \cup \mathcal{D}_{0}^{\mathrm{cal}}}),
\end{align*}
where $\bar{\sigma}$ is understood to be the permutation obtained by restricting $\sigma$ on $\{n+1\} \cup \mathcal{D}_{0}^{\mathrm{cal}}$.
The proof is completed by noting the data exchangeability assumption says
\begin{align*}
   \{\hat{r}'(X_{\sigma(i)})\}_{i \in \{n+1\} \cup \mathcal{D}_{0}^{\mathrm{cal}}}
  \overset{d}{=}
  \{\hat{r}(X_{i})\}_{i \in \{n+1\} \cup \mathcal{D}_{0}^{\mathrm{cal}}},
\end{align*}
and hence
\begin{align*}
  \{\hat{r}(X_{i})\}_{i \in \{n+1\} \cup \mathcal{D}_{0}^{\mathrm{cal}}}
  \overset{d}{=}
  \bar{\sigma}(\{\hat{r}(X_i)\}_{i \in \{n+1\} \cup \mathcal{D}_{0}^{\mathrm{cal}}}).
\end{align*}
\end{proof}

\begin{proof}[Proof of Theorem~\ref{theorem:ratio-exchangeable-tuning}]
This follows from essentially the same argument as in the proof of Theorem \ref{theorem:ratio-exchangeable}. Note that the only difference between Algorithm \ref{alg:weighted-pvalues-tuning} and Algorithm \ref{alg:weighted-pvalues} is the automatic model selection. Hence, it suffices to show that the identity of the selected model is invariant to permutations, and then the rest of the proof follows that of Theorem \ref{theorem:ratio-exchangeable}. 
Note that the model selection is based on the unordered sets of conformity scores $\hat{s}_0^m(\{n+1\} \cup \mathcal{D}_{0})$ and $\hat{s}_0^m(\mathcal{D}_{1})$, which implies that the tuning process is invariant to permutations of $\{n+1\} \cup \mathcal{D}_{0}$ and $m_0*'=m_0*$. The same argument also yields $m_1*'=m_1*$.
\end{proof}

\begin{proof}[Proof of Theorem~\ref{theorem:ratio-cv+}]
For simplicity, imagine $\mathcal{D}_0 \cup \{n+1\} = \{1,\ldots,n,n+1\}$, so that $n=|\mathcal{D}^0|$ is the number of inliers. (This slight abuse of notation is allowed because we do not need to explicitly index the labeled outliers in the proof.)
Suppose each fold $\mathcal{D}_0^k$ has $m=(n+1)/K$ points.
For simplicity, assume $m$ is an integer (this assumption could be relaxed at the cost of heavier notation, as in~\cite{barber2019predictive}).
Define $R \in \mathbb{R}^{(n+1)}$ such that, for each $i \in \{n+1\} \cup \mathcal{D}_0$,
\begin{equation*}
    R_{i}=\hat{r}(X_{i}; k(i)).
\end{equation*}
Define the comparison matrix $A \in \{0,1\}^{(n+1)\times(n+1)}$ with entries
\begin{align*}
    A_{i,j} = 1\left\{ R_{i} > R_{j}\right\},
\end{align*}
and the set of ``strange'' points, for any fixed $\alpha \in (0,1)$,
\begin{equation*}
    \mathcal{S}(A)=\bigg\{i \in \{1, \dots, n+1\}: \sum_{j=1}^{n+1}A_{i,j} \geq (1-\alpha)(1+n) \bigg\}.
\end{equation*}
The proof involves three main steps.

\textbf{Step 1:} The event $\hat{u}(X_{n+1}) \leq \alpha$ is equivalent to $n+1 \in \mathcal{S}(A)$. By definition,
\begin{align*}
    \hat{u}(X_{n+1}) \leq \alpha & \Leftrightarrow
    \sum_{i=1}^{n+1}  \mathbbm{1}\left[ \hat{r}(X_{i}; k(i)) \leq \hat{r}(X_{n+1}; k(n+1)) \right] \leq \alpha (1+n)\\
    & \Leftrightarrow
    \sum_{i=1}^{n+1}  \mathbbm{1}\left[ \hat{r}(X_{i}; k(i)) > \hat{r}(X_{n+1}; k(n+1)) \right] > (1-\alpha) (1+n)\\
    & \Leftrightarrow
    \sum_{i=1}^{n+1}  A_{i,n+1} \geq (1-\alpha) (1+n).
\end{align*}

\textbf{Step 2:} The number of strange points is deterministically bounded from above by $|\mathcal{S}(A)| \leq \alpha (n+1)$.
Note that one can think of the matrix $A$ as keeping score of all pairwise games between $n+1$ distinct players in a chess tournament.
Is it easy to see that any $j \in \{1,\ldots,n+1\}$ belongs to the set $\mathcal{S}(A)$ of strange points if and only if player $j$ wins against at least $(1-\alpha)(n+1)$ other players. Therefore, only those players $j$ with the largest $\alpha(n+1)$ values of $R_j$ can be strange, which implies $|\mathcal{S}(A)| \leq \alpha (n+1)$.

\textbf{Step 3:} For any permutation $\sigma$ of $\mathcal{D}_0 \cup \{n+1\} = \{1,\ldots,n,n+1\}$ which does not mix points assigned to different folds, the distribution of the matrix $A$ satisfies:
\begin{align*}
    A_{\sigma(i)\sigma(j)}
    & \stackrel{d}{=} A_{i,j}.
\end{align*}
First, note that the design of Algorithm~\ref{alg_clra_cv+} ensures permuting the data points indexed by $\mathcal{D}_0 \cup \{n+1\}$ does not alter any of the machine-learning models $\hat{s}_1^k$ learnt from the labeled outliers, for $k \in \{1,\ldots,K\}$. Further, the $\hat{s}_0^k$ are also invariant as long as the permutation $\sigma$ does not mix together points assigned to different folds, because the training algorithm $\mathcal{A}_0$ is assumed to be invariant to the order of its input.
In fact, the only effect of permuting the data points in $\mathcal{D}_0 \cup \{n+1\}$ with $\sigma$ is to permute accordingly the preliminary p-values $\hat{u}_0(X_i; k(i))$ given by~\eqref{eq:cv++_pval_0}, as the latter is an invariant function of $X_i$. At the same time, the preliminary p-values $\hat{u}_1(X_i; k(i))$ given by~\eqref{eq:cv++_pval_1} are permuted similarly, as the latter is also an invariant function of $X_i$.
Therefore, if we let $R'$ indicate the vector of scores $R_{i}=\hat{r}(X_{i}; k(i))$ computed under a permutation $\sigma$ of the data points indexed by $\mathcal{D}_0 \cup \{n+1\}$, it follows that $R'_{i} = R_{\sigma(i)}$ for all $i \in \mathcal{D}_0 \cup \{n+1\}$, and therefore $ A'_{i,j} = A'_{\sigma(i),\sigma(j)}$.
Finally, by the data exchangeability assumption, under the null hypothesis, $A_{i,j}$ has the same distribution as $A'_{i,j}$.

The proof is completed by noting that Steps 2--3 imply that, for all $i \in \{1,\dots,n+1\}$,
\begin{align*}
    \P{n+1 \in \mathcal{S}(A)}
  & = \P{i \in \mathcal{S}(A)}  \\
  & = \frac{1}{n+1} \sum_{j =1}^{n+1} \P{j \in \mathcal{S}(A)} \\
  & = \frac{\E{|\mathcal{S}(A)|}}{n+1} \\
  & \leq \frac{\alpha(n+1)}{(n+1)} = \alpha.
\end{align*}
Therefore, by Step 1, $\P{\hat{u}(X_{n+1}) \leq \alpha \mid Y_{n+1} = 0} \leq \alpha$.
\end{proof}

\subsection{TCV+}

\begin{proof}[Proof of Theorem~\ref{theorem:general_tcv+}]
Given that $Y_{n+1} \notin \hat{\mathcal{C}}_{\alpha}(X_{n+1})$  if and only if $\hat{u}(X_{n+1}; Y_{n+1}) \leq \alpha$, by definition of $\hat{\mathcal{C}}_{\alpha}(X_{n+1})$, it suffices to prove $\P{\hat{u}(X_{n+1}; Y_{n+1}) \leq \alpha} \leq \alpha$. This can be shown by proceeding as in the proof of Theorem~\ref{theorem:ratio-cv+}.
For simplicity, imagine each fold $\mathcal{D}^k$ has $m=(n+1)/K$ points and $m$ is an integer (this assumption could be relaxed at the cost of heavier notation, as in~\cite{barber2019predictive}).
For $y=Y_{n+1}$, define $R \in \mathbb{R}^{(n+1)}$ such that, for each $i \in \{n+1\} \cup \mathcal{D}$,
\begin{equation*}
    R_{i}=\hat{s}^y(X_{i}; k(i)).
\end{equation*}
Define the comparison matrix $A \in \{0,1\}^{(n+1)\times(n+1)}$ with entries
\begin{align*}
    A_{i,j} = 1\left\{ R_{i} > R_{j}\right\},
\end{align*}
and the set of ``strange'' points, for any fixed $\alpha \in (0,1)$,
\begin{equation*}
    \mathcal{S}(A)=\bigg\{i \in \{1, \dots, n+1\}: \sum_{j=1}^{n+1}A_{i,j} \geq (1-\alpha)(1+n) \bigg\}.
\end{equation*}
The event $\hat{u}(X_{n+1}; Y_{n+1}) \leq \alpha$ is equivalent to $n+1 \in \mathcal{S}(A)$, and the probability of the latter event can be bounded from above by $\alpha$ exactly as in the proof of Theorem~\ref{theorem:ratio-cv+}.
\end{proof}

\begin{proof}[Proof of Theorem~\ref{theorem:general_tcv+-label-conditional}]
The proof is similar to that of Theorem~\ref{theorem:general_tcv+}.
Fix any possible label $y$ and condition on the event that $Y_{n+1}$.
Given that $Y_{n+1} \notin \hat{\mathcal{C}}_{\alpha}(X_{n+1})$  if and only if $\hat{u}(X_{n+1}; Y_{n+1}) \leq \alpha$, it suffices to prove $\P{\hat{u}(X_{n+1}; Y_{n+1}) \leq \alpha \mid Y_{n+1} = y} \leq \alpha$. This can be shown by proceeding as in the proof of Theorem~\ref{theorem:ratio-cv+}.
Let $n_y$ denote the cardinality of $\mathcal{D}^k$.
For simplicity, imagine each fold $\mathcal{D}^k_y$ has $m=(n_y+1)/K$ points and $m$ is an integer (this assumption could be relaxed at the cost of heavier notation, as in~\cite{barber2019predictive}).
With a slight abuse of notation, we refer to the test point index $\{n+1\}$ as $\{n_y+1\}$.
Define $R \in \mathbb{R}^{(n+1)}$ such that, for each $i \in \{n_y+1\} \cup \mathcal{D}$,
\begin{equation*}
    R_{i}=\hat{s}^y(X_{i}; k(i)).
\end{equation*}
Define the comparison matrix $A \in \{0,1\}^{(n_y+1)\times(n_y+1)}$ with entries
\begin{align*}
    A_{i,j} = 1\left\{ R_{i} > R_{j}\right\},
\end{align*}
and the set of ``strange'' points, for any fixed $\alpha \in (0,1)$,
\begin{equation*}
    \mathcal{S}(A)=\bigg\{i \in \{1, \dots, n_y+1\}: \sum_{j=1}^{n_y+1}A_{i,j} \geq (1-\alpha)(1+n_y) \bigg\}.
\end{equation*}
The event $\hat{u}(X_{n_y+1}; Y_{n_y+1}) \leq \alpha$ is equivalent to $n_y+1 \in \mathcal{S}(A)$, and the probability of the latter event can be bounded from above by $\alpha$ as in the proof of Theorem~\ref{theorem:ratio-cv+}, because the test point is exchangeable with the calibration data in $\mathcal{D}_{y}$ conditional on its true label being $y$.
\end{proof}

\subsection{FDR control}

\begin{proof}[Proof of Theorem~\ref{theorem:fdr}]
Note that all the $\tilde{u}_{i}(X_j)$ are measurable with respect to $\Phi_i$, which implies $\tilde{R}_i$ is also measurable with respect to $\Phi_i$.
Further, if a test point $i$ is an inlier, it is easy to see $\hat{u}(X_i)$ is still super-uniform conditional on $\Phi_i$ because $\Phi_i$ is invariant to permutations of $\mathcal{D}^0_{\mathrm{cal}} \cup \{i\}$.
Thus, this result is a special case of Theorem~\ref{theorem:fdr-general}, proved below.
\end{proof}

\begin{theorem} \label{theorem:fdr-general}
Suppose the three-step algorithm described in Section~\ref{sec:fdr-3steps} is applied with conformal p-values $\hat{u}(X_i)$ that are super uniform under the null hypothesis conditional on $\Phi_i$; i.e., for all $i \in \mathcal{D}^{\mathrm{test}}$ such that $Y_i=0$ and all $\alpha \in(0,1)$,
\begin{align*}
  \P{\hat{u}(X_i) \leq \alpha \mid \Phi_i, Y_i = 0} \leq \alpha.
\end{align*}
Assume also $\tilde{R}_i$ is measurable with respect to $\Phi_i$. Then, the algorithm controls the FDR below $\alpha m_0/m$.
\end{theorem}

\begin{proof}[Proof of Theorem~\ref{theorem:fdr-general}]
Our proof of Theorem~\ref{theorem:fdr-general} follows the same strategy as~\cite{fithian2020conditional}; in fact, our method can be seen as a special case of their theory.
Nonetheless, it is more convenient and instructive to revisit the full proof from scratch, as opposed to explicitly connecting the details of our method to the general results in~\cite{fithian2020conditional}.
For any $i \in \mathcal{D}^{\mathrm{test}}$, define the variable $V_i$ such that $V_i=1$ if $Y_i=0$ and the corresponding null hypothesis is falsely rejected, and $V_i=0$ otherwise.
Denote the total number of final rejections as $R = |\mathcal{R}|$.
Then, the FDR can be written as
\begin{align*}
  \text{FDR}
  & = \sum_{i : Y_i=0} \E{ \frac{V_i}{R \lor 1} } \leq \alpha \frac{m_0}{m},
\end{align*}
where the last inequality is proved below.

Let $R^*=R(\hat{u}(X_i) \leftarrow 0)$ denote the hypothetical total number of rejections obtained by fixing $\hat{u}(X_i)=0$ prior to applying our procedure.
If $X_i$ is an inlier, then
\begin{align*}
    \E{\frac{V_i}{R \lor 1} }
    & = \E{ \frac{\I{i \in \mathcal{R}^+} \cdot \I{\epsilon_i \leq R/\tilde{R}_i}}{R \lor 1} } \\
    & = \E{ \frac{\I{i \in \mathcal{R}^+} \cdot \I{\epsilon_i \leq R^*/\tilde{R}_i}}{R^* \lor 1} }\\
     & = \E{ \frac{\I{i \in \mathcal{R}^+}}{\tilde{R}_i} }\\
     & = \E{ \E{ \frac{ \I{ \hat{u}(X_i) \leq \alpha\tilde{R}_i / m }}{\tilde{R}_i} \mid \Phi_i } }  \\
     & = \E{ \E{ \I{ \hat{u}(X_i) \leq \alpha\tilde{R}_i / m } \mid \Phi_i } \frac{1}{\tilde{R}_i} }  \\
     & \leq \E{ \alpha\tilde{R}_i / m \frac{1}{\tilde{R}_i} }   \\
     & = \frac{\alpha}{m}.
\end{align*}
Note that $\epsilon_{-i}$ denotes all $\epsilon_j$ variables for $j \in \mathcal{R}_+ \setminus \{i\}$.
The inequality above follows from the fact that $\tilde{R}_i$ is measurable with respect to $\Phi_i$ and $\hat{u}(X_i)$ was assumed to be super-uniform conditional on $\Phi_i$.
\end{proof}

\begin{proof}[Proof of Theorem~\ref{theorem:fdr-cv++}]
All $\tilde{u}_{i}(X_j)$ are measurable with respect to $\Phi_i$, which implies $\tilde{R}_i$ is also measurable with respect to $\Phi_i$.
Further, if $X_i$ is an inlier, it is easy to see the original $\hat{u}(X_i)$ computed by Algorithm~\ref{alg_clra_cv+} is still super-uniform conditional on $\Phi_i$ because $\Phi_i$ is invariant to permutations of $\mathcal{D}^0 \cup \{i\}$.
Thus, this result is also a special case of Theorem~\ref{theorem:fdr-general}.
\end{proof}

\subsection{Asymptotic power analysis}

\begin{proof}[Proof of Theorem~\ref{theorem:power}]
This proof follows the same strategy as that of Theorem~2 in~\cite{liang2022locally}, but we nonetheless report all details here for completeness.
To simplify the notation, we prove the equivalent result with the inverse weights $w = (\hat{u}_1(X_{n+1}), \ldots, \hat{u}_1(X_{n+m}))$ replaced by $\tilde{w}^*$, which is defined as in~\eqref{eq:power-tilde-w}.
Note that this constant rescaling of the inverse weights is allowed because the $m$ elements of $\tilde{w}^*$ maintain the same ordering as those of $w$.

Let $\mathcal{F}$ be the collection of random variables $\{Z_{1,i}\}_{i\in \mathcal{D}^{\mathrm{test}}}$, $\mathcal{D}_0$, and $\mathcal{D}_1$.
Define $F_{1,i}(t)$ is the cumulative distribution function of $\hat{u}_0(X_i)$ conditional on $Y_i=1$ and on all the information in $\mathcal{F}$.
Then, the expected number of true discoveries obtained by applying $\delta(t; v)$ with a fixed threshold $t$ and inverse weights $v_i$ is:
\begin{align} \label{eq:psi-expectation}
  \Psi(t;v)
  &= \E{ \sum_{i \in \mathcal{D}^{\mathrm{test}}} \P{Y_i=1 \mid Z_{1,i}} \cdot F_{1,i}(v_i \cdot t) },
\end{align}
because
\begin{align*}
  \Psi(t;v)
  &= \E{\sum_{i \in \mathcal{D}^{\mathrm{test}}} Y_i \cdot \delta_i(t; v_i) }\\
  &= \E{ \E{ \sum_{i \in \mathcal{D}^{\mathrm{test}}} Y_i \cdot \delta_i(t; v_i) \mid \mathcal{F} }}\\
  &= \E{ \sum_{i \in \mathcal{D}^{\mathrm{test}}} \P{Y_i=1 \mid \mathcal{F}} \cdot \P{\hat{u}_0(X_i) \leq v_i \cdot t \mid Y_i=1, \mathcal{F} } }\\
  &= \E{ \sum_{i \in \mathcal{D}^{\mathrm{test}}} \P{Y_i=1 \mid Z_{1,i}} \cdot F_{1,i}(v_i \cdot t) }.
\end{align*}

First, we prove that if the same threshold $t$ is applied with weights $w$ and with constant unit weights, then $\delta(t; w)$ will have larger power:
\begin{align} \label{eq:power-more-discoveries}
  \Psi(t;w) \geq \Psi(t;1).
\end{align}
To prove~\eqref{eq:power-more-discoveries}, note that, assuming
\begin{align*}
\sum_{i \in \mathcal{D}^{\mathrm{test}}} a_i F_{1,i}(t/x_i) \geq \sum_{i \in \mathcal{D}^{\mathrm{test}}} a_i F_{1,i}\left( \frac{t \sum_{j=1}^m a_j}{\sum_{j=1}^m a_jx_j} \right),
\end{align*}
for any $0\leq a_i\leq 1$ and $\min_{1 \leq i \leq m}\tilde w_i^{-1} \leq x_i \leq \max_{1 \leq i \leq m}\tilde w_i^{-1}$, we can prove:
\begin{align} \label{eq:power-F1-lemma}
  \sum_{i \in \mathcal{D}^{\mathrm{test}}}{ \P{Y_i = 1 \mid  Z_{1,i}} F_{1,i}(\tilde w_it )}
  \geq \sum_{i \in \mathcal{D}^{\mathrm{test}}}{ \P{Y_i = 1 \mid  Z_{1,i}} F_{1,i}(t)}.
\end{align}
In fact,
\begin{align*}
    & \sum_{i \in \mathcal{D}^{\mathrm{test}}}{ \P{Y_i = 1 \mid  Z_{1,i}} F_{1,i}(\tilde w_it )}\\
& = \sum_{i \in \mathcal{D}^{\mathrm{test}}}{ \P{Y_i = 1 \mid  Z_{1,i}} F_{1,i}(t/\tilde w_i^{-1} )}\\
    & \geq \sum_{i \in \mathcal{D}^{\mathrm{test}}} \P{Y_i = 1 \mid  Z_{1,i}} F_{1,i}\left(\frac{\sum_{j=1}^m{ \P{Y_j = 1 \mid  s^1_j} }t}{\sum_{j=1}^m \P{Y_j = 1 \ s^1_j} \tilde w_j^{-1}}\right)\\
    & =\sum_{i \in \mathcal{D}^{\mathrm{test}}} \P{Y_i = 1 \mid  Z_{1,i}} F_{1,i}\left(t\cdot \frac{\sum_{j=1}^m{ \P{Y_j = 0 \mid  s^1_j} }}{\sum_{j=1}^m \P{Y_j = 0 \mid  s^1_j} w_j} \cdot
    \frac{\sum_{j=1}^m{ \P{Y_j = 1 \mid  s^1_j} }}{\sum_{j=1}^m \P{Y_j = 1 \mid  s^1_j} w_j^{-1}} \right)\\
    & \geq \sum_{i \in \mathcal{D}^{\mathrm{test}}}{ \P{Y_i = 1 \mid  Z_{1,i}} F_{1,i}(t)},
\end{align*}
where the last inequality follows from Assumption~\ref{assumption:informative-weights}.
Combining~\eqref{eq:power-F1-lemma} with~\eqref{eq:psi-expectation} proves~\eqref{eq:power-more-discoveries}.

Next, we prove that, for any fixed threshold $t$,
\begin{align} \label{eq:power-fdr}
  Q(t; \tilde{w}) \leq Q(t; 1).
\end{align}
This is shown easily by recalling the definition of $\tilde{w}$ in~\eqref{eq:power-tilde-w}:
\begin{align*}
  Q(t; \tilde{w})
  & = \frac{\sum_{i \in \mathcal{D}^{\mathrm{test}}} \tilde{w}_i t \cdot \P{Y_i = 0 \mid Z_{1,i}} }{ \sum_{i \in \mathcal{D}^{\mathrm{test}}} \tilde{w}_i t \cdot \P{Y_i = 0 \mid Z_{1,i}} + \sum_{i \in \mathcal{D}^{\mathrm{test}}} F_{1,i}(\tilde{w}_i t \mid Z_{1,i}) \cdot \P{Y_i = 1 \mid Z_{1,i}}} \\
  & = \frac{\sum_{i \in \mathcal{D}^{\mathrm{test}}} t \cdot \P{Y_i = 0 \mid Z_{1,i}} }{ \sum_{i \in \mathcal{D}^{\mathrm{test}}} t \cdot \P{Y_i = 0 \mid Z_{1,i}} + \sum_{i \in \mathcal{D}^{\mathrm{test}}} F_{1,i}(\tilde{w}_i t \mid Z_{1,i}) \cdot \P{Y_i = 1 \mid Z_{1,i}}} \\
  & \leq \frac{\sum_{i \in \mathcal{D}^{\mathrm{test}}} t \cdot \P{Y_i = 0 \mid Z_{1,i}} }{ \sum_{i \in \mathcal{D}^{\mathrm{test}}} t \cdot \P{Y_i = 0 \mid Z_{1,i}} + \sum_{i \in \mathcal{D}^{\mathrm{test}}} F_{1,i}(t \mid Z_{1,i}) \cdot \P{Y_i = 1 \mid Z_{1,i}}} \\
  & = Q(t; 1).
\end{align*}
where the inequality follows from~\eqref{eq:power-F1-lemma}.

Now, let us apply~\eqref{eq:power-more-discoveries} and~\eqref{eq:power-fdr} setting the generic threshold $t$ equal to $t^{\alpha}_{\mathrm{oracle}}(1) = \sup \{ t: Q(t; (1,\ldots,1)) \leq \alpha\}$, which gives:
\begin{align*}
  & Q(t^{\alpha}_{\mathrm{oracle}}(1); \tilde{w}) \leq Q(t^{\alpha}_{\mathrm{oracle}}(1); 1) \leq \alpha,
  & \Psi(t^{\alpha}_{\mathrm{oracle}}(1);\tilde{w}) \geq \Psi(t^{\alpha}_{\mathrm{oracle}}(1);1).
\end{align*}
The proof is completed by noting that, by definition of $t^{\alpha}_{\mathrm{oracle}}(\tilde{w})$,
\begin{align*}
  & Q(t^{\alpha}_{\mathrm{oracle}}(\tilde{w}); \tilde{w}) \leq \alpha,
  & \Psi(t^{\alpha}_{\mathrm{oracle}}(t^{\alpha}_{\mathrm{oracle}}(\tilde{w}));\tilde{w}) \geq \Psi(t^{\alpha}_{\mathrm{oracle}}(1);\tilde{w}).
\end{align*}

\end{proof}

\clearpage

\section{Supplementary figures} \label{app:figures}

\subsection{Method schematics}

\begin{sidewaysfigure}
  \centering
  \includegraphics[width=\linewidth]{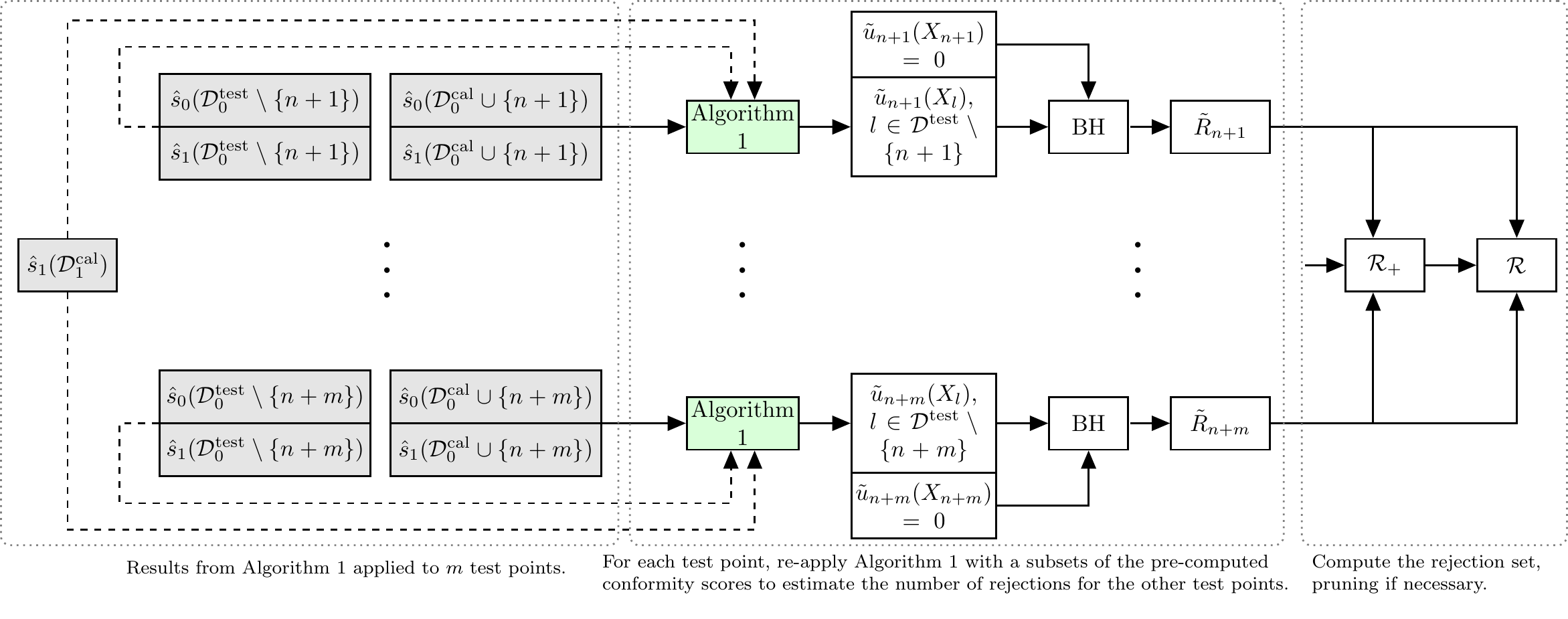}
  \caption{Graphical representation of the conditional FDR calibration method with integrative split-conformal p-values described in Section~\ref{sec:fdr-3steps}. The operations in the green nodes are not computationally expensive because they utilize pre-computed conformity scores.}
  \label{fig:split_fdr_diagram}
\end{sidewaysfigure}


\begin{figure}[H]
    \centering
    \includegraphics[width=\linewidth]{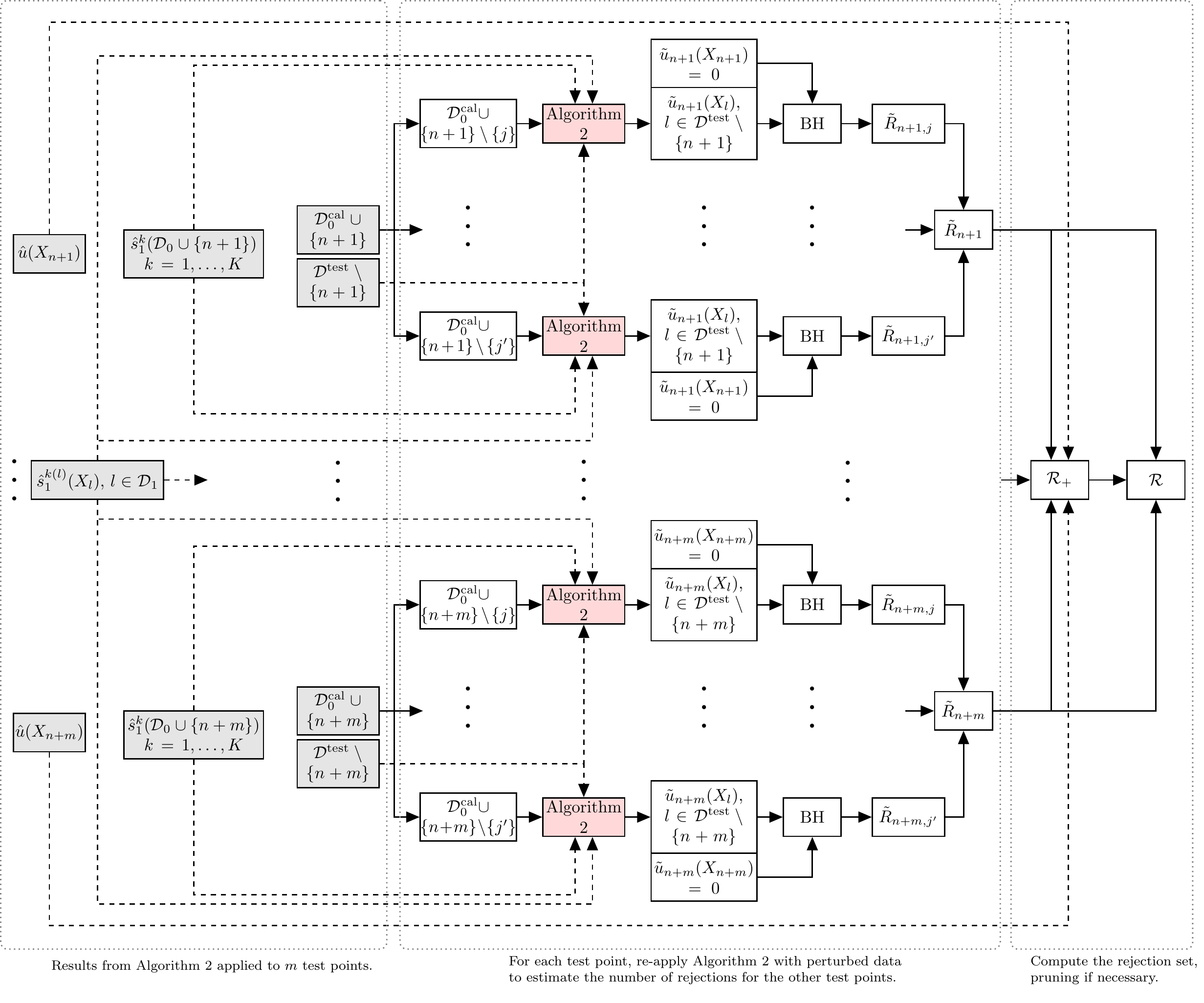}
    \caption{Graphical representation of the conditional FDR calibration method with integrative TCV+ p-values described in Section~\ref{sec:fdr-3steps-cv++}. The red nodes are computationally expensive.}
    \label{fig:cv_loo_diagram}
\end{figure}

\subsection{Numerical experiments with synthetic data}

\subsubsection{Demonstration of automatic model tuning}

\begin{figure}[H]
    \centering
    \includegraphics[width=\linewidth]{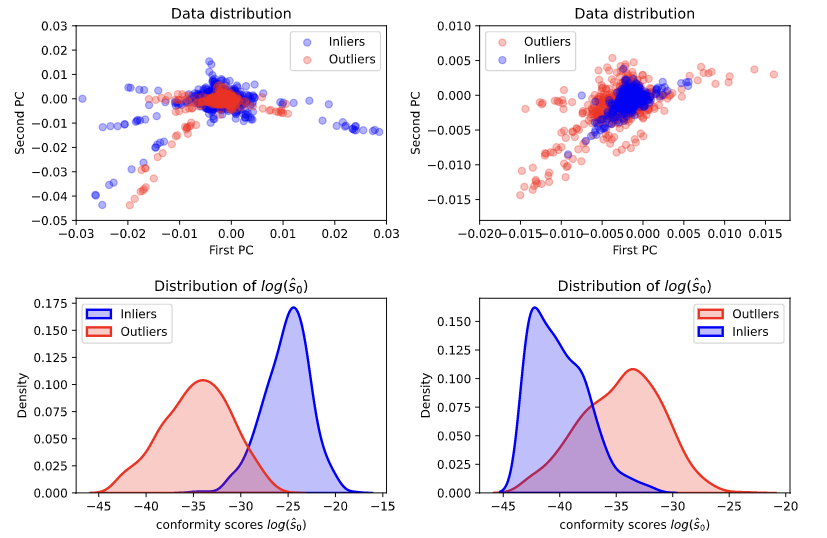}
    \caption{Empirical demonstration of the need to adaptively tune the sign of conformity scores calculated by one-class classifiers, as discussed in Section~\ref{sec:auto-tuning}. Left: the inlier data are generated from the Gaussian mixture model described in Section~\ref{sec:exp-split}, with $p=100$ and $a=2$, while the outlier data are generated with $a=1$. Right: the data are generated from the same model as on the left, swapping the inlier/outlier labels. Top: first two principal components of the data. Bottom: distributions of  $\log(\hat{s}_0(X_i))$ for inlier and outlier data points $X_i$.}
    \label{fig:auto-tuning}
\end{figure}

\subsubsection{Split-conformal integrative p-values}

\begin{figure}[H]
    \centering
    \includegraphics[width=0.9\linewidth]{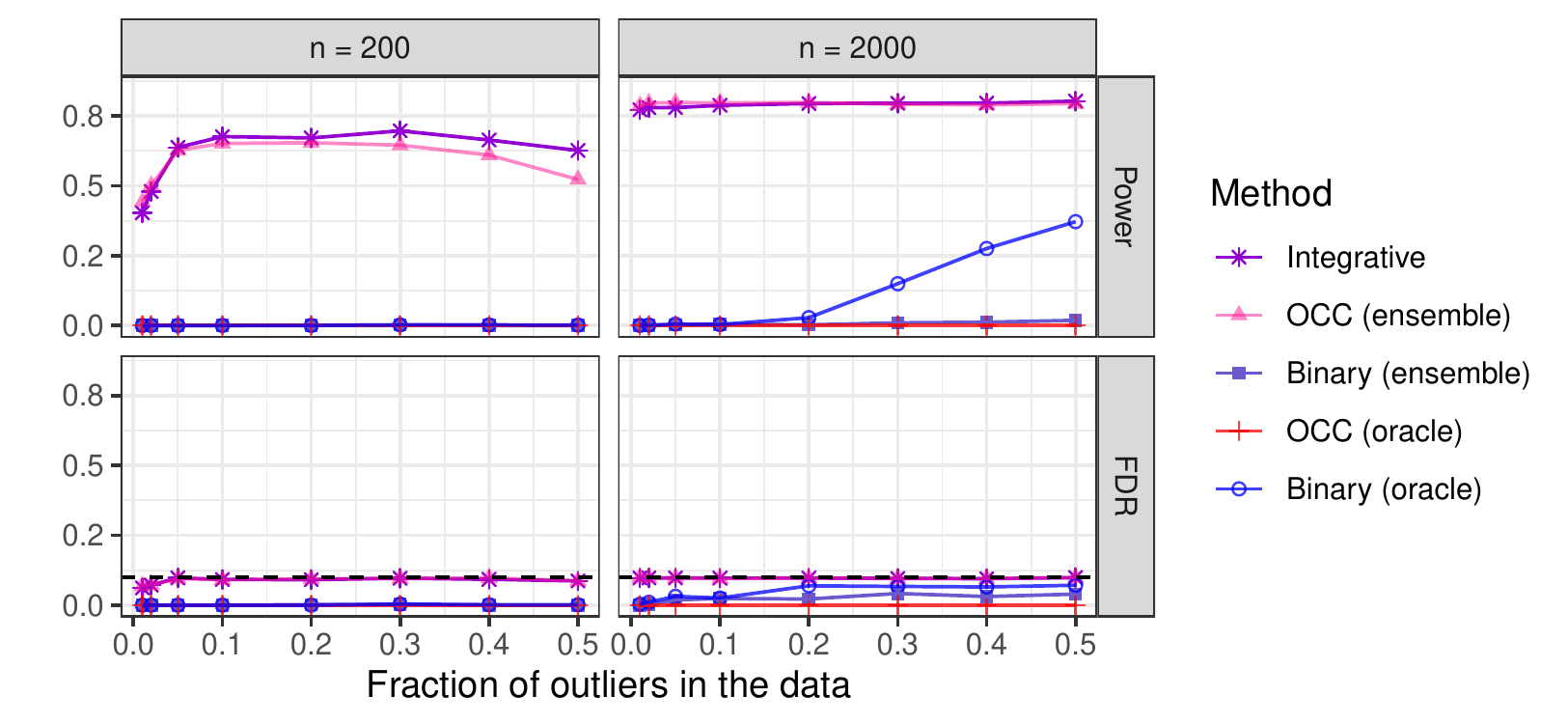}
    \caption{Performance of Storey's BH applied to conformal p-values computed with different methods, on simulated data. The results are shown as a function of the fraction of outliers in the labeled data, for different sample sizes. Other details are as in Figure~\ref{fig:exp-1-n}.}
    \label{fig:exp-2-bh}
\end{figure}

\begin{figure}[H]
    \centering
    \includegraphics[width=0.9\linewidth]{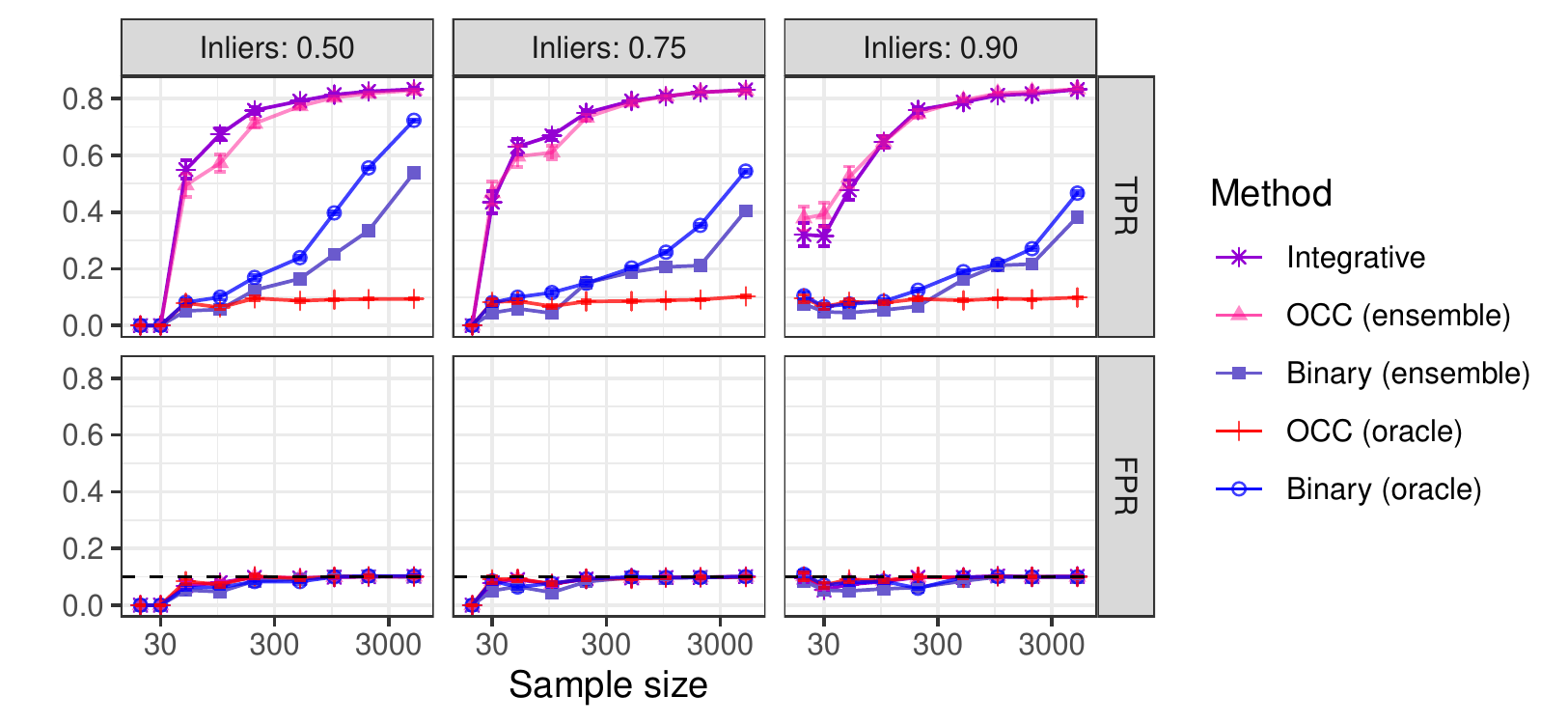}
    \caption{Performance of conformal p-values computed with different methods, on simulated data. The performance is measured in terms of true positive rate (TPR) and false positive rate (FPR). The horizontal dashed line corresponds to the nominal 10\% FPR level.  Other details are as in Figure~\ref{fig:exp-1-n}.}
    \label{fig:exp-1-fixed}
\end{figure}

\begin{figure}[H]
    \centering
    \includegraphics[width=0.9\linewidth]{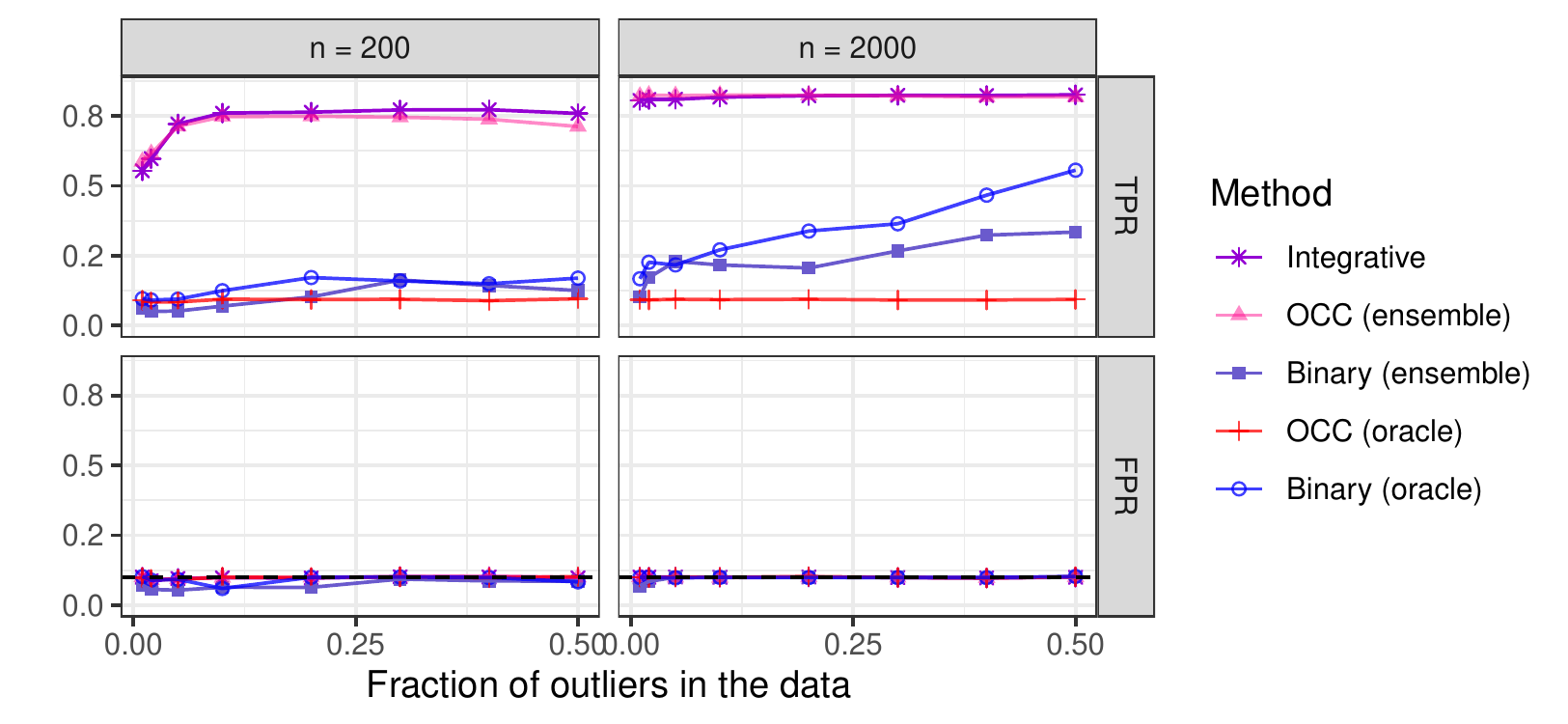}
    \caption{Performance of conformal p-values computed with different methods, on simulated data. The performance is measured in terms of true positive rate (TPR) and false positive rate (FPR). The horizontal dashed line corresponds to the nominal 10\% FPR level.  Other details are as in Figure~\ref{fig:exp-2-bh}.}
    \label{fig:exp-2-fixed}
\end{figure}

\begin{figure}[H]
    \centering
    \includegraphics[width=\linewidth]{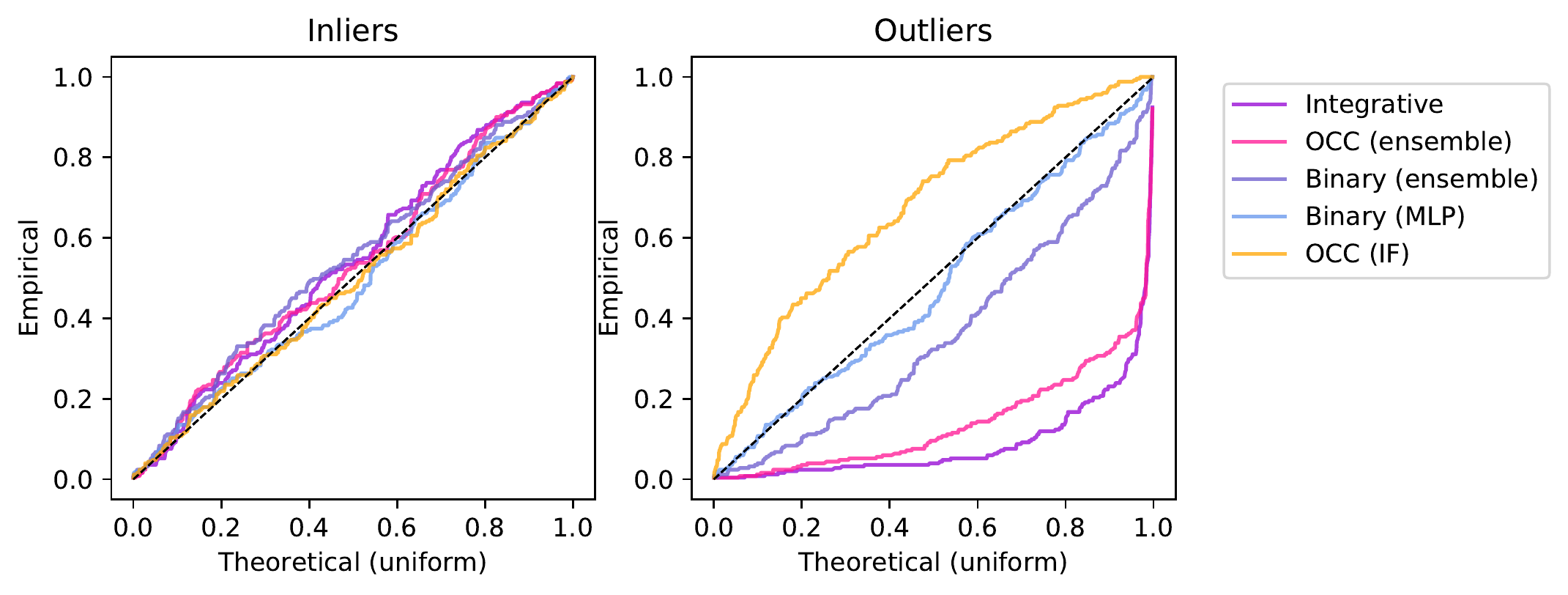}
    \caption{QQ-plot comparing the empirical distribution of conformal p-values, computed with different methods on synthetic data, to the theoretical uniform distribution, separately for true inliers (left) and outliers (right). In the underlying experiments, the labeled sample size is 500 and the proportion of data points which are outliers is 50\%. Other details are as in Figure~\ref{fig:exp-1-n}.}
    \label{fig:exp-1-qq}
\end{figure}

\begin{figure}[H]
    \centering
    \includegraphics[width=0.9\linewidth]{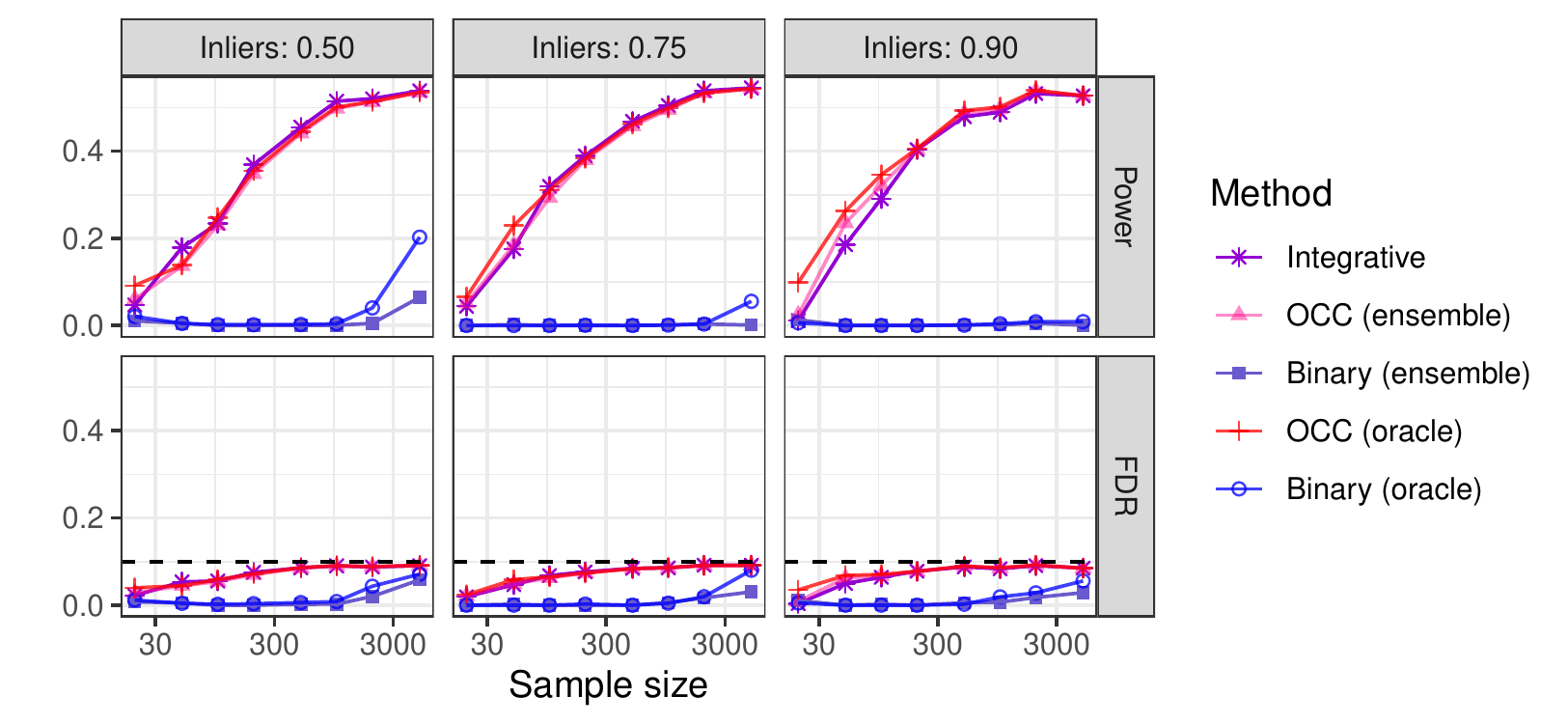}
    \caption{Performance of Storey's BH applied to conformal p-values computed with different methods, on simulated data. The data are simulated from a Gaussian mixture model as in Section~\ref{sec:exp-split}, but with parameter $a=1.25$ instead of $a=0.7$. Other details are as in Figure~\ref{fig:exp-1-n}.}
    \label{fig:exp-3-n}
\end{figure}

\begin{figure}[H]
    \centering
    \includegraphics[width=0.7\linewidth]{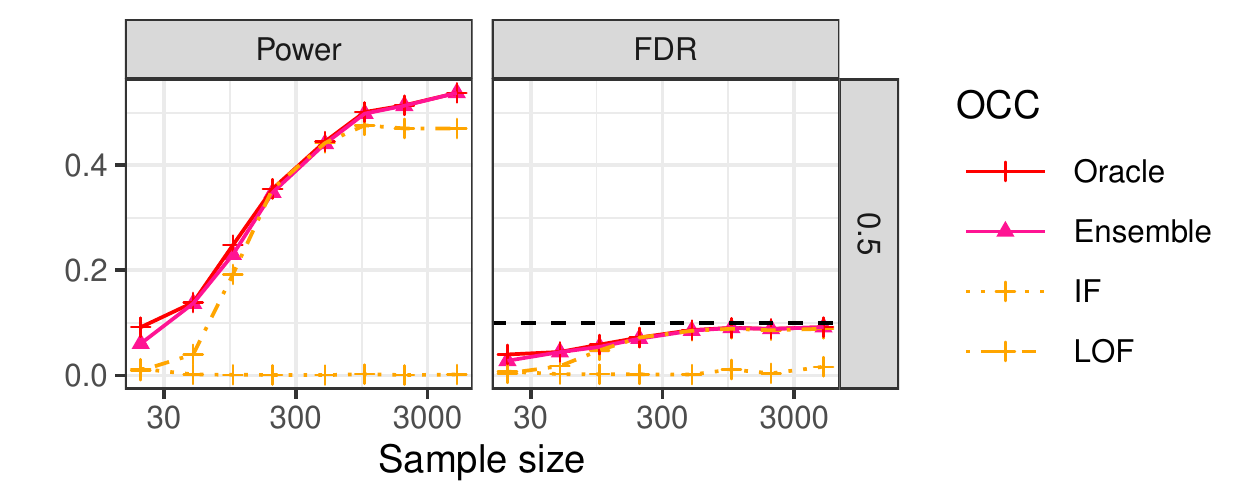}
    \caption{Performance of Storey's BH applied to conformal p-values based on different underlying machine-learning models, on simulated data. Integrative conformal p-values without weighting (ensemble method) are compared to standard conformal p-values based on two different one-class classification models (IF and LOF), as well as to those based on the most powerful one-class classification model selected by an imaginary oracle. Other details are as in Figure~\ref{fig:exp-3-n}.}
    \label{fig:exp-3-oracle}
\end{figure}

\begin{figure}[H]
    \centering
    \includegraphics[width=0.9\linewidth]{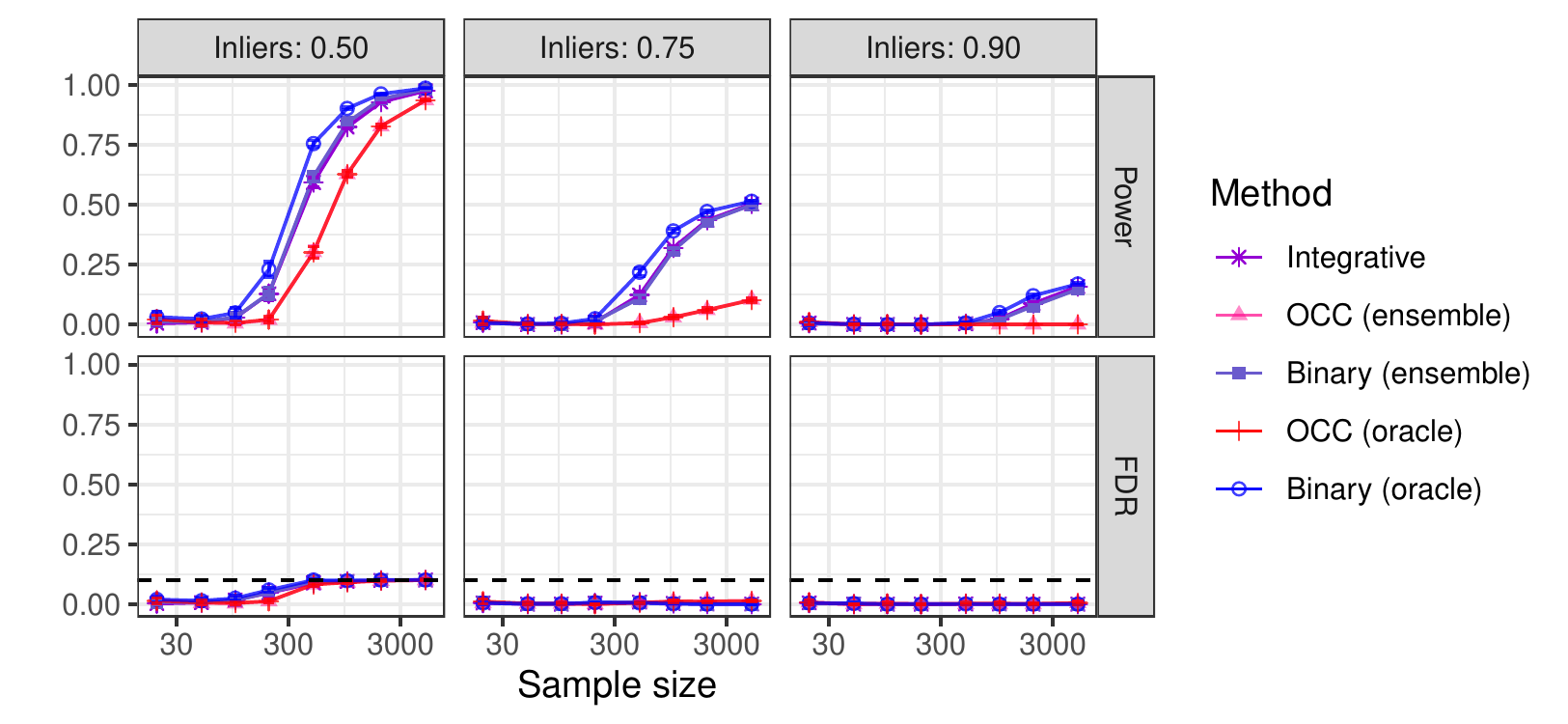}
    \caption{Performance of Storey's BH applied to conformal p-values computed with different methods, on simulated data. The data are simulated from a logistic regression model with random parameters. Other details are as in Figure~\ref{fig:exp-1-n}.}
    \label{fig:exp-4-n}
\end{figure}

\begin{figure}[H]
    \centering
    \includegraphics[width=0.9\linewidth]{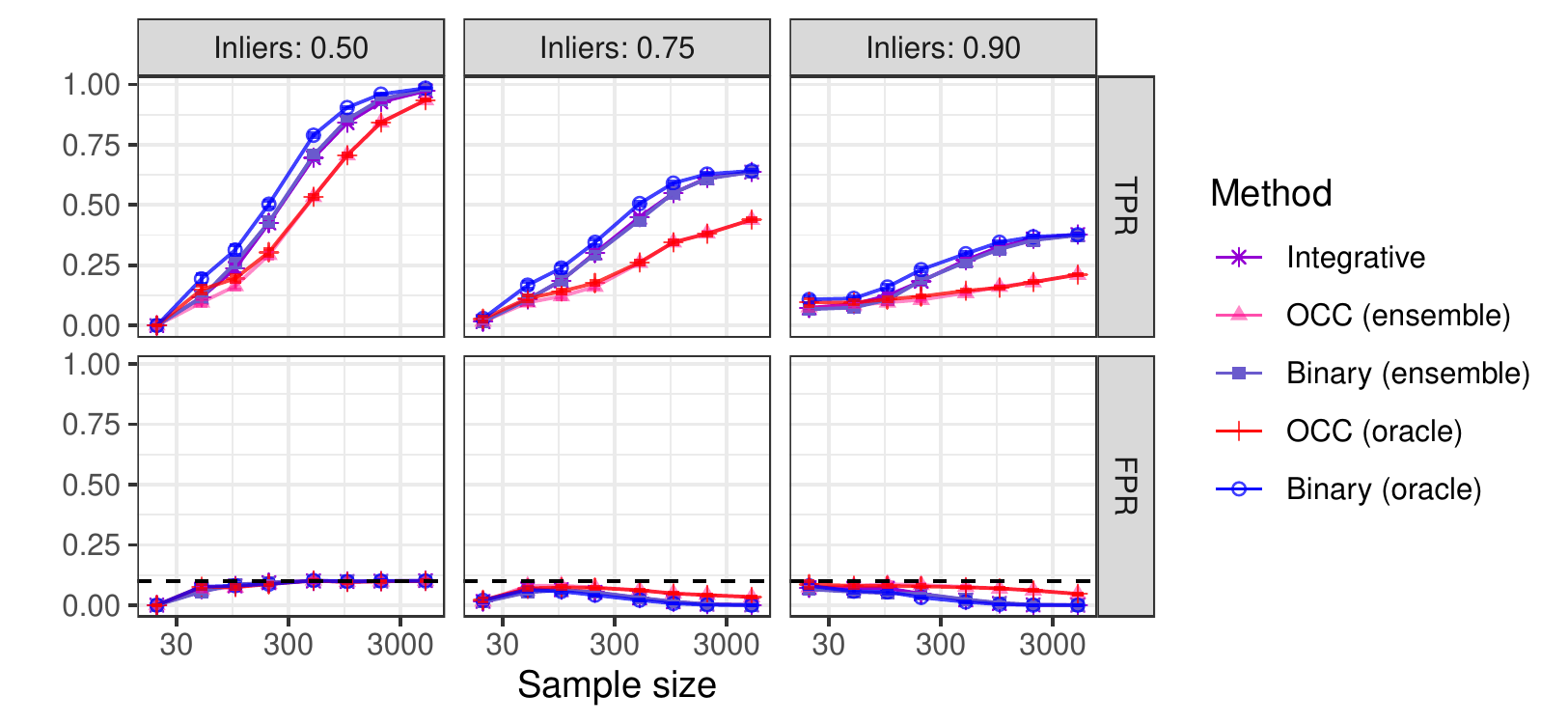}
    \caption{Performance of conformal p-values computed with different methods, on simulated data. The performance is measured in terms of true positive rate (TPR) and false positive rate (FPR). The data are simulated from a logistic regression model with random parameters. Other details are as in Figure~\ref{fig:exp-1-fixed}.}
    \label{fig:exp-4-n-fixed}
\end{figure}

\subsubsection{Integrative p-values via TCV+} \label{app:figures-cv}

\begin{figure}[H]
    \centering
    \includegraphics[width=0.95\linewidth]{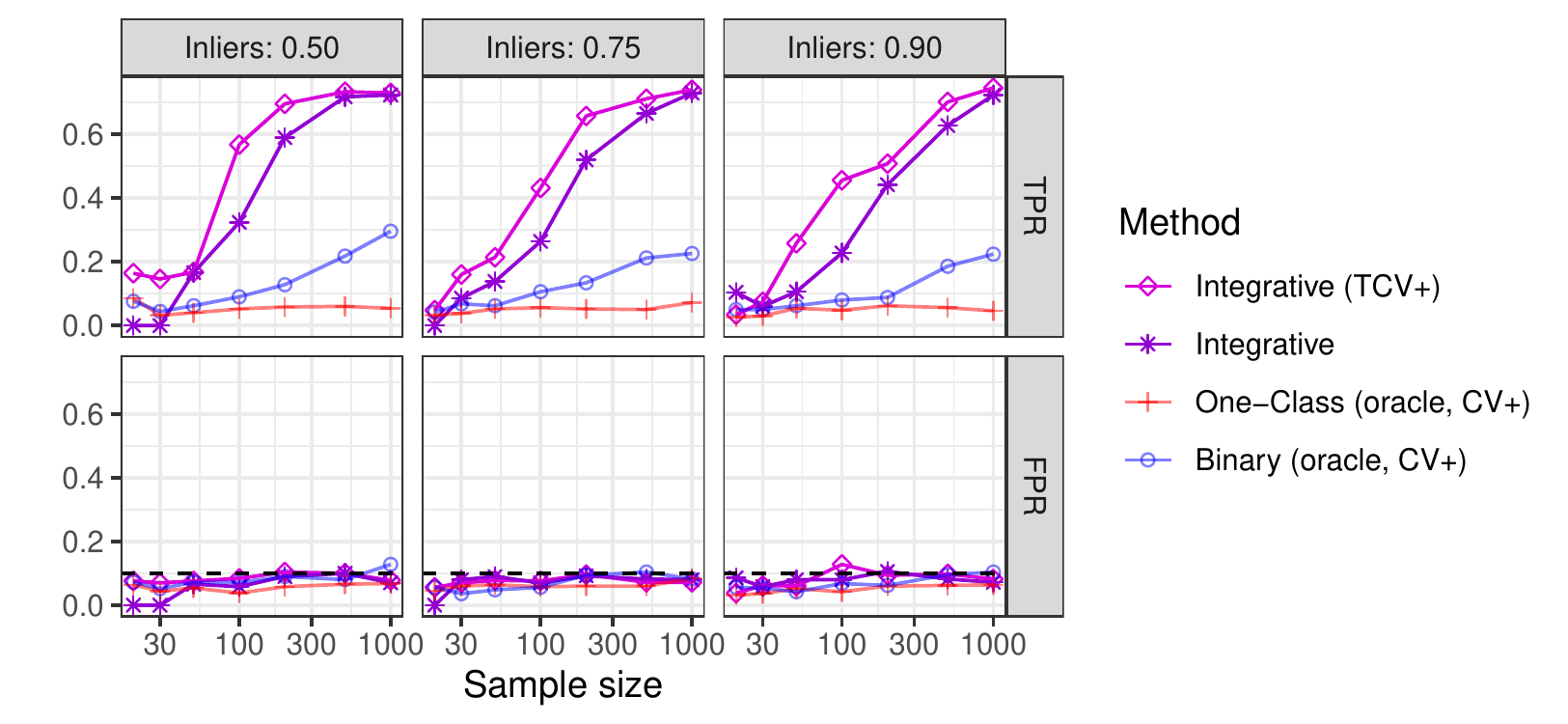}
    \caption{Performance on simulated data of integrative conformal p-values computed with either TCV+ or sample splitting. The performance is measured in terms of true positive rate (TPR) and false positive rate (FPR). The horizontal dashed line corresponds to the nominal 10\% FPR level. Other details are as in Figure~\ref{fig:exp-cv}.}
    \label{fig:exp-cv-fixed}
\end{figure}

\begin{figure}[H]
    \centering
    \includegraphics[width=0.95\linewidth]{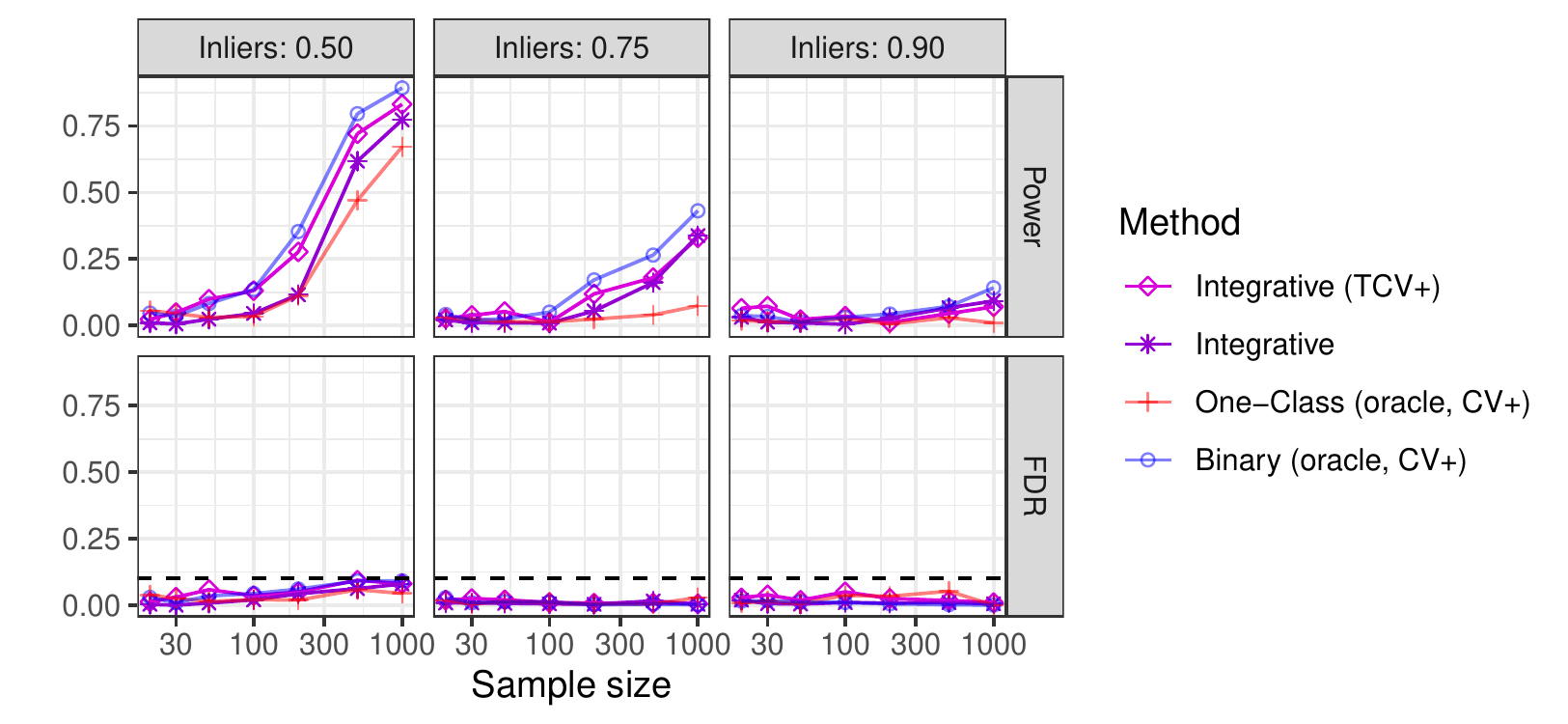}
    \caption{Performance on simulated data of Storey's BH applied to integrative conformal p-values computed with either TCV+ or sample splitting. The data are simulated from a logistic regression model with random parameters. Other details are as in Figure~\ref{fig:exp-cv}.}
    \label{fig:exp-cv-binomial}
\end{figure}

\begin{figure}[H]
    \centering
    \includegraphics[width=0.95\linewidth]{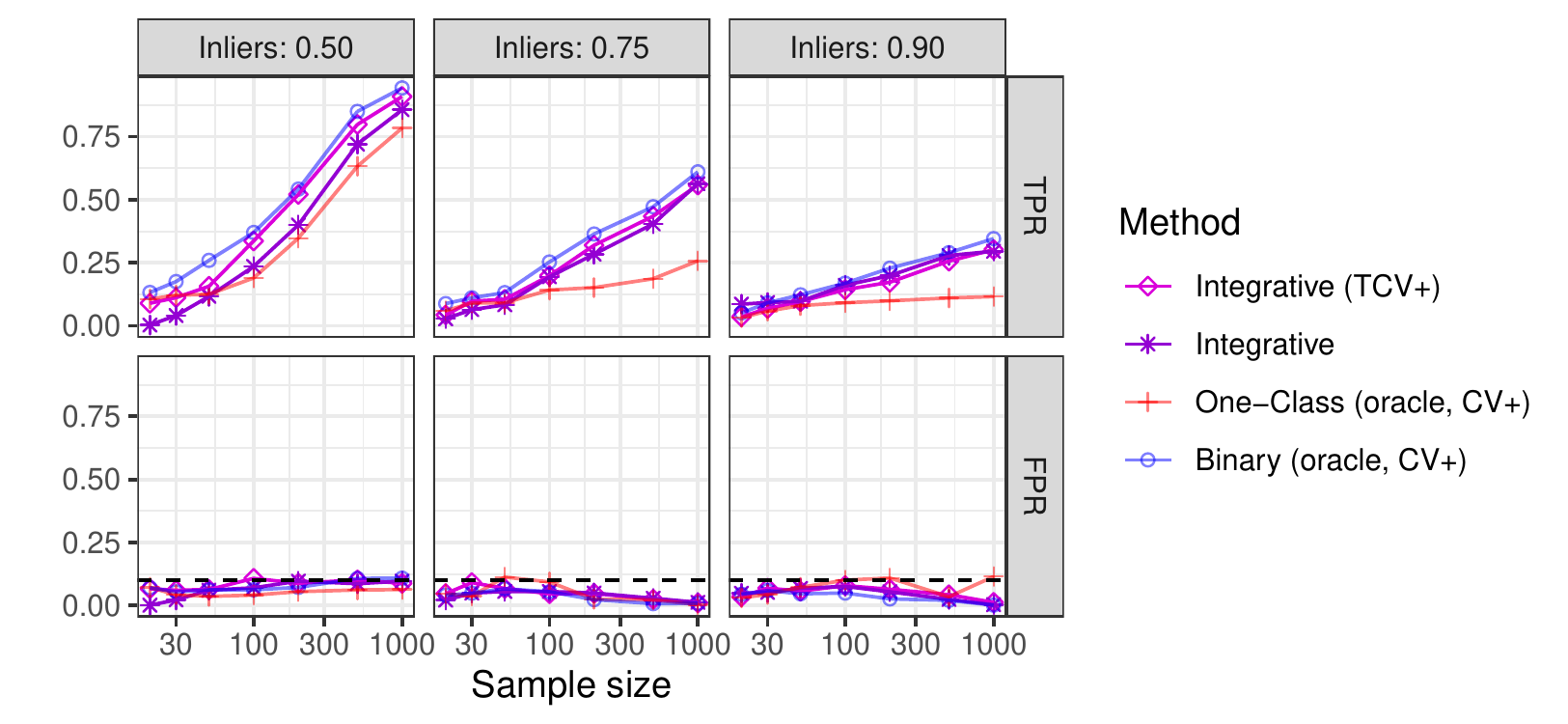}
    \caption{Performance on simulated data of integrative conformal p-values computed with either TCV+ or sample splitting. The performance is measured in terms of true positive rate (TPR) and false positive rate (FPR). The horizontal dashed line corresponds to the nominal 10\% FPR level.  The data are simulated from a logistic regression model with random parameters. Other details are as in Figure~\ref{fig:exp-cv-binomial}.}
    \label{fig:exp-cv-fixed-binomial}
\end{figure}

\subsection{Numerical experiments with real data}

\begin{figure}[H]
    \centering
    \includegraphics[width=0.75\linewidth]{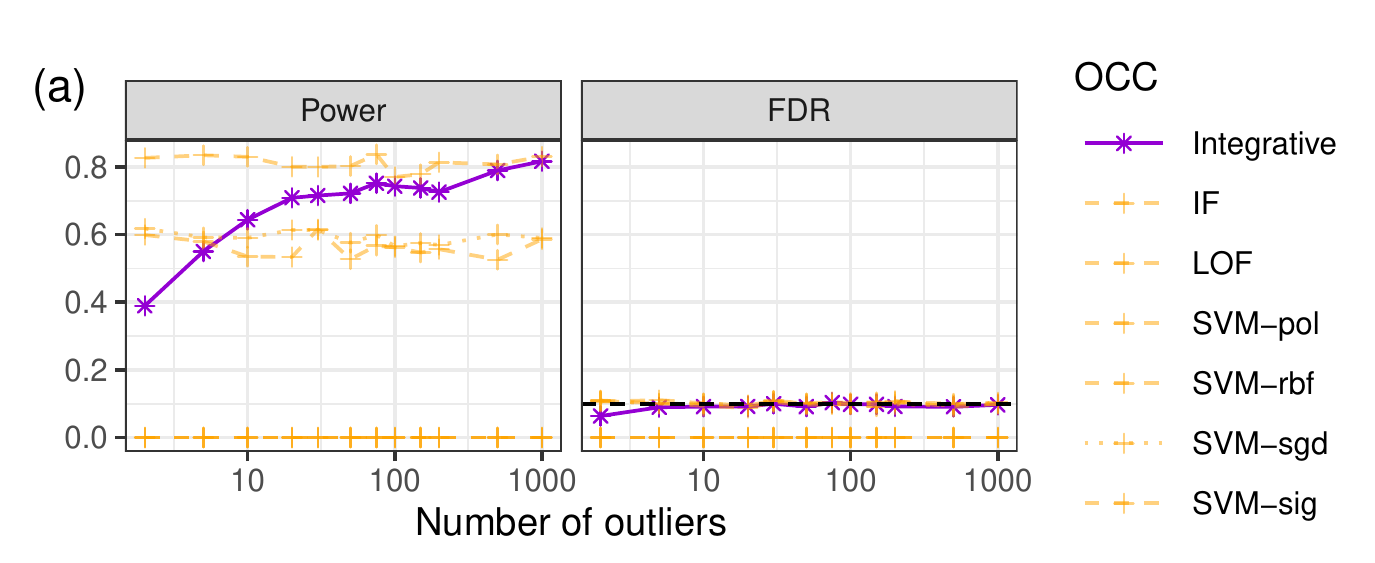}
    \includegraphics[width=0.75\linewidth]{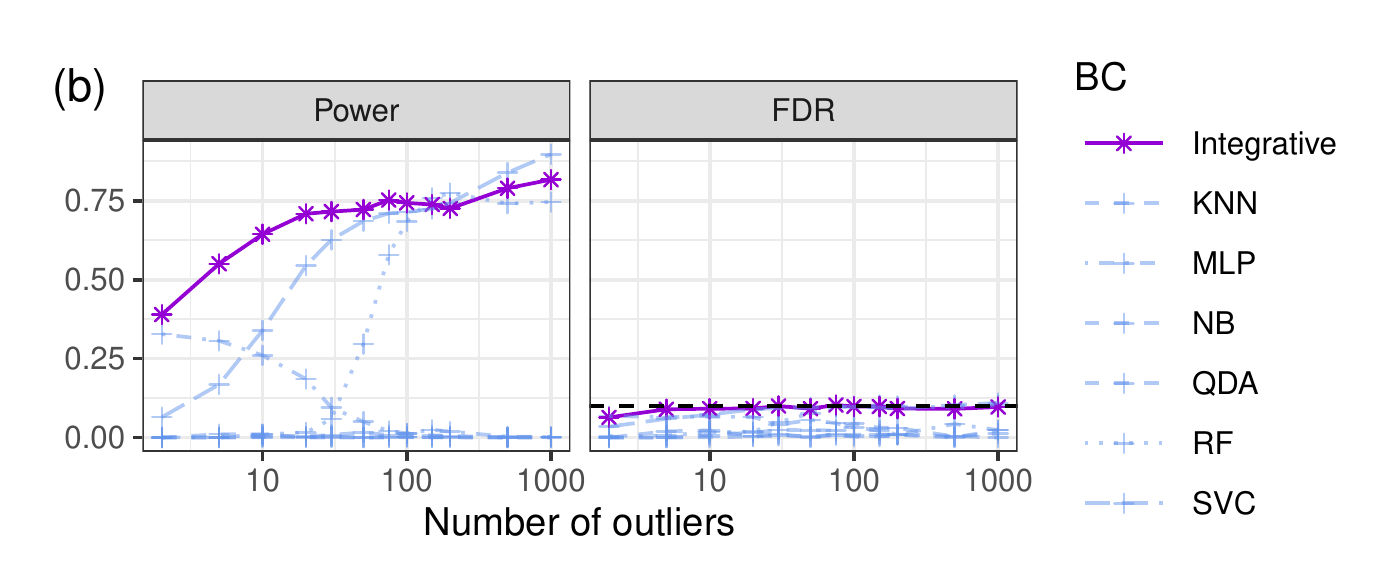}
    \caption{Performance of Storey's BH applied to conformal p-values based on different underlying machine-learning models, on flower image classification data. Integrative conformal p-values are compared to standard conformal p-values based on six different one-class classification models (a) and six different binary classification models (b). Other details are as in Figure~\ref{fig:exp-flowers}.}
    \label{fig:exp-flowers-oracle}
\end{figure}

\begin{figure}[H]
    \centering
    \includegraphics[width=0.82\linewidth]{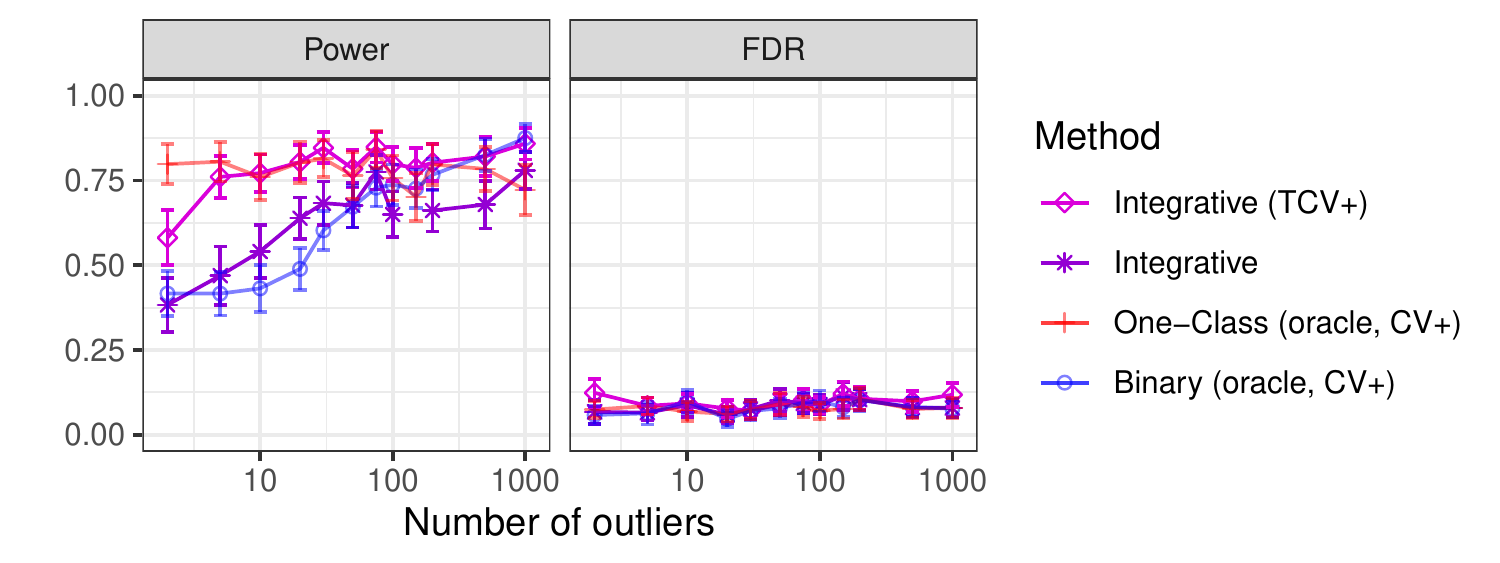}
    \caption{Performance on flower classification data of Storey's BH applied to integrative conformal p-values computed with either TCV+ or sample splitting. To serve as benchmarks, standard conformal p-values based on one-class or binary classification models tuned by an ideal oracle are computed using cross-validation+. Other details are as in Figure~\ref{fig:exp-flowers}.}
    \label{fig:exp-flowers-cv}
\end{figure}

\begin{figure}[H]
    \centering
    \includegraphics[width=0.75\linewidth]{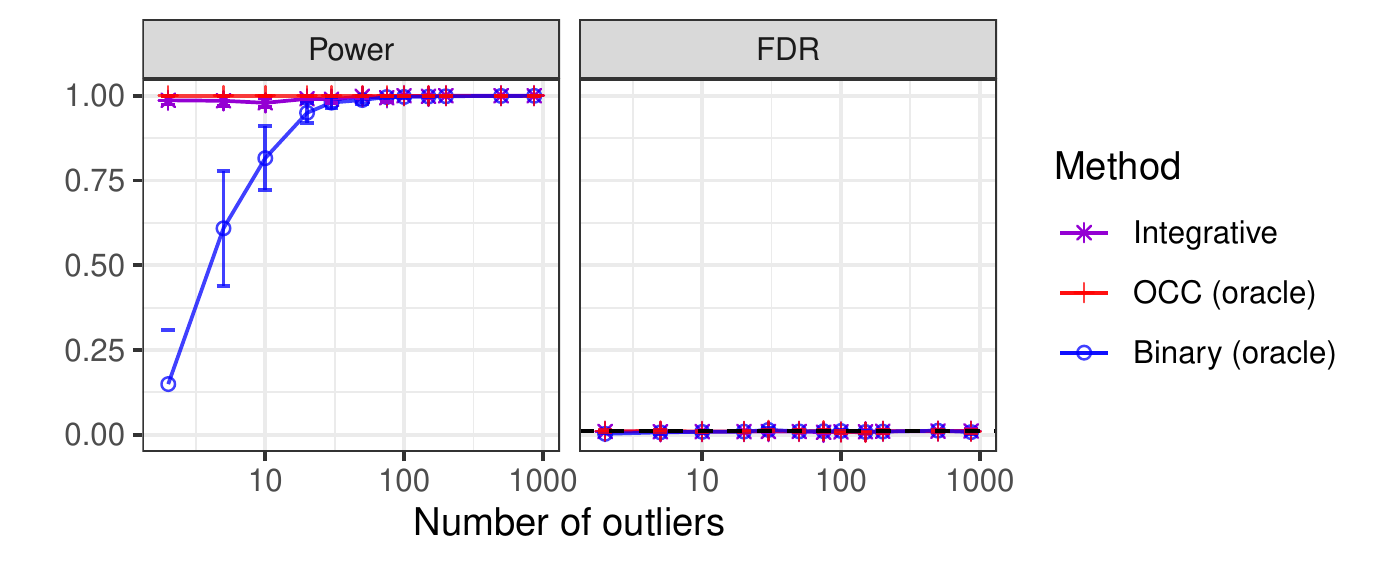}
    \caption{Performance of Storey's BH applied to conformal p-values based on different underlying machine-learning models, on car image classification data. Other details are as in Figure~\ref{fig:exp-flowers}.}
    \label{fig:exp-cars}
\end{figure}

\begin{figure}[H]
    \centering
    \includegraphics[width=0.75\linewidth]{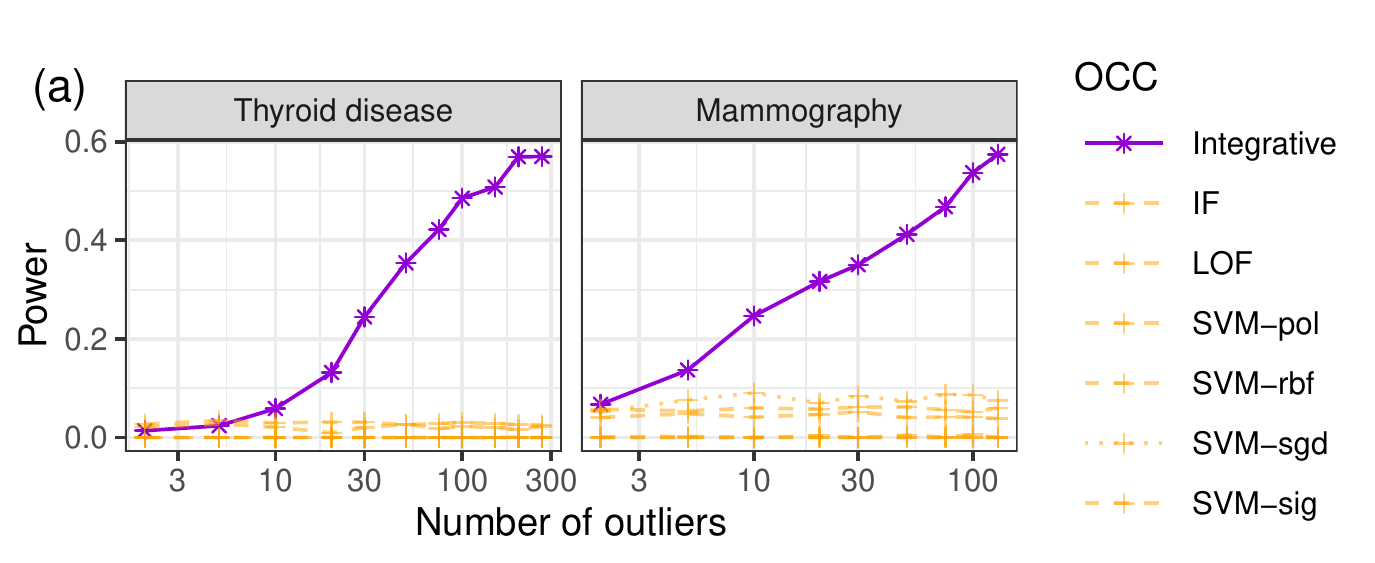}
    \includegraphics[width=0.75\linewidth]{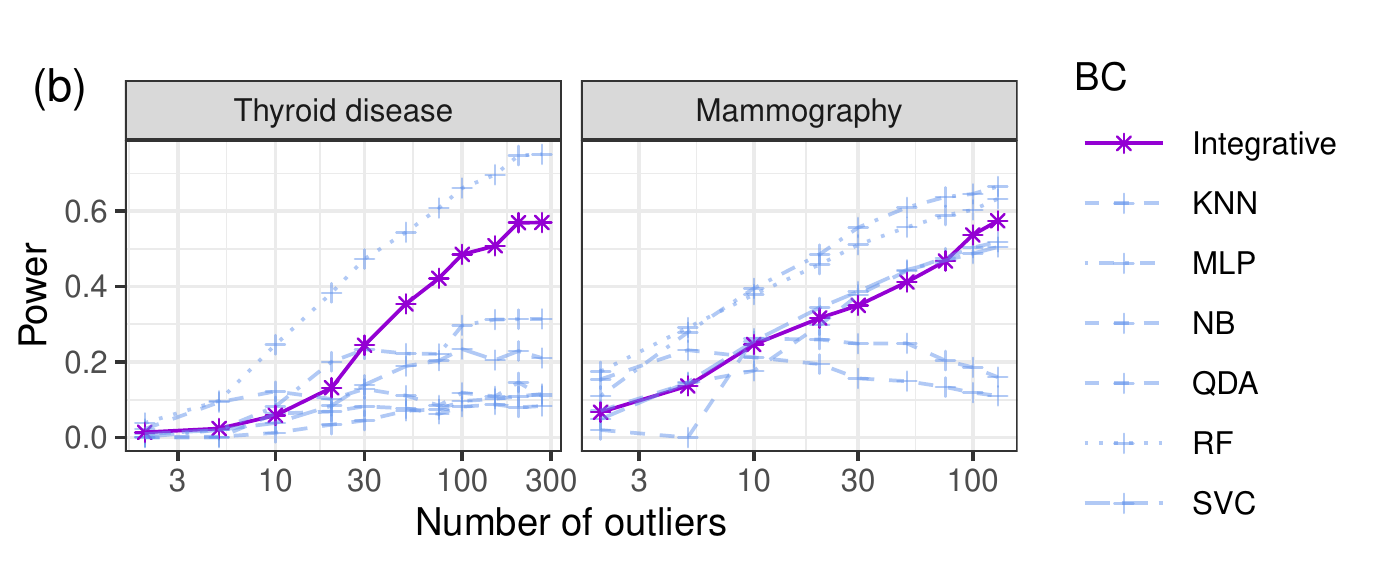}
    \caption{Performance of Storey's BH applied to conformal p-values based on different underlying machine-learning models, on two medical diagnostics data sets. Integrative conformal p-values are compared to standard conformal p-values based on six different one-class classification models (a) and six different binary classification models (b). Other details are as in Figure~\ref{fig:exp-tabular}.}
    \label{fig:exp-flowers-oracle}
    \label{fig:exp-tabular-oracle}
\end{figure}

\end{document}